\documentclass[prl, twocolumn, superscriptaddress]{revtex4-2}
\usepackage{bm, amsmath, amsfonts, amssymb, braket}
\usepackage{subfigure}
\usepackage{color}
\usepackage{graphicx}
\usepackage{dcolumn} 

\usepackage{hyperref}
\usepackage{comment}
\usepackage{color}

\newcommand{\beq}{\begin{equation}}
\newcommand{\eneq}{\end{equation}}

\begin{document}

\title{Anomalous crystal shapes of topological crystalline insulators}

\author{Yutaro Tanaka}
\affiliation{
 Department of Physics, Tokyo Institute of Technology, 2-12-1 Ookayama, Meguro-ku, Tokyo 152-8551, Japan
}
\author{Tiantian Zhang}
\affiliation{
 Department of Physics, Tokyo Institute of Technology, 2-12-1 Ookayama, Meguro-ku, Tokyo 152-8551, Japan
}
\affiliation{
	TIES, Tokyo Institute of Technology, 2-12-1 Ookayama, Meguro-ku, Tokyo 152-8551, Japan
}

\author{Makio Uwaha}
\affiliation{
 Science Division, Center for General Education, Aichi Institute of Technology, 1247 Yachigusa, Yakusa-cho, Toyota, Aichi 470-0392, Japan
}
\author{Shuichi Murakami}
\affiliation{
 Department of Physics, Tokyo Institute of Technology, 2-12-1 Ookayama, Meguro-ku, Tokyo 152-8551, Japan
}
\affiliation{
 TIES, Tokyo Institute of Technology, 2-12-1 Ookayama, Meguro-ku, Tokyo 152-8551, Japan
}

\begin{abstract}
Understanding crystal shapes is a fundamental subject in surface science. It is now well studied how chemical bondings determine crystal shapes via  dependence of surface energies on surface orientations. Meanwhile, discoveries of topological materials have led us to a new paradigm in surface science, and one can expect that topological surface states may affect surface energies and crystal facets in an unconventional way.  
Here we show that  the surface energy of glide-symmetric topological crystalline insulators (TCI) depends on the surface orientation in a singular way via the parity of the Miller index. This singular surface energy of the TCI affects equilibrium crystal shapes, resulting in emergence of unique crystal facets of the TCI. This singular dependence of the topological surface states is unique to the TCI protected by the glide symmetry in contrast to a TCI protected by a mirror symmetry. In addition, we show that such singular surface states of the TCI protected by the glide symmetries can be realized in KHgSb with  first-principles calculations. Our results provide a basis for designs and manipulations of crystal facets by utilizing symmetry and topology.
\end{abstract}
\maketitle


{\it Introduction.}---One of the fascinating phenomena in crystal physics is characteristic crystal facets.
The surface energy and the crystal facets affect morphologies of materials \cite{lovette2008crystal,  auyeung2014dna,  yang2014titania, liu2014titanium}, and therefore they are vital factors in controlling properties of nanomaterials 
\cite{sun2002shape, yang2008anatase, ringe2011wulff, tran2016surface, anderson2017predicting, wang2019crystal}. 
In particular, the surface energies determine equilibrium crystal shapes \cite{wulffconst, wulffconst2, PhysRev.82.87}, which can be realized in nanocrystals \cite{Marks_1994, barmparis2015nanoparticle, Xia_2009}, and the surface energies are mainly determined by chemical bondings in crystals \cite{Haiss_2001, GALANAKIS20021, Ma_2007, Eberhart_2010}. 

We expect that exotic surface states of topological crystalline insulators (TCIs)  \cite{PhysRevLett.106.106802, hsieh2012topological, PhysRevB.90.165114} lead to unconventional contributions to surface energies and to unique crystal facets. 
Although the crystal shapes of TCIs have been observed \cite{li2013single, safdar2014crystal}, it is not well understood how the topological surface states affect the crystal shapes.
Among TCIs with various crystal symmetries \cite{bradlyn2017topological,  po2017symmetry, PhysRevX.7.041069, song2018quantitative, PhysRevX.8.031070,  schindler2018higher, song2019topological, fang2019new, PhysRevB.101.115120, PhysRevResearch.2.013300}, those with  nonsymmorphic symmetries are particularly interesting because of the presence of fractional translations, such as glide mirror and screw rotations  \cite{PhysRevB.91.155120, PhysRevB.91.161105, lu2016symmetry, PhysRevB.93.195138, PhysRevB.96.064102, wieder2018wallpaper, PhysRevB.100.165202, kim2020glide, PhysRevResearch.2.043274}.
Here we focus on a glide-symmetric TCI with time-reversal ($\mathcal{T}$) symmetry \cite{PhysRevB.90.085304, PhysRevB.93.195413, wang2016hourglass}.

In this Letter, we show that the emergence of the topological surface states depends on the surface orientation in a singular way because of the nonsymmorphic nature of the glide symmetry. 
In addition, we obtain equilibrium crystal shapes of the TCI  from the surface energies. 
We discover that the crystal shapes of the TCI are affected by the topological surface states, and the TCI has unique facets, unlike the trivial insulator.

\begin{figure*}
\includegraphics[width=2.\columnwidth]{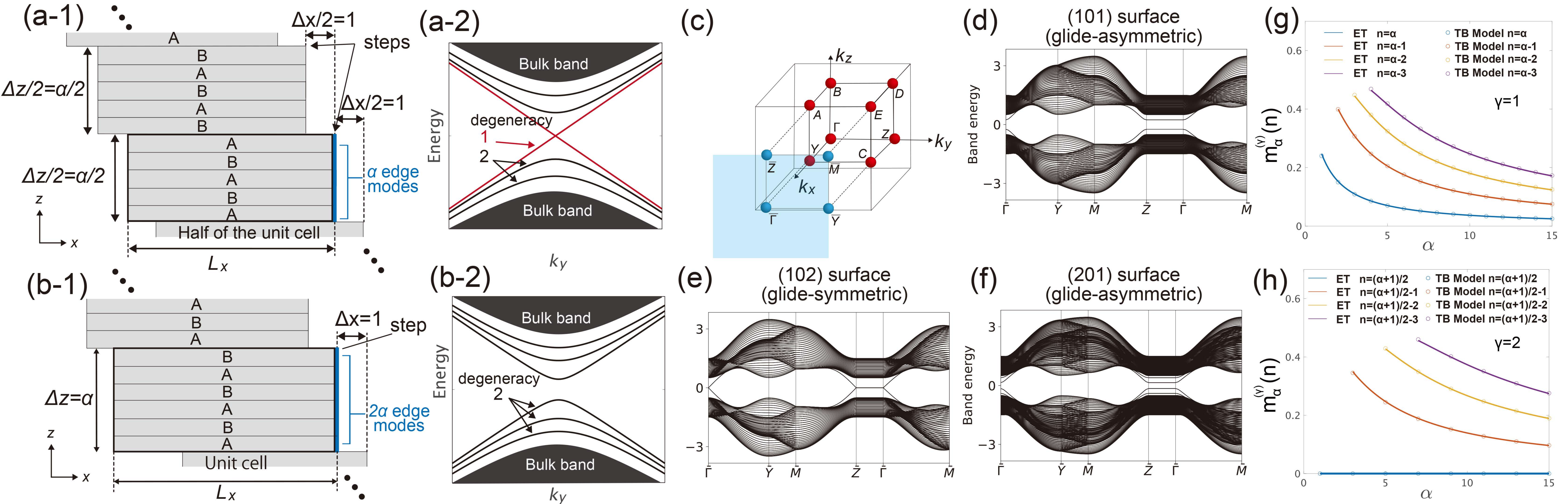}
\caption{
(a,b) The steps of the $(\alpha 0 \gamma)$ surface consisting of the layers $L_{A}$ and $L_{B}$ with $\gamma=2$ in (a-1) and $\gamma=1$ in (b-1), respectively. When $\gamma=2$, the $\alpha$ edge modes in each step lead to the surface bands in (a-2). When $\gamma=1$, the $2\alpha$ edge modes lead to the surface bands in (b-2).  
(c) The bulk Brillouin zone and the (100) surface Brillouin zone. 
(d-f) The band structures of Eq.~(\ref{eq:3dbloch_TCI}) with the parameters  $t_{1}=t_2=t_3=m=1$, $t_{AB}=0.5$, $t'_{AB}=0$. 
(g,h) Comparison of the surface Dirac masses at $\bar{Z}$ point from the effective theory (ET) by Eqs.~(\ref{Diracmass_equaiton_gamma_even}) and (\ref{Diracmass_equaiton_gamma_odd}) (solid lines) with those from the numerical diagonalization of the simple tight-binding (TB) model $\mathcal{H}^{(1)}_{\rm TCI}(\boldsymbol{k})$ (dots), where the parameter is $\delta=0.247$ in Eqs.~(\ref{Diracmass_equaiton_gamma_even}) and (\ref{Diracmass_equaiton_gamma_odd}).
} 
\label{fig:surface_bands_various_Miller_index}
\end{figure*}

{\it Analysis in terms of crystal symmetry.}---Here we consider nonmagnetic TCIs protected by glide symmetry 
$
\hat{G}_{y}=\{ M_{y}|\tfrac{1}{2}\hat{\boldsymbol{z}} \},
$
$i.e.$, a mirror reflection $M_{y}$ with respect to the $xz$ plane followed by translation by a half of a lattice vector $\hat{\boldsymbol{z}}$ along the $z$-direction. 
Henceforth we take the lattice constants to be unity and let $\hat{\boldsymbol{x}}$, $\hat{\boldsymbol{y}}$ and $\hat{\boldsymbol{z}}$ denote the primitive vectors. 
Let us first discuss surfaces with a Miller index $(\alpha \beta \gamma)$,  
which can be written as $\alpha x +\beta y +\gamma z=d$, where $d$ is a constant.
Under $\hat{G}_{y}$, this plane is transformed into
$
\alpha x -\beta y +\gamma( z-\frac{1}{2})=d.
$
The $(\alpha \beta \gamma)$ surface is glide-symmetric, if this plane is identical with a plane $\alpha (x-a)+ \beta (y-b) +\gamma( z-c)=d$, where $a$, $b$, $c$ are integers. 
Therefore, when the Miller index satisfies $\beta=0$ and $\gamma \equiv 0$ (mod 2),
the surface is glide symmetric.
On the other hand, when $\beta=0$ and $\gamma \equiv 1$ (mod 2), the surface is not glide symmetric.

{\it Layer constructions.}---To construct a surface theory, we use layer constructions \cite{song2018quantitative, PhysRevB.100.165202, kim2020glide, song2019topological}, where two-dimensional topological insulator layers are periodically located along the out-of-plane direction. 
Here we consider the simplest layer constructions for a glide-symmetric TCI, which consists of two kinds of layers $L_{A}$ at $z=n$ and $L_{B}$  at $z=n+\tfrac{1}{2}$ ($n$: integer), where $L_A$ and $L_B$ can be interchanged by $\hat{G}_{y}$.
Next, we introduce weak interlayer couplings without closing  the gap, while preserving ${G}_{y}$ symmetry. Through these procedures, we obtain a three-dimensional (3D) TCI phase protected by ${G}_{y}$ symmetry \cite{song2018quantitative, PhysRevB.100.165202, PhysRevB.94.125405, PhysRevX.7.011020, PhysRevB.96.205106}. 

Henceforth, we consider ($\alpha \beta \gamma$) surfaces with $\beta=0$ and $\gamma=1,2$ in order to see differences between glide-symmetric ($\gamma = 2$) and glide-asymmetric ($\gamma = 1$) cases. Here the surface becomes equally spaced steps consisting of the layers $L_{A}$ and $L_{B}$ [Figs.~\ref{fig:surface_bands_various_Miller_index}(a-1) and \ref{fig:surface_bands_various_Miller_index}(b-1)].  We can classify the configurations of the steps into two types in terms of the number of layers in each step.
In the glide-symmetric case with $\gamma=2$, $\alpha$ is odd because $\alpha$ and $\gamma$ should be mutually coprime.
In such a case, a surface step includes $\alpha$ ($=$ odd) layers.  On the other hand, in the glide-asymmetric case with $\gamma=1$, a single step includes $2\alpha$ ($=$ even) layers. The odd (even) number of layers in each step leads to gapless (gapped) states.  

From our effective surface theory (see Supplemental Material No.~1 and No.~7 \cite{sup}), the energy with $\gamma=2$ is given by 
$
E^{\pm}_{n}=\pm \sqrt{v^{2}k^{2}_{y}+(m^{(\gamma=2)}_{\alpha}(n))^2}
$
, where $m^{(\gamma=2)}_{\alpha}(n)$ is a Dirac mass 
\begin{equation}\label{Diracmass_equaiton_gamma_even}
m^{(\gamma=2)}_{\alpha}(n)=2\delta \cos \left( \frac{\pi n}{\alpha +1}\right)\ \ \left( n=1,2,\cdots  \tfrac{\alpha+1}{2} \right),
\end{equation}
with $\delta$ being a real parameter. For the glide-symmetric case with $\gamma=2$, the bands with $n=1,2,\cdots (\alpha-1)/2$ are doubly degenerate, and the band with $n=(\alpha+1)/2$
forms the gapless surface Dirac cone [Fig.~\ref{fig:surface_bands_various_Miller_index}(a-2)]. 
On the other hand, in the glide-asymmetric case with $\gamma=1$, the energy is   
$
E^{\pm}_{n}=\pm \sqrt{v^{2}k^{2}_{y}+( m^{(\gamma=1)}_{\alpha}(n) )^2}
$
\cite{sup}, where the Dirac mass is 
\begin{equation}\label{Diracmass_equaiton_gamma_odd}
m^{(\gamma=1)}_{\alpha}(n)=2\delta \cos \left( \frac{\pi n}{2\alpha +1} \right)\ \ (n=1,2,\cdots , \alpha).
\end{equation} 
In this case, all the bands are doubly degenerate, and gapless states do not appear [Fig.~\ref{fig:surface_bands_various_Miller_index}(b-2)]. 


Next, 
we calculate surface states of the following simple tight-binding model on a simple orthorhombic lattice: 
\begin{align}\label{eq:3dbloch_TCI}
\mathcal{H}^{(1)}_{\rm TCI}(\boldsymbol{k})=&(-m+t_{1} \cos k_{x}+t_{1}\cos k_{y})\mu_{0}\tau_{3}\nonumber \\
&+t_{2}\sin k_y \mu_{3}\tau_{1}\sigma_{1}+t_{3} \sin k_{x} \mu_{0} \tau_{1}\sigma_{3}\nonumber \\
&+(t_{AB}+2t'_{AB} \cos k_{x})\cos \tfrac{k_{z}}{2}\mu_{1}\tau_{0},
\end{align}
where $\sigma_{i}$, $\tau_{i}$, and $\mu_{i}$ $(i=1,2,3)$ are the Pauli matrices, and $\sigma_{0}$, $\tau_{0}$,  and $\mu_{0}$ are the $2\times 2$ identity matrices. 
This model is constructed by stacking layers of the two-dimensional topological insulators, as shown in Supplemental Material No.~2 \cite{sup}, and its topological invariant is given in Supplemental Material No.~3 \cite{sup}.
Figure \ref{fig:surface_bands_various_Miller_index}(c) shows the bulk Brillouin zone  and the (100) surface Brillouin zone. 
In the following, we calculate this model using the PythTB package \cite{PythTB}, and Fig.~\ref{fig:surface_bands_various_Miller_index}(d-f) shows the band structures in the slab geometries. 
For the (101) and (201) surfaces with odd $\gamma $ [Figs.~\ref{fig:surface_bands_various_Miller_index}(d) and \ref{fig:surface_bands_various_Miller_index}(f)], the surface states are gapped. On the other hand, for the (102) surface with even $\gamma$ [Fig.~\ref{fig:surface_bands_various_Miller_index}(e)], the surface states are gapless. 


Next, we quantitatively compare the energy gaps of $\mathcal{H}^{(1)}_{\rm TCI}(\boldsymbol{k})$ with the Dirac mass in Eqs.~(\ref{Diracmass_equaiton_gamma_even}) and (\ref{Diracmass_equaiton_gamma_odd}).
Figures~\ref{fig:surface_bands_various_Miller_index}(g) and \ref{fig:surface_bands_various_Miller_index}(h) show a half of the energy gap at $\bar{Z}$ point for various surface bands.
This value is to be compared with the Dirac mass in Eqs.~(\ref{Diracmass_equaiton_gamma_even}) and (\ref{Diracmass_equaiton_gamma_odd}). 
At $\bar{Z}$ point, the $i$-th energy closest to the zero energy is given by $m_{\alpha}^{(\gamma =2)}(n)$ with $n=(\alpha+1)/2-(i-1)$ when $\gamma=2$. When $\gamma=1$, it is given by  $m_{\alpha}^{(\gamma=1)}(n)$ with $n=\alpha -(i-1)$. 
The values of the Dirac mass perfectly agree with those from  $\mathcal{H}^{(1)}_{\rm TCI}(\boldsymbol{k})$ [Figs.~\ref{fig:surface_bands_various_Miller_index}(g) and \ref{fig:surface_bands_various_Miller_index}(h)]. 
The behaviors of the surface states discussed so far for the nonmagnetic TCI also hold true in the magnetic TCI (Supplemental Material No.~4 \cite{sup}).

{\it Crystal shape of TCI.}---From the surface energies $E_{\rm surf}^{(\alpha \beta \gamma)}$, we can calculate the equilibrium crystal shape of the TCI by using the Wulff construction \cite{wulffconst,sup}.
The model $\mathcal{H}^{(1)}_{\rm TCI} (\boldsymbol{k})$ is a minimal model for understanding the behaviors of the topological surface states.
 On the other hand, the surface energies of this model are quite anisotropic, which leads to the crystal shape with a very small thickness in the $z$-direction (see Supplemental Material No.~3 \cite{sup}), and it is difficult to see the effects of the topological surface states on the crystal shape. 
 Thus, we need  another model with a more isotropic crystal shape and introduce the following tight-binding model on a simple orthorhombic lattice:
\begin{align}\label{eq:second_model_TCI}
		\mathcal{H}^{(2)}_{\rm TCI}(\boldsymbol{k})=&\biggl( -m+\sum_{j=x,y,z}t_{j}\cos k_{j} \biggr) \mu_{0} \tau_{3}\nonumber \\
		&+v_{x}\sin k_{x} \mu_{0}\tau_{1}\sigma_{3}+v_{y}\sin k_{y}\mu_{3}\tau_{1}\sigma_{1} \nonumber \\
		&+v_{z} \sin k_{z} \mu_{0}\tau_{1}\sigma_{2}\nonumber \\
		&+(v_{ab1}+v_{ab2}\cos k_{x})\sin \tfrac{k_{z}}{2} \mu_{1}\tau_{1}\sigma_{1}.
\end{align}
Both $\mathcal{H}^{(1)}_{\rm TCI}(\boldsymbol{k})$ and $\mathcal{H}^{(2)}_{\rm TCI}(\boldsymbol{k})$ have the same TCI phase, and the similar dependence of the surface states on the surface orientations (see Supplemental Material No.~5 \cite{sup}). 

\begin{figure}
	\includegraphics[width=1.\columnwidth]{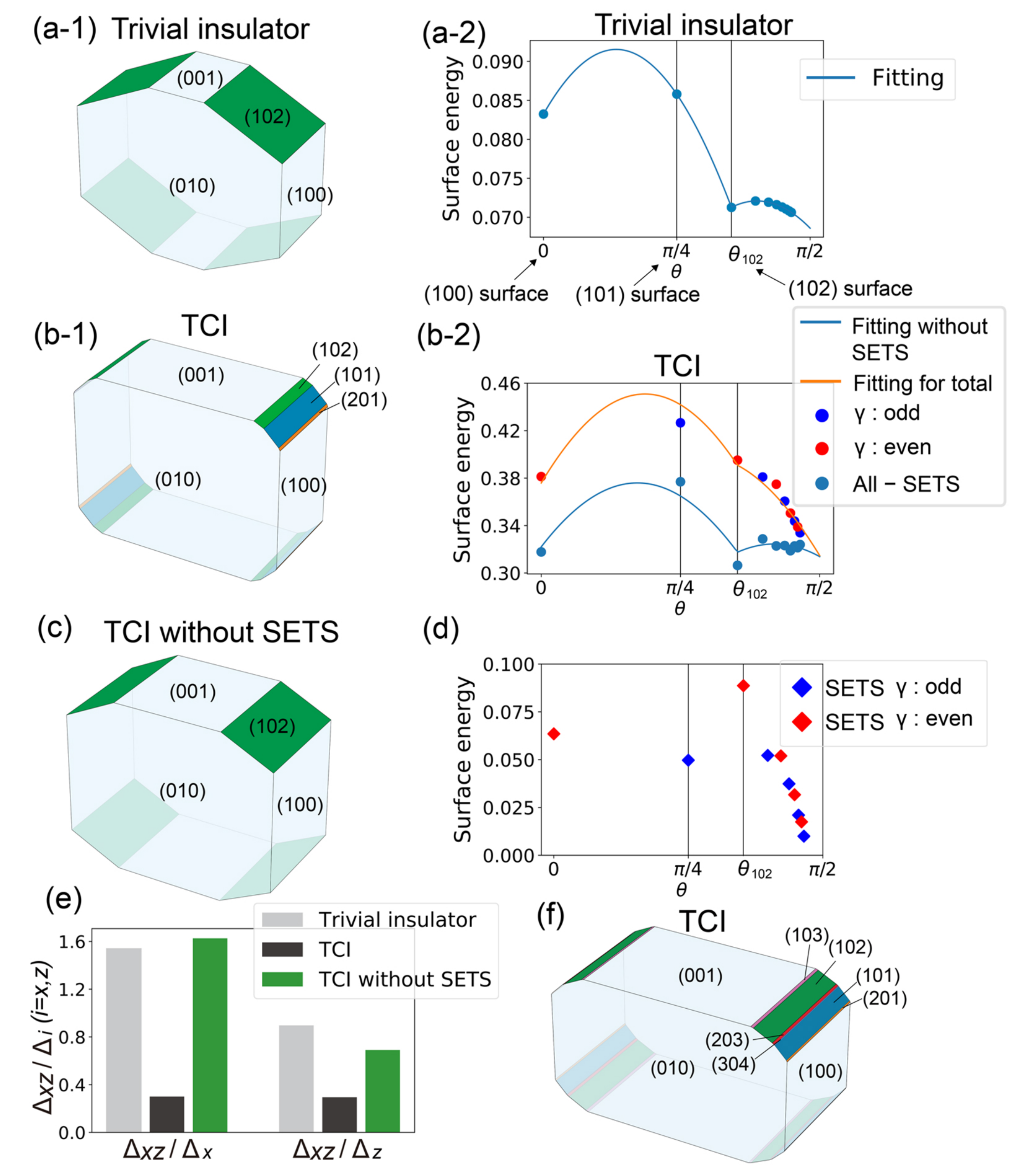}
	\caption{(a,b) Equilibrium crystal shapes and surface energies of $\mathcal{H}^{(2)}_{\rm TCI}(\boldsymbol{k})$ with the parameters $t_{x}=t_{y}=t_{z}=1$, $v_{ab1}=0.8$, $v_{ab2}=1.2$. The other parameters are $m=6$, $v_{x}=v_{z}=0.4$, $v_{y}=0.6$ in (a) and $m=2$, $v_{x}=v_{y}=v_{z}=0.4$ in (b). The $(10\gamma)$ surface energy from $\gamma=0$ to $\gamma=9$. 
The dots are determined by $E^{(\alpha \beta \gamma)}_{\rm surf}$. (a-2) The surface energy can be fitted with $F(\theta)$ with $\Delta_x = 0.02218$, $\Delta_z = 0.03806$, and $\Delta_{xz}=0.03414$. (b-2) We fit the total surface energy (orange) and the total surface energy without the SETS (blue) with $F(\theta)$ with $\Delta_{x}=0.2445$, $\Delta_{z}=0.2490$, $\Delta_{xz}=0.07343$ in the former case and $\Delta_{x}=0.08238$, $\Delta_{z}=0.1940$, $\Delta_{xz}=0.1340$ in the latter case. 
 (c) The equilibrium crystal shape from the surface energies of (b-2) without the SETS.  
(d) The SETS in the TCI. (e) Comparison of $\Delta_{xz}/\Delta_{i}$ ($i=x,z$) for the trivial insulator in (a-2), the TCI [orange fitting in (b-2)], and the TCI without the SETS [blue fitting in (b-2)].
 (f) The equilibrium crystal shape when $\mathcal{H}^{(2)}_{\rm TCI}(\boldsymbol{k})$ has the more distant hopping term with $m=2$, $t_{x}=t_{y}=t_{z}=1$, $v_{x}=0.4$, $v_{y}=0.6$, $v_{z}=0.2$, $v_{ab1}=0.8$, $v_{ab2}=0.9$, and $v_{ab3}=0.8$. }
\label{fig:surface_energy_Wulff_const}
\end{figure} 

We calculate the surface energies $E_{\rm surf}^{(0 1 0)}$ and $E_{\rm surf}^{(\alpha 0 \gamma)} $ of this model for various surface orientations up to  maximum absolute values of the Miller index $(\alpha_{\rm max}, \gamma_{\rm max})=(3, 9)$, where $E_{\rm surf}^{(\alpha \beta \gamma)} $  is defined by Eq.~(S.3) in the Supplemental Material No.~1 \cite{sup}.
According to the Wulff construction \cite{wulffconst}, we can obtain the equilibrium crystal shape minimizing the total surface energy 
as the following 3D region:
 \begin{gather}
 	\mathcal{W}=\bigcap_{\boldsymbol{n}_{\alpha \beta \gamma} \in S^{2}} \Gamma_{\boldsymbol{n}_{\alpha \beta \gamma}}, \\
 	\Gamma_{\boldsymbol{n}_{\alpha \beta \gamma}}=\left\{ \boldsymbol{r} \in \mathbb{R}^{3}\ |\  \boldsymbol{r} \cdot \boldsymbol{n}_{\alpha \beta \gamma} \leq E^{(\alpha \beta \gamma)}_{\rm surf} \right\},
 \end{gather}
 where $\boldsymbol{n}_{\alpha \beta \gamma}$ is the outward unit normal vector to the $(\alpha \beta \gamma)$ surface, and $S^{2}$ is the unit sphere. By using the WulffPack  package \cite{WulffPack}, we obtain this shape from $E_{\rm surf}^{(\alpha \beta \gamma)}$. 
Figures \ref{fig:surface_energy_Wulff_const}(a-1) and \ref{fig:surface_energy_Wulff_const}(b-1) show the equilibrium crystal shapes of $\mathcal{H}^{(2)}_{\rm TCI}(\boldsymbol{k})$ with the trivial insulator phase and the TCI phase, respectively.
We show the dependence of the surface energies on the surface orientations as a function of the angle $\theta$ between the surface and the (100) plane, defined by $\tan \theta=\gamma / \alpha$
[Figs.~\ref{fig:surface_energy_Wulff_const}(a-2) and \ref{fig:surface_energy_Wulff_const}(b-2)]. 
To analyze the results, we introduce the following trial function:
\begin{align}\label{eq:fitting_secb}
	F(\theta)=&\Delta_{x}\cos|\theta|+\Delta_{z}\sin|\theta|\nonumber \\
	&+\Delta_{xz}\sin|\theta-\theta_{102}|+\Delta_{xz}\sin|\theta+\theta_{102}|,
\end{align}
where $\theta_{102}$ is the angle satisfying $\cos \theta_{102}=1/\sqrt{5}$ and $\sin \theta_{102}=2/\sqrt{5}$.  
This function can be obtained by the analysis in terms of the numbers of dangling  bonds on the surface
(see Supplemental Material No.~6 \cite{sup}).
The surface energy of the trivial insulator can be fitted perfectly with $F(\theta)$ [Fig.~\ref{fig:surface_energy_Wulff_const}(a-2)], and therefore the facets of the trivial insulator are determined mainly by the surface energy from chemical bonding (SECB).

Next, we discuss the surface energy of the TCI in Fig.~\ref{fig:surface_energy_Wulff_const}(b-2).
The appearance of the (101) and the (201) facets in Fig.~\ref{fig:surface_energy_Wulff_const}(b-1) suggests a new mechanism other than the SECB.  
Here, we  attribute it to the surface energy obtained only from the bands forming the Dirac cones. We refer to this as the surface energy from topological surface states (SETS), which is defined by Eq.~(S.4) in Supplemental Material No.~1 \cite{sup}.
Figure~\ref{fig:surface_energy_Wulff_const}(b-2) also shows the total energy minus the SETS.
We fit the total surface energy and the total surface energy without the SETS with Eq.~(\ref{eq:fitting_secb}), and in the latter fitting a sharp dip at $\theta=\theta_{102}$ appears, which is similar to the surface energy in the trivial insulator in Fig.~\ref{fig:surface_energy_Wulff_const}(b-1).
In addition, we calculate the equilibrium crystal shape from the surface energies of the TCI without the SETS [Fig.~\ref{fig:surface_energy_Wulff_const}(c)]. This result shows that the (101) and the (201) facets are due to the SETS.
Figure~\ref{fig:surface_energy_Wulff_const}(d) also shows the SETS in the TCI. 

We confirm the above interpretation by analyzing the fitting data for $\Delta_{x}$, $\Delta_{z}$, and $\Delta_{xz}$. 
Figure~\ref{fig:surface_energy_Wulff_const}(e) shows comparison of $\Delta_{xz}/\Delta_{i}$ ($i=x,z$) for the cases corresponding to Figs.~\ref{fig:surface_energy_Wulff_const}(a-1), \ref{fig:surface_energy_Wulff_const}(b-1), and \ref{fig:surface_energy_Wulff_const}(c). 
In the trivial insulator, $\Delta_{x}$, $\Delta_{z}$, and $\Delta_{xz}$ are of a similar order of magnitude, corresponding to hoppings almost  isotropically distributed in the $xz$ plane. 
In the TCI phase, the hopping parameters used in our calculation are almost the same as the trivial insulator phase, but unexpectedly, $\Delta_{x}$, $\Delta_{z}$, and $\Delta_{xz}$ become quite anisotropic [Fig.~\ref{fig:surface_energy_Wulff_const}(e)]. 
We see from Fig.~\ref{fig:surface_energy_Wulff_const}(e) that it restores the isotropic behavior by subtracting the SETS. A more detailed discussion is in Supplemental Material No.~5 \cite{sup}.
Thus, we conclude that the unique crystal shape of the TCI  are due to the SETS. 
Figure~\ref{fig:surface_energy_Wulff_const}(f) also shows the crystal shape when we add a more distant hopping term $v_{ab3}\cos 2k_{x} \sin \tfrac{k_{z}}{2} \mu_1 \tau_1 \sigma_1 $ to $\mathcal{H}^{(2)}_{\rm TCI}(\boldsymbol{k})$. In this case, the  additional facets appear because of the interplay between the SETS and the SECB from the bonds in various directions.  

\begin{figure}
	\includegraphics[width=1.0\columnwidth]{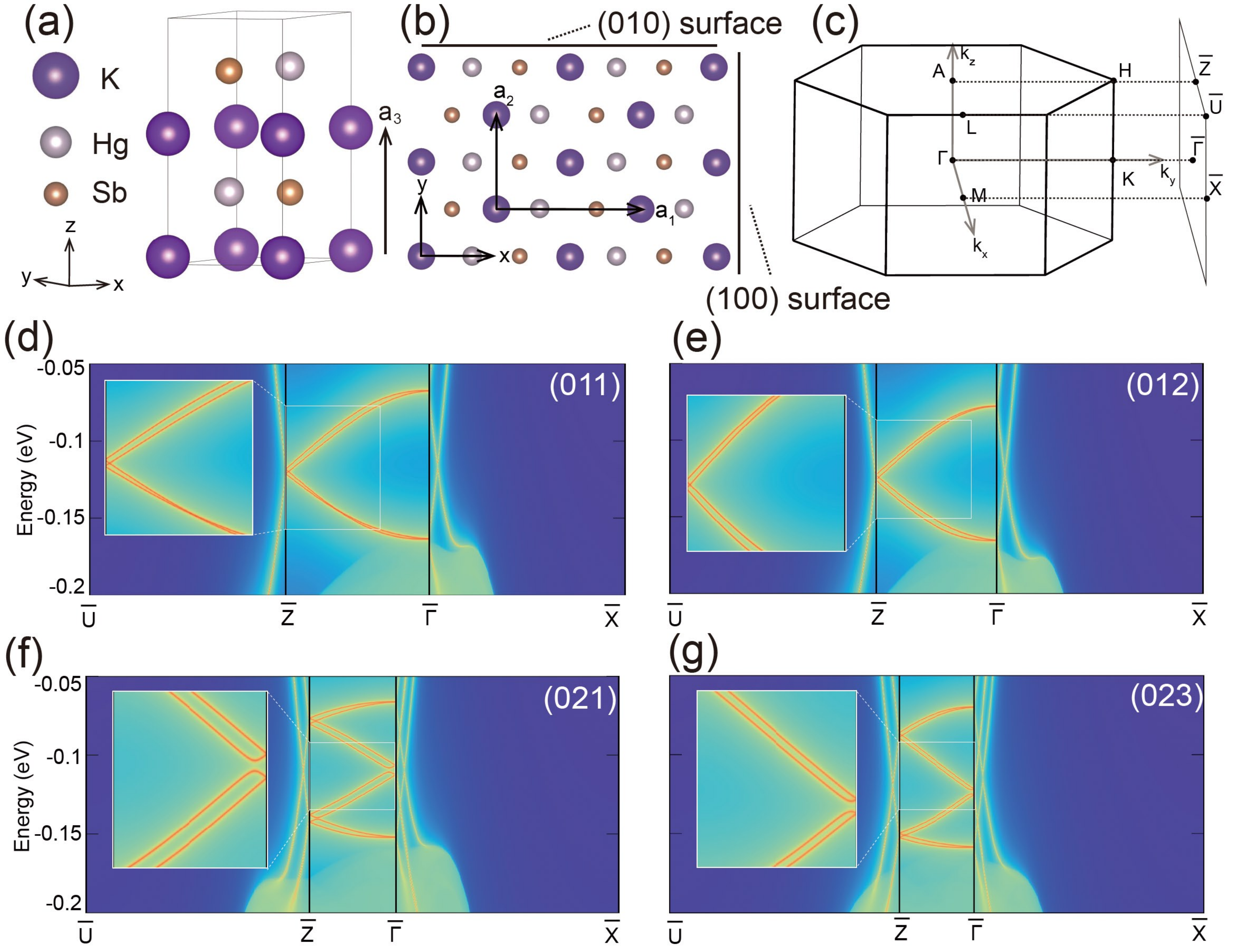}
	\caption{
	(a,b) Crystal structure of KHgSb. (c) The bulk Brillouin zone and the (010) surface Brillouin zone for KHgSb with space group \#186.  (d-g) Surface states of KHgSb for various surfaces. 
}\label{ss}
\end{figure}

{\it Material realization.}---Here we show that these singular surface states can be realized in KHgSb proposed as a $\hat{G}_{x}$-symmetric TCI with space group \#194 
\cite{wang2016hourglass}, where $\hat{G}_{x}=\{  M_{x}|\tfrac{1}{2}\hat{\boldsymbol{z}} \}$ with $M_{x}$ being a mirror reflection with the $yz$ mirror plane. 
It 
has a phase transition to \#186 when $T$ $<$ 150K \cite{chen2017robustness}, and here
we will discuss the low-temperature phase with \#186. 
Figures \ref{ss}(a) and \ref{ss}(b) show the crystal structures of KHgSb. 
Figure \ref{ss}(c) is the Brillouin zone and the (010) surface Brillouin zone.  
To make it easier to see whether the surfaces are glide invariant or not, we double the original hexagonal unit cell and take an orthorhombic unit cell with the lattice vectors $\boldsymbol{a}_{1}$=$(A,0,0)$, $\boldsymbol{a}_{2}=(0,B,0)$, and $\boldsymbol{a}_{3}=(0,0,C)$, where $A$, $B$, and $C$ are lattice constants.
The enlarged unit cell leads to a translation symmetry given by $\hat{T}=\{E|\tfrac{1}{2}\boldsymbol{a}_{1}+ \tfrac{1}{2} \boldsymbol{a}_{2} \}$.
The index $(\alpha \beta \gamma)$ in this orthorhombicc lattice is different from the conventional Miller index in hexagonal crystals. 

Next, we consider  the conditions for the $(\alpha \beta \gamma)$ surface to be glide symmetric in these symmetry settings. Let a plane $P$ be the $(\alpha \beta \gamma)$ plane. 
Under $\hat{G}_{x}$, 
the plane $P$ is transformed into 
$\tilde{P}$. 
The plane $P$ is glide symmetric when $\tilde{P}$ is equivalent to either $P$ or 
$P'$, where $P'$ is the plane transformed from $P$ via $\hat{T}$. In the former case, the index satisfies (i) $\alpha=0$ and $\gamma \equiv 0$ (mod 2), and in the latter case, the index satisfies (ii) $\alpha= 0$ and $\beta -\gamma \equiv 0$ (mod 2). When the index satisfies (i) or (ii), the $(\alpha \beta \gamma)$ surface is symmetric under  the glide $\hat{G}_{x}$ symmetry. 
 
The surface states on four different surfaces are shown in Fig.~\ref{ss}(d-g), which have gapless hourglass surface states for the (011) and (012) surfaces because of the presence of the glide symmetry, but gapped surface states for the (021) and (023) surfaces because of the absence of the glide symmetry. These results are consistent with our analysis in terms of symmetries in the previous paragraph and can be observed by angle-resolved photoemission spectroscopy. Because these results are due to the glide symmetry,  a similar result is expected in the high-temperature phase with space group \#194. On the other hand, the gaps of the (021) and the (023) surface states are 2.4meV and 2.6meV, respectively. Therefore, the surface gap is smaller than temperatures in crystal growth, and the effects of these small band gaps cannot be reflected in crystal shapes in this material. 

The smallness of the gaps on the glide-asymmetric surface states in KHgSb is attributed to weak interlayer coupling of KHgSb \cite{chen2017robustness}.
The gaps on the glide-asymmetric surfaces are proportional to the strength of the interlayer couplings, as seen from the calculation of $\mathcal{H}_{\rm TCI}^{(1)}(\boldsymbol{k})$. Thus, the glide-symmetric TCIs with stronger interlayer coupling are suitable for experiments. In addition, nanocrystals are suitable for experimental realization of equilibrium crystal shapes.

\textit{Summary and Discussion.}---
Here we discuss another type of TCI, $i.e.$ a TCI with surface states protected by a mirror $M_{y}$ symmetry.
The surface states can affect the equilibrium crystal shapes. On the other hand, in this letter, we theoretically show novel dependence of presence/absence of gapless topological surface states on the parity of the surface Miller index.  This is unique to the TCI protected by glide symmetry. As we show in Supplemental Material No.~8 \cite{sup}, the gapless surface states survive surface reconstructions. 
We discovered that the topological surface  states significantly affect the facets realized in the TCI by calculations of the model.  
The facets of the TCI can be determined by the interplay between the surface energy from chemical bonding and that from the topological surface states. 

\begin{acknowledgments}
\textit{Acknowledgments.}---
We thank Toru Hirahara, Ryo Okugawa, Kazuki Yokomizo, and Ryo Takahashi for fruitful discussions. This work was supported by Japan Society for the Promotion of Science (JSPS) KAKENHI Grants No.~JP18H03678, No.~JP18K03500, No.~JP20H04633, No.~JP21J22264, No.~JP21K13865 and JP22H00108, and by Elements Strategy Initiative to Form Core Research Center (TIES), from MEXT Grant No. JPMXP0112101001. 
\end{acknowledgments}


\begin{thebibliography}{54}%
\makeatletter
\providecommand \@ifxundefined [1]{%
 \@ifx{#1\undefined}
}%
\providecommand \@ifnum [1]{%
 \ifnum #1\expandafter \@firstoftwo
 \else \expandafter \@secondoftwo
 \fi
}%
\providecommand \@ifx [1]{%
 \ifx #1\expandafter \@firstoftwo
 \else \expandafter \@secondoftwo
 \fi
}%
\providecommand \natexlab [1]{#1}%
\providecommand \enquote  [1]{``#1''}%
\providecommand \bibnamefont  [1]{#1}%
\providecommand \bibfnamefont [1]{#1}%
\providecommand \citenamefont [1]{#1}%
\providecommand \href@noop [0]{\@secondoftwo}%
\providecommand \href [0]{\begingroup \@sanitize@url \@href}%
\providecommand \@href[1]{\@@startlink{#1}\@@href}%
\providecommand \@@href[1]{\endgroup#1\@@endlink}%
\providecommand \@sanitize@url [0]{\catcode `\\12\catcode `\$12\catcode
  `\&12\catcode `\#12\catcode `\^12\catcode `\_12\catcode `\%12\relax}%
\providecommand \@@startlink[1]{}%
\providecommand \@@endlink[0]{}%
\providecommand \url  [0]{\begingroup\@sanitize@url \@url }%
\providecommand \@url [1]{\endgroup\@href {#1}{\urlprefix }}%
\providecommand \urlprefix  [0]{URL }%
\providecommand \Eprint [0]{\href }%
\providecommand \doibase [0]{https://doi.org/}%
\providecommand \selectlanguage [0]{\@gobble}%
\providecommand \bibinfo  [0]{\@secondoftwo}%
\providecommand \bibfield  [0]{\@secondoftwo}%
\providecommand \translation [1]{[#1]}%
\providecommand \BibitemOpen [0]{}%
\providecommand \bibitemStop [0]{}%
\providecommand \bibitemNoStop [0]{.\EOS\space}%
\providecommand \EOS [0]{\spacefactor3000\relax}%
\providecommand \BibitemShut  [1]{\csname bibitem#1\endcsname}%
\let\auto@bib@innerbib\@empty
\bibitem [{\citenamefont {Lovette}\ \emph {et~al.}(2008)\citenamefont
  {Lovette}, \citenamefont {Browning}, \citenamefont {Griffin}, \citenamefont
  {Sizemore}, \citenamefont {Snyder},\ and\ \citenamefont
  {Doherty}}]{lovette2008crystal}%
  \BibitemOpen
  \bibfield  {author} {\bibinfo {author} {\bibfnamefont {M.~A.}\ \bibnamefont
  {Lovette}}, \bibinfo {author} {\bibfnamefont {A.~R.}\ \bibnamefont
  {Browning}}, \bibinfo {author} {\bibfnamefont {D.~W.}\ \bibnamefont
  {Griffin}}, \bibinfo {author} {\bibfnamefont {J.~P.}\ \bibnamefont
  {Sizemore}}, \bibinfo {author} {\bibfnamefont {R.~C.}\ \bibnamefont
  {Snyder}},\ and\ \bibinfo {author} {\bibfnamefont {M.~F.}\ \bibnamefont
  {Doherty}},\ }\bibfield  {title} {\bibinfo {title} {{Crystal shape
  engineering}},\ }\href {https://pubs.acs.org/doi/abs/10.1021/ie800900f}
  {\bibfield  {journal} {\bibinfo  {journal} {Ind. \& Eng. Chem. Res.}\
  }\textbf {\bibinfo {volume} {47}},\ \bibinfo {pages} {9812} (\bibinfo {year}
  {2008})}\BibitemShut {NoStop}%
\bibitem [{\citenamefont {Auyeung}\ \emph {et~al.}(2014)\citenamefont
  {Auyeung}, \citenamefont {Li}, \citenamefont {Senesi}, \citenamefont
  {Schmucker}, \citenamefont {Pals}, \citenamefont {de~La~Cruz},\ and\
  \citenamefont {Mirkin}}]{auyeung2014dna}%
  \BibitemOpen
  \bibfield  {author} {\bibinfo {author} {\bibfnamefont {E.}~\bibnamefont
  {Auyeung}}, \bibinfo {author} {\bibfnamefont {T.~I.}\ \bibnamefont {Li}},
  \bibinfo {author} {\bibfnamefont {A.~J.}\ \bibnamefont {Senesi}}, \bibinfo
  {author} {\bibfnamefont {A.~L.}\ \bibnamefont {Schmucker}}, \bibinfo {author}
  {\bibfnamefont {B.~C.}\ \bibnamefont {Pals}}, \bibinfo {author}
  {\bibfnamefont {M.~O.}\ \bibnamefont {de~La~Cruz}},\ and\ \bibinfo {author}
  {\bibfnamefont {C.~A.}\ \bibnamefont {Mirkin}},\ }\bibfield  {title}
  {\bibinfo {title} {{DNA-mediated nanoparticle crystallization into Wulff
  polyhedra}},\ }\href {https://www.nature.com/articles/nature12739} {\bibfield
   {journal} {\bibinfo  {journal} {Nature}\ }\textbf {\bibinfo {volume}
  {505}},\ \bibinfo {pages} {73} (\bibinfo {year} {2014})}\BibitemShut
  {NoStop}%
\bibitem [{\citenamefont {Yang}\ \emph {et~al.}(2014)\citenamefont {Yang},
  \citenamefont {Yang}, \citenamefont {Wu}, \citenamefont {Li}, \citenamefont
  {Liu}, \citenamefont {Zhao}, \citenamefont {Yu}, \citenamefont {Gong},\ and\
  \citenamefont {Yang}}]{yang2014titania}%
  \BibitemOpen
  \bibfield  {author} {\bibinfo {author} {\bibfnamefont {S.}~\bibnamefont
  {Yang}}, \bibinfo {author} {\bibfnamefont {B.~X.}\ \bibnamefont {Yang}},
  \bibinfo {author} {\bibfnamefont {L.}~\bibnamefont {Wu}}, \bibinfo {author}
  {\bibfnamefont {Y.~H.}\ \bibnamefont {Li}}, \bibinfo {author} {\bibfnamefont
  {P.}~\bibnamefont {Liu}}, \bibinfo {author} {\bibfnamefont {H.}~\bibnamefont
  {Zhao}}, \bibinfo {author} {\bibfnamefont {Y.~Y.}\ \bibnamefont {Yu}},
  \bibinfo {author} {\bibfnamefont {X.~Q.}\ \bibnamefont {Gong}},\ and\
  \bibinfo {author} {\bibfnamefont {H.~G.}\ \bibnamefont {Yang}},\ }\bibfield
  {title} {\bibinfo {title} {{Titania single crystals with a curved surface}},\
  }\href {https://www.nature.com/articles/ncomms6355} {\bibfield  {journal}
  {\bibinfo  {journal} {Nat. Commun.}\ }\textbf {\bibinfo {volume} {5}},\
  \bibinfo {pages} {5355} (\bibinfo {year} {2014})}\BibitemShut {NoStop}%
\bibitem [{\citenamefont {Liu}\ \emph {et~al.}(2014{\natexlab{a}})\citenamefont
  {Liu}, \citenamefont {Yang}, \citenamefont {Pan}, \citenamefont {Yang},
  \citenamefont {Lu},\ and\ \citenamefont {Cheng}}]{liu2014titanium}%
  \BibitemOpen
  \bibfield  {author} {\bibinfo {author} {\bibfnamefont {G.}~\bibnamefont
  {Liu}}, \bibinfo {author} {\bibfnamefont {H.~G.}\ \bibnamefont {Yang}},
  \bibinfo {author} {\bibfnamefont {J.}~\bibnamefont {Pan}}, \bibinfo {author}
  {\bibfnamefont {Y.~Q.}\ \bibnamefont {Yang}}, \bibinfo {author}
  {\bibfnamefont {G.~Q.}\ \bibnamefont {Lu}},\ and\ \bibinfo {author}
  {\bibfnamefont {H.-M.}\ \bibnamefont {Cheng}},\ }\bibfield  {title} {\bibinfo
  {title} {{Titanium dioxide crystals with tailored facets}},\ }\href
  {https://pubs.acs.org/doi/10.1021/cr400621z} {\bibfield  {journal} {\bibinfo
  {journal} {Chem. Rev.}\ }\textbf {\bibinfo {volume} {114}},\ \bibinfo {pages}
  {9559} (\bibinfo {year} {2014}{\natexlab{a}})}\BibitemShut {NoStop}%
\bibitem [{\citenamefont {Sun}\ and\ \citenamefont {Xia}(2002)}]{sun2002shape}%
  \BibitemOpen
  \bibfield  {author} {\bibinfo {author} {\bibfnamefont {Y.}~\bibnamefont
  {Sun}}\ and\ \bibinfo {author} {\bibfnamefont {Y.}~\bibnamefont {Xia}},\
  }\bibfield  {title} {\bibinfo {title} {{Shape-controlled synthesis of gold
  and silver nanoparticles}},\ }\href
  {https://www.science.org/doi/full/10.1126/science.1077229} {\bibfield
  {journal} {\bibinfo  {journal} {Science}\ }\textbf {\bibinfo {volume}
  {298}},\ \bibinfo {pages} {2176} (\bibinfo {year} {2002})}\BibitemShut
  {NoStop}%
\bibitem [{\citenamefont {Yang}\ \emph {et~al.}(2008)\citenamefont {Yang},
  \citenamefont {Sun}, \citenamefont {Qiao}, \citenamefont {Zou}, \citenamefont
  {Liu}, \citenamefont {Smith}, \citenamefont {Cheng},\ and\ \citenamefont
  {Lu}}]{yang2008anatase}%
  \BibitemOpen
  \bibfield  {author} {\bibinfo {author} {\bibfnamefont {H.~G.}\ \bibnamefont
  {Yang}}, \bibinfo {author} {\bibfnamefont {C.~H.}\ \bibnamefont {Sun}},
  \bibinfo {author} {\bibfnamefont {S.~Z.}\ \bibnamefont {Qiao}}, \bibinfo
  {author} {\bibfnamefont {J.}~\bibnamefont {Zou}}, \bibinfo {author}
  {\bibfnamefont {G.}~\bibnamefont {Liu}}, \bibinfo {author} {\bibfnamefont
  {S.~C.}\ \bibnamefont {Smith}}, \bibinfo {author} {\bibfnamefont {H.~M.}\
  \bibnamefont {Cheng}},\ and\ \bibinfo {author} {\bibfnamefont {G.~Q.}\
  \bibnamefont {Lu}},\ }\bibfield  {title} {\bibinfo {title} {{Anatase
  ${\mathrm{TiO}}_{2}$ single crystals with a large percentage of reactive
  facets}},\ }\href {https://www.nature.com/articles/nature06964} {\bibfield
  {journal} {\bibinfo  {journal} {Nature}\ }\textbf {\bibinfo {volume} {453}},\
  \bibinfo {pages} {638} (\bibinfo {year} {2008})}\BibitemShut {NoStop}%
\bibitem [{\citenamefont {Ringe}\ \emph {et~al.}(2011)\citenamefont {Ringe},
  \citenamefont {Van~Duyne},\ and\ \citenamefont {Marks}}]{ringe2011wulff}%
  \BibitemOpen
  \bibfield  {author} {\bibinfo {author} {\bibfnamefont {E.}~\bibnamefont
  {Ringe}}, \bibinfo {author} {\bibfnamefont {R.~P.}\ \bibnamefont
  {Van~Duyne}},\ and\ \bibinfo {author} {\bibfnamefont {L.}~\bibnamefont
  {Marks}},\ }\bibfield  {title} {\bibinfo {title} {{Wulff construction for
  alloy nanoparticles}},\ }\href {https://pubs.acs.org/doi/10.1021/nl2018146}
  {\bibfield  {journal} {\bibinfo  {journal} {Nano Lett.}\ }\textbf {\bibinfo
  {volume} {11}},\ \bibinfo {pages} {3399} (\bibinfo {year}
  {2011})}\BibitemShut {NoStop}%
\bibitem [{\citenamefont {Tran}\ \emph {et~al.}(2016)\citenamefont {Tran},
  \citenamefont {Xu}, \citenamefont {Radhakrishnan}, \citenamefont {Winston},
  \citenamefont {Sun}, \citenamefont {Persson},\ and\ \citenamefont
  {Ong}}]{tran2016surface}%
  \BibitemOpen
  \bibfield  {author} {\bibinfo {author} {\bibfnamefont {R.}~\bibnamefont
  {Tran}}, \bibinfo {author} {\bibfnamefont {Z.}~\bibnamefont {Xu}}, \bibinfo
  {author} {\bibfnamefont {B.}~\bibnamefont {Radhakrishnan}}, \bibinfo {author}
  {\bibfnamefont {D.}~\bibnamefont {Winston}}, \bibinfo {author} {\bibfnamefont
  {W.}~\bibnamefont {Sun}}, \bibinfo {author} {\bibfnamefont {K.~A.}\
  \bibnamefont {Persson}},\ and\ \bibinfo {author} {\bibfnamefont {S.~P.}\
  \bibnamefont {Ong}},\ }\bibfield  {title} {\bibinfo {title} {{Surface
  energies of elemental crystals}},\ }\href
  {https://www.nature.com/articles/sdata201680} {\bibfield  {journal} {\bibinfo
   {journal} {Sci. Data}\ }\textbf {\bibinfo {volume} {3}},\ \bibinfo {pages}
  {160080} (\bibinfo {year} {2016})}\BibitemShut {NoStop}%
\bibitem [{\citenamefont {Anderson}\ \emph {et~al.}(2017)\citenamefont
  {Anderson}, \citenamefont {Gebbie-Rayet}, \citenamefont {Hill}, \citenamefont
  {Farida}, \citenamefont {Attfield}, \citenamefont {Cubillas}, \citenamefont
  {Blatov}, \citenamefont {Proserpio}, \citenamefont {Akporiaye}, \citenamefont
  {Arstad},\ and\ \citenamefont {Gale}}]{anderson2017predicting}%
  \BibitemOpen
  \bibfield  {author} {\bibinfo {author} {\bibfnamefont {M.~W.}\ \bibnamefont
  {Anderson}}, \bibinfo {author} {\bibfnamefont {J.~T.}\ \bibnamefont
  {Gebbie-Rayet}}, \bibinfo {author} {\bibfnamefont {A.~R.}\ \bibnamefont
  {Hill}}, \bibinfo {author} {\bibfnamefont {N.}~\bibnamefont {Farida}},
  \bibinfo {author} {\bibfnamefont {M.~P.}\ \bibnamefont {Attfield}}, \bibinfo
  {author} {\bibfnamefont {P.}~\bibnamefont {Cubillas}}, \bibinfo {author}
  {\bibfnamefont {V.~A.}\ \bibnamefont {Blatov}}, \bibinfo {author}
  {\bibfnamefont {D.~M.}\ \bibnamefont {Proserpio}}, \bibinfo {author}
  {\bibfnamefont {D.}~\bibnamefont {Akporiaye}}, \bibinfo {author}
  {\bibfnamefont {B.}~\bibnamefont {Arstad}},\ and\ \bibinfo {author}
  {\bibfnamefont {J.~D.}\ \bibnamefont {Gale}},\ }\bibfield  {title} {\bibinfo
  {title} {{Predicting crystal growth via a unified kinetic three-dimensional
  partition model}},\ }\href {https://www.nature.com/articles/nature21684}
  {\bibfield  {journal} {\bibinfo  {journal} {Nature}\ }\textbf {\bibinfo
  {volume} {544}},\ \bibinfo {pages} {456} (\bibinfo {year}
  {2017})}\BibitemShut {NoStop}%
\bibitem [{\citenamefont {Wang}\ \emph {et~al.}(2019)\citenamefont {Wang},
  \citenamefont {Liu},\ and\ \citenamefont {Wang}}]{wang2019crystal}%
  \BibitemOpen
  \bibfield  {author} {\bibinfo {author} {\bibfnamefont {S.}~\bibnamefont
  {Wang}}, \bibinfo {author} {\bibfnamefont {G.}~\bibnamefont {Liu}},\ and\
  \bibinfo {author} {\bibfnamefont {L.}~\bibnamefont {Wang}},\ }\bibfield
  {title} {\bibinfo {title} {{Crystal facet engineering of photoelectrodes for
  photoelectrochemical water splitting}},\ }\href
  {https://pubs.acs.org/doi/10.1021/acs.chemrev.8b00584} {\bibfield  {journal}
  {\bibinfo  {journal} {Chem. Rev.}\ }\textbf {\bibinfo {volume} {119}},\
  \bibinfo {pages} {5192} (\bibinfo {year} {2019})}\BibitemShut {NoStop}%
\bibitem [{\citenamefont {Wulff}(1901)}]{wulffconst}%
  \BibitemOpen
  \bibfield  {author} {\bibinfo {author} {\bibfnamefont {G.}~\bibnamefont
  {Wulff}},\ }\bibfield  {title} {\bibinfo {title} {{Zur Frage der
  Geschwindigkeit des Wachstums und der Aufl\"osung der Krystallflagen}},\
  }\href@noop {} {\bibfield  {journal} {\bibinfo  {journal} {Z. Kristallogr.}\
  }\textbf {\bibinfo {volume} {34}},\ \bibinfo {pages} {449} (\bibinfo {year}
  {1901})}\BibitemShut {NoStop}%
\bibitem [{\citenamefont {von Laue}(1943)}]{wulffconst2}%
  \BibitemOpen
  \bibfield  {author} {\bibinfo {author} {\bibfnamefont {M.}~\bibnamefont {von
  Laue}},\ }\bibfield  {title} {\bibinfo {title} {{Der Wulffsche Satz f\"ur die
  Gleichgewichtsform von Kristallen}},\ }\href@noop {} {\bibfield  {journal}
  {\bibinfo  {journal} {Z. Kristallogr.}\ }\textbf {\bibinfo {volume} {105}},\
  \bibinfo {pages} {124} (\bibinfo {year} {1943})}\BibitemShut {NoStop}%
\bibitem [{\citenamefont {Herring}(1951)}]{PhysRev.82.87}%
  \BibitemOpen
  \bibfield  {author} {\bibinfo {author} {\bibfnamefont {C.}~\bibnamefont
  {Herring}},\ }\bibfield  {title} {\bibinfo {title} {{Some Theorems on the
  Free Energies of Crystal Surfaces}},\ }\href
  {https://doi.org/10.1103/PhysRev.82.87} {\bibfield  {journal} {\bibinfo
  {journal} {Phys. Rev.}\ }\textbf {\bibinfo {volume} {82}},\ \bibinfo {pages}
  {87} (\bibinfo {year} {1951})}\BibitemShut {NoStop}%
\bibitem [{\citenamefont {Marks}(1994)}]{Marks_1994}%
  \BibitemOpen
  \bibfield  {author} {\bibinfo {author} {\bibfnamefont {L.~D.}\ \bibnamefont
  {Marks}},\ }\bibfield  {title} {\bibinfo {title} {{Experimental studies of
  small particle structures}},\ }\href
  {https://doi.org/10.1088/0034-4885/57/6/002} {\bibfield  {journal} {\bibinfo
  {journal} {Rep. Prog. Phys.}\ }\textbf {\bibinfo {volume} {57}},\ \bibinfo
  {pages} {603} (\bibinfo {year} {1994})}\BibitemShut {NoStop}%
\bibitem [{\citenamefont {Barmparis}\ \emph {et~al.}(2015)\citenamefont
  {Barmparis}, \citenamefont {Lodziana}, \citenamefont {Lopez},\ and\
  \citenamefont {Remediakis}}]{barmparis2015nanoparticle}%
  \BibitemOpen
  \bibfield  {author} {\bibinfo {author} {\bibfnamefont {G.~D.}\ \bibnamefont
  {Barmparis}}, \bibinfo {author} {\bibfnamefont {Z.}~\bibnamefont {Lodziana}},
  \bibinfo {author} {\bibfnamefont {N.}~\bibnamefont {Lopez}},\ and\ \bibinfo
  {author} {\bibfnamefont {I.~N.}\ \bibnamefont {Remediakis}},\ }\bibfield
  {title} {\bibinfo {title} {{Nanoparticle shapes by using Wulff constructions
  and first-principles calculations}},\ }\href
  {https://www.beilstein-journals.org/bjnano/articles/6/35} {\bibfield
  {journal} {\bibinfo  {journal} {Beilstein J. Nanotechnol.}\ }\textbf
  {\bibinfo {volume} {6}},\ \bibinfo {pages} {361} (\bibinfo {year}
  {2015})}\BibitemShut {NoStop}%
\bibitem [{\citenamefont {Xia}\ \emph {et~al.}(2009)\citenamefont {Xia},
  \citenamefont {Xiong}, \citenamefont {Lim},\ and\ \citenamefont
  {Skrabalak}}]{Xia_2009}%
  \BibitemOpen
  \bibfield  {author} {\bibinfo {author} {\bibfnamefont {Y.}~\bibnamefont
  {Xia}}, \bibinfo {author} {\bibfnamefont {Y.}~\bibnamefont {Xiong}}, \bibinfo
  {author} {\bibfnamefont {B.}~\bibnamefont {Lim}},\ and\ \bibinfo {author}
  {\bibfnamefont {S.}~\bibnamefont {Skrabalak}},\ }\bibfield  {title} {\bibinfo
  {title} {{Shape-Controlled Synthesis of Metal Nanocrystals: Simple Chemistry
  Meets Complex Physics?}},\ }\href
  {https://doi.org/https://doi.org/10.1002/anie.200802248} {\bibfield
  {journal} {\bibinfo  {journal} {Angew. Chem. Int. Ed.}\ }\textbf {\bibinfo
  {volume} {48}},\ \bibinfo {pages} {60} (\bibinfo {year} {2009})}\BibitemShut
  {NoStop}%
\bibitem [{\citenamefont {Haiss}(2001)}]{Haiss_2001}%
  \BibitemOpen
  \bibfield  {author} {\bibinfo {author} {\bibfnamefont {W.}~\bibnamefont
  {Haiss}},\ }\bibfield  {title} {\bibinfo {title} {{Surface stress of clean
  and adsorbate-covered solids}},\ }\href
  {https://doi.org/10.1088/0034-4885/64/5/201} {\bibfield  {journal} {\bibinfo
  {journal} {Rep. Prog. Phys.}\ }\textbf {\bibinfo {volume} {64}},\ \bibinfo
  {pages} {591} (\bibinfo {year} {2001})}\BibitemShut {NoStop}%
\bibitem [{\citenamefont {Galanakis}\ \emph {et~al.}(2002)\citenamefont
  {Galanakis}, \citenamefont {Papanikolaou},\ and\ \citenamefont
  {Dederichs}}]{GALANAKIS20021}%
  \BibitemOpen
  \bibfield  {author} {\bibinfo {author} {\bibfnamefont {I.}~\bibnamefont
  {Galanakis}}, \bibinfo {author} {\bibfnamefont {N.}~\bibnamefont
  {Papanikolaou}},\ and\ \bibinfo {author} {\bibfnamefont {P.}~\bibnamefont
  {Dederichs}},\ }\bibfield  {title} {\bibinfo {title} {{Applicability of the
  broken-bond rule to the surface energy of the fcc metals}},\ }\href
  {https://doi.org/https://doi.org/10.1016/S0039-6028(02)01547-9} {\bibfield
  {journal} {\bibinfo  {journal} {Surf. Sci.}\ }\textbf {\bibinfo {volume}
  {511}},\ \bibinfo {pages} {1} (\bibinfo {year} {2002})}\BibitemShut {NoStop}%
\bibitem [{\citenamefont {Ma}\ and\ \citenamefont {Xu}(2007)}]{Ma_2007}%
  \BibitemOpen
  \bibfield  {author} {\bibinfo {author} {\bibfnamefont {F.}~\bibnamefont
  {Ma}}\ and\ \bibinfo {author} {\bibfnamefont {K.-W.}\ \bibnamefont {Xu}},\
  }\bibfield  {title} {\bibinfo {title} {{Using dangling bond density to
  characterize the surface energy of nanomaterials}},\ }\href
  {https://doi.org/https://doi.org/10.1002/sia.2562} {\bibfield  {journal}
  {\bibinfo  {journal} {Surf. Interface Anal.}\ }\textbf {\bibinfo {volume}
  {39}},\ \bibinfo {pages} {611} (\bibinfo {year} {2007})}\BibitemShut
  {NoStop}%
\bibitem [{\citenamefont {Eberhart}\ and\ \citenamefont
  {Horner}(2010)}]{Eberhart_2010}%
  \BibitemOpen
  \bibfield  {author} {\bibinfo {author} {\bibfnamefont {J.~G.}\ \bibnamefont
  {Eberhart}}\ and\ \bibinfo {author} {\bibfnamefont {S.}~\bibnamefont
  {Horner}},\ }\bibfield  {title} {\bibinfo {title} {{Bond-Energy and
  Surface-Energy Calculations in Metals}},\ }\href
  {https://doi.org/10.1021/ed100189v} {\bibfield  {journal} {\bibinfo
  {journal} {J. Chem. Educ.}\ }\textbf {\bibinfo {volume} {87}},\ \bibinfo
  {pages} {608} (\bibinfo {year} {2010})}\BibitemShut {NoStop}%
\bibitem [{\citenamefont {Fu}(2011)}]{PhysRevLett.106.106802}%
  \BibitemOpen
  \bibfield  {author} {\bibinfo {author} {\bibfnamefont {L.}~\bibnamefont
  {Fu}},\ }\bibfield  {title} {\bibinfo {title} {{Topological Crystalline
  Insulators}},\ }\href {https://doi.org/10.1103/PhysRevLett.106.106802}
  {\bibfield  {journal} {\bibinfo  {journal} {Phys. Rev. Lett.}\ }\textbf
  {\bibinfo {volume} {106}},\ \bibinfo {pages} {106802} (\bibinfo {year}
  {2011})}\BibitemShut {NoStop}%
\bibitem [{\citenamefont {Hsieh}\ \emph {et~al.}(2012)\citenamefont {Hsieh},
  \citenamefont {Lin}, \citenamefont {Liu}, \citenamefont {Duan}, \citenamefont
  {Bansil},\ and\ \citenamefont {Fu}}]{hsieh2012topological}%
  \BibitemOpen
  \bibfield  {author} {\bibinfo {author} {\bibfnamefont {T.~H.}\ \bibnamefont
  {Hsieh}}, \bibinfo {author} {\bibfnamefont {H.}~\bibnamefont {Lin}}, \bibinfo
  {author} {\bibfnamefont {J.}~\bibnamefont {Liu}}, \bibinfo {author}
  {\bibfnamefont {W.}~\bibnamefont {Duan}}, \bibinfo {author} {\bibfnamefont
  {A.}~\bibnamefont {Bansil}},\ and\ \bibinfo {author} {\bibfnamefont
  {L.}~\bibnamefont {Fu}},\ }\bibfield  {title} {\bibinfo {title} {{Topological
  crystalline insulators in the SnTe material class}},\ }\href
  {https://www.nature.com/articles/ncomms1969} {\bibfield  {journal} {\bibinfo
  {journal} {Nat. Commun.}\ }\textbf {\bibinfo {volume} {3}},\ \bibinfo {pages}
  {982} (\bibinfo {year} {2012})}\BibitemShut {NoStop}%
\bibitem [{\citenamefont {Shiozaki}\ and\ \citenamefont
  {Sato}(2014)}]{PhysRevB.90.165114}%
  \BibitemOpen
  \bibfield  {author} {\bibinfo {author} {\bibfnamefont {K.}~\bibnamefont
  {Shiozaki}}\ and\ \bibinfo {author} {\bibfnamefont {M.}~\bibnamefont
  {Sato}},\ }\bibfield  {title} {\bibinfo {title} {{Topology of crystalline
  insulators and superconductors}},\ }\href
  {https://doi.org/10.1103/PhysRevB.90.165114} {\bibfield  {journal} {\bibinfo
  {journal} {Phys. Rev. B}\ }\textbf {\bibinfo {volume} {90}},\ \bibinfo
  {pages} {165114} (\bibinfo {year} {2014})}\BibitemShut {NoStop}%
\bibitem [{\citenamefont {Li}\ \emph {et~al.}(2013)\citenamefont {Li},
  \citenamefont {Shao}, \citenamefont {Li}, \citenamefont {McCall},
  \citenamefont {Wang},\ and\ \citenamefont {Zhang}}]{li2013single}%
  \BibitemOpen
  \bibfield  {author} {\bibinfo {author} {\bibfnamefont {Z.}~\bibnamefont
  {Li}}, \bibinfo {author} {\bibfnamefont {S.}~\bibnamefont {Shao}}, \bibinfo
  {author} {\bibfnamefont {N.}~\bibnamefont {Li}}, \bibinfo {author}
  {\bibfnamefont {K.}~\bibnamefont {McCall}}, \bibinfo {author} {\bibfnamefont
  {J.}~\bibnamefont {Wang}},\ and\ \bibinfo {author} {\bibfnamefont
  {S.}~\bibnamefont {Zhang}},\ }\bibfield  {title} {\bibinfo {title} {{Single
  crystalline nanostructures of topological crystalline insulator SnTe with
  distinct facets and morphologies}},\ }\href
  {https://pubs.acs.org/doi/abs/10.1021/nl4030193} {\bibfield  {journal}
  {\bibinfo  {journal} {Nano Lett.}\ }\textbf {\bibinfo {volume} {13}},\
  \bibinfo {pages} {5443} (\bibinfo {year} {2013})}\BibitemShut {NoStop}%
\bibitem [{\citenamefont {Safdar}\ \emph {et~al.}(2014)\citenamefont {Safdar},
  \citenamefont {Wang}, \citenamefont {Mirza}, \citenamefont {Wang},\ and\
  \citenamefont {He}}]{safdar2014crystal}%
  \BibitemOpen
  \bibfield  {author} {\bibinfo {author} {\bibfnamefont {M.}~\bibnamefont
  {Safdar}}, \bibinfo {author} {\bibfnamefont {Q.}~\bibnamefont {Wang}},
  \bibinfo {author} {\bibfnamefont {M.}~\bibnamefont {Mirza}}, \bibinfo
  {author} {\bibfnamefont {Z.}~\bibnamefont {Wang}},\ and\ \bibinfo {author}
  {\bibfnamefont {J.}~\bibnamefont {He}},\ }\bibfield  {title} {\bibinfo
  {title} {{Crystal shape engineering of topological crystalline insulator SnTe
  microcrystals and nanowires with huge thermal activation energy gap}},\
  }\href {https://pubs.acs.org/doi/abs/10.1021/cg5002122} {\bibfield  {journal}
  {\bibinfo  {journal} {Crystal growth \& design}\ }\textbf {\bibinfo {volume}
  {14}},\ \bibinfo {pages} {2502} (\bibinfo {year} {2014})}\BibitemShut
  {NoStop}%
\bibitem [{\citenamefont {Bradlyn}\ \emph {et~al.}(2017)\citenamefont
  {Bradlyn}, \citenamefont {Elcoro}, \citenamefont {Cano}, \citenamefont
  {Vergniory}, \citenamefont {Wang}, \citenamefont {Felser}, \citenamefont
  {Aroyo},\ and\ \citenamefont {Bernevig}}]{bradlyn2017topological}%
  \BibitemOpen
  \bibfield  {author} {\bibinfo {author} {\bibfnamefont {B.}~\bibnamefont
  {Bradlyn}}, \bibinfo {author} {\bibfnamefont {L.}~\bibnamefont {Elcoro}},
  \bibinfo {author} {\bibfnamefont {J.}~\bibnamefont {Cano}}, \bibinfo {author}
  {\bibfnamefont {M.}~\bibnamefont {Vergniory}}, \bibinfo {author}
  {\bibfnamefont {Z.}~\bibnamefont {Wang}}, \bibinfo {author} {\bibfnamefont
  {C.}~\bibnamefont {Felser}}, \bibinfo {author} {\bibfnamefont
  {M.}~\bibnamefont {Aroyo}},\ and\ \bibinfo {author} {\bibfnamefont {B.~A.}\
  \bibnamefont {Bernevig}},\ }\bibfield  {title} {\bibinfo {title}
  {{Topological quantum chemistry}},\ }\href
  {https://www.nature.com/articles/nature23268} {\bibfield  {journal} {\bibinfo
   {journal} {Nature}\ }\textbf {\bibinfo {volume} {547}},\ \bibinfo {pages}
  {298} (\bibinfo {year} {2017})}\BibitemShut {NoStop}%
\bibitem [{\citenamefont {Po}\ \emph {et~al.}(2017)\citenamefont {Po},
  \citenamefont {Vishwanath},\ and\ \citenamefont {Watanabe}}]{po2017symmetry}%
  \BibitemOpen
  \bibfield  {author} {\bibinfo {author} {\bibfnamefont {H.~C.}\ \bibnamefont
  {Po}}, \bibinfo {author} {\bibfnamefont {A.}~\bibnamefont {Vishwanath}},\
  and\ \bibinfo {author} {\bibfnamefont {H.}~\bibnamefont {Watanabe}},\
  }\bibfield  {title} {\bibinfo {title} {{Symmetry-based indicators of band
  topology in the 230 space groups}},\ }\href
  {https://doi.org/10.1038/s41467-017-00133-2} {\bibfield  {journal} {\bibinfo
  {journal} {Nat. Commun.}\ }\textbf {\bibinfo {volume} {8}},\ \bibinfo {pages}
  {50} (\bibinfo {year} {2017})}\BibitemShut {NoStop}%
\bibitem [{\citenamefont {Kruthoff}\ \emph {et~al.}(2017)\citenamefont
  {Kruthoff}, \citenamefont {de~Boer}, \citenamefont {van Wezel}, \citenamefont
  {Kane},\ and\ \citenamefont {Slager}}]{PhysRevX.7.041069}%
  \BibitemOpen
  \bibfield  {author} {\bibinfo {author} {\bibfnamefont {J.}~\bibnamefont
  {Kruthoff}}, \bibinfo {author} {\bibfnamefont {J.}~\bibnamefont {de~Boer}},
  \bibinfo {author} {\bibfnamefont {J.}~\bibnamefont {van Wezel}}, \bibinfo
  {author} {\bibfnamefont {C.~L.}\ \bibnamefont {Kane}},\ and\ \bibinfo
  {author} {\bibfnamefont {R.-J.}\ \bibnamefont {Slager}},\ }\bibfield  {title}
  {\bibinfo {title} {{Topological Classification of Crystalline Insulators
  through Band Structure Combinatorics}},\ }\href
  {https://doi.org/10.1103/PhysRevX.7.041069} {\bibfield  {journal} {\bibinfo
  {journal} {Phys. Rev. X}\ }\textbf {\bibinfo {volume} {7}},\ \bibinfo {pages}
  {041069} (\bibinfo {year} {2017})}\BibitemShut {NoStop}%
\bibitem [{\citenamefont {Song}\ \emph {et~al.}(2018)\citenamefont {Song},
  \citenamefont {Zhang}, \citenamefont {Fang},\ and\ \citenamefont
  {Fang}}]{song2018quantitative}%
  \BibitemOpen
  \bibfield  {author} {\bibinfo {author} {\bibfnamefont {Z.}~\bibnamefont
  {Song}}, \bibinfo {author} {\bibfnamefont {T.}~\bibnamefont {Zhang}},
  \bibinfo {author} {\bibfnamefont {Z.}~\bibnamefont {Fang}},\ and\ \bibinfo
  {author} {\bibfnamefont {C.}~\bibnamefont {Fang}},\ }\bibfield  {title}
  {\bibinfo {title} {{Quantitative mappings between symmetry and topology in
  solids}},\ }\href {https://www.nature.com/articles/s41467-018-06010-w}
  {\bibfield  {journal} {\bibinfo  {journal} {Nat. Commun}\ }\textbf {\bibinfo
  {volume} {9}},\ \bibinfo {pages} {3530} (\bibinfo {year} {2018})}\BibitemShut
  {NoStop}%
\bibitem [{\citenamefont {Khalaf}\ \emph {et~al.}(2018)\citenamefont {Khalaf},
  \citenamefont {Po}, \citenamefont {Vishwanath},\ and\ \citenamefont
  {Watanabe}}]{PhysRevX.8.031070}%
  \BibitemOpen
  \bibfield  {author} {\bibinfo {author} {\bibfnamefont {E.}~\bibnamefont
  {Khalaf}}, \bibinfo {author} {\bibfnamefont {H.~C.}\ \bibnamefont {Po}},
  \bibinfo {author} {\bibfnamefont {A.}~\bibnamefont {Vishwanath}},\ and\
  \bibinfo {author} {\bibfnamefont {H.}~\bibnamefont {Watanabe}},\ }\bibfield
  {title} {\bibinfo {title} {{Symmetry Indicators and Anomalous Surface States
  of Topological Crystalline Insulators}},\ }\href
  {https://doi.org/10.1103/PhysRevX.8.031070} {\bibfield  {journal} {\bibinfo
  {journal} {Phys. Rev. X}\ }\textbf {\bibinfo {volume} {8}},\ \bibinfo {pages}
  {031070} (\bibinfo {year} {2018})}\BibitemShut {NoStop}%
\bibitem [{\citenamefont {Schindler}\ \emph {et~al.}(2018)\citenamefont
  {Schindler}, \citenamefont {Cook}, \citenamefont {Vergniory}, \citenamefont
  {Wang}, \citenamefont {Parkin}, \citenamefont {Bernevig},\ and\ \citenamefont
  {Neupert}}]{schindler2018higher}%
  \BibitemOpen
  \bibfield  {author} {\bibinfo {author} {\bibfnamefont {F.}~\bibnamefont
  {Schindler}}, \bibinfo {author} {\bibfnamefont {A.~M.}\ \bibnamefont {Cook}},
  \bibinfo {author} {\bibfnamefont {M.~G.}\ \bibnamefont {Vergniory}}, \bibinfo
  {author} {\bibfnamefont {Z.}~\bibnamefont {Wang}}, \bibinfo {author}
  {\bibfnamefont {S.~S.}\ \bibnamefont {Parkin}}, \bibinfo {author}
  {\bibfnamefont {B.~A.}\ \bibnamefont {Bernevig}},\ and\ \bibinfo {author}
  {\bibfnamefont {T.}~\bibnamefont {Neupert}},\ }\bibfield  {title} {\bibinfo
  {title} {{Higher-order topological insulators}},\ }\href
  {https://doi.org/10.1126/sciadv.aat0346} {\bibfield  {journal} {\bibinfo
  {journal} {Sci. Adv.}\ }\textbf {\bibinfo {volume} {4}},\ \bibinfo {pages}
  {eaat0346} (\bibinfo {year} {2018})}\BibitemShut {NoStop}%
\bibitem [{\citenamefont {Song}\ \emph {et~al.}(2019)\citenamefont {Song},
  \citenamefont {Huang}, \citenamefont {Qi}, \citenamefont {Fang},\ and\
  \citenamefont {Hermele}}]{song2019topological}%
  \BibitemOpen
  \bibfield  {author} {\bibinfo {author} {\bibfnamefont {Z.}~\bibnamefont
  {Song}}, \bibinfo {author} {\bibfnamefont {S.-J.}\ \bibnamefont {Huang}},
  \bibinfo {author} {\bibfnamefont {Y.}~\bibnamefont {Qi}}, \bibinfo {author}
  {\bibfnamefont {C.}~\bibnamefont {Fang}},\ and\ \bibinfo {author}
  {\bibfnamefont {M.}~\bibnamefont {Hermele}},\ }\bibfield  {title} {\bibinfo
  {title} {{Topological states from topological crystals}},\ }\href
  {https://advances.sciencemag.org/content/5/12/eaax2007} {\bibfield  {journal}
  {\bibinfo  {journal} {Sci. Adv.}\ }\textbf {\bibinfo {volume} {5}},\ \bibinfo
  {pages} {eaax2007} (\bibinfo {year} {2019})}\BibitemShut {NoStop}%
\bibitem [{\citenamefont {Fang}\ and\ \citenamefont {Fu}(2019)}]{fang2019new}%
  \BibitemOpen
  \bibfield  {author} {\bibinfo {author} {\bibfnamefont {C.}~\bibnamefont
  {Fang}}\ and\ \bibinfo {author} {\bibfnamefont {L.}~\bibnamefont {Fu}},\
  }\bibfield  {title} {\bibinfo {title} {{New classes of topological
  crystalline insulators having surface rotation anomaly}},\ }\href
  {https://advances.sciencemag.org/content/5/12/eaat2374} {\bibfield  {journal}
  {\bibinfo  {journal} {Sci. Adv.}\ }\textbf {\bibinfo {volume} {5}},\ \bibinfo
  {pages} {eaat2374} (\bibinfo {year} {2019})}\BibitemShut {NoStop}%
\bibitem [{\citenamefont {Tanaka}\ \emph
  {et~al.}(2020{\natexlab{a}})\citenamefont {Tanaka}, \citenamefont
  {Takahashi},\ and\ \citenamefont {Murakami}}]{PhysRevB.101.115120}%
  \BibitemOpen
  \bibfield  {author} {\bibinfo {author} {\bibfnamefont {Y.}~\bibnamefont
  {Tanaka}}, \bibinfo {author} {\bibfnamefont {R.}~\bibnamefont {Takahashi}},\
  and\ \bibinfo {author} {\bibfnamefont {S.}~\bibnamefont {Murakami}},\
  }\bibfield  {title} {\bibinfo {title} {{Appearance of hinge states in
  second-order topological insulators via the cutting procedure}},\ }\href
  {https://doi.org/10.1103/PhysRevB.101.115120} {\bibfield  {journal} {\bibinfo
   {journal} {Phys. Rev. B}\ }\textbf {\bibinfo {volume} {101}},\ \bibinfo
  {pages} {115120} (\bibinfo {year} {2020}{\natexlab{a}})}\BibitemShut
  {NoStop}%
\bibitem [{\citenamefont {Takahashi}\ \emph {et~al.}(2020)\citenamefont
  {Takahashi}, \citenamefont {Tanaka},\ and\ \citenamefont
  {Murakami}}]{PhysRevResearch.2.013300}%
  \BibitemOpen
  \bibfield  {author} {\bibinfo {author} {\bibfnamefont {R.}~\bibnamefont
  {Takahashi}}, \bibinfo {author} {\bibfnamefont {Y.}~\bibnamefont {Tanaka}},\
  and\ \bibinfo {author} {\bibfnamefont {S.}~\bibnamefont {Murakami}},\
  }\bibfield  {title} {\bibinfo {title} {{Bulk-edge and bulk-hinge
  correspondence in inversion-symmetric insulators}},\ }\href
  {https://doi.org/10.1103/PhysRevResearch.2.013300} {\bibfield  {journal}
  {\bibinfo  {journal} {Phys. Rev. Research}\ }\textbf {\bibinfo {volume}
  {2}},\ \bibinfo {pages} {013300} (\bibinfo {year} {2020})}\BibitemShut
  {NoStop}%
\bibitem [{\citenamefont {Shiozaki}\ \emph {et~al.}(2015)\citenamefont
  {Shiozaki}, \citenamefont {Sato},\ and\ \citenamefont
  {Gomi}}]{PhysRevB.91.155120}%
  \BibitemOpen
  \bibfield  {author} {\bibinfo {author} {\bibfnamefont {K.}~\bibnamefont
  {Shiozaki}}, \bibinfo {author} {\bibfnamefont {M.}~\bibnamefont {Sato}},\
  and\ \bibinfo {author} {\bibfnamefont {K.}~\bibnamefont {Gomi}},\ }\bibfield
  {title} {\bibinfo {title} {{${Z}_{2}$ topology in nonsymmorphic crystalline
  insulators: M\"obius twist in surface states}},\ }\href
  {https://doi.org/10.1103/PhysRevB.91.155120} {\bibfield  {journal} {\bibinfo
  {journal} {Phys. Rev. B}\ }\textbf {\bibinfo {volume} {91}},\ \bibinfo
  {pages} {155120} (\bibinfo {year} {2015})}\BibitemShut {NoStop}%
\bibitem [{\citenamefont {Fang}\ and\ \citenamefont
  {Fu}(2015)}]{PhysRevB.91.161105}%
  \BibitemOpen
  \bibfield  {author} {\bibinfo {author} {\bibfnamefont {C.}~\bibnamefont
  {Fang}}\ and\ \bibinfo {author} {\bibfnamefont {L.}~\bibnamefont {Fu}},\
  }\bibfield  {title} {\bibinfo {title} {{New classes of three-dimensional
  topological crystalline insulators: Nonsymmorphic and magnetic}},\ }\href
  {https://doi.org/10.1103/PhysRevB.91.161105} {\bibfield  {journal} {\bibinfo
  {journal} {Phys. Rev. B}\ }\textbf {\bibinfo {volume} {91}},\ \bibinfo
  {pages} {161105(R)} (\bibinfo {year} {2015})}\BibitemShut {NoStop}%
\bibitem [{\citenamefont {Lu}\ \emph {et~al.}(2016)\citenamefont {Lu},
  \citenamefont {Fang}, \citenamefont {Fu}, \citenamefont {Johnson},
  \citenamefont {Joannopoulos},\ and\ \citenamefont
  {Solja{\v{c}}i{\'c}}}]{lu2016symmetry}%
  \BibitemOpen
  \bibfield  {author} {\bibinfo {author} {\bibfnamefont {L.}~\bibnamefont
  {Lu}}, \bibinfo {author} {\bibfnamefont {C.}~\bibnamefont {Fang}}, \bibinfo
  {author} {\bibfnamefont {L.}~\bibnamefont {Fu}}, \bibinfo {author}
  {\bibfnamefont {S.~G.}\ \bibnamefont {Johnson}}, \bibinfo {author}
  {\bibfnamefont {J.~D.}\ \bibnamefont {Joannopoulos}},\ and\ \bibinfo {author}
  {\bibfnamefont {M.}~\bibnamefont {Solja{\v{c}}i{\'c}}},\ }\bibfield  {title}
  {\bibinfo {title} {{Symmetry-protected topological photonic crystal in three
  dimensions}},\ }\href {https://www.nature.com/articles/nphys3611} {\bibfield
  {journal} {\bibinfo  {journal} {Nat. Phys.}\ }\textbf {\bibinfo {volume}
  {12}},\ \bibinfo {pages} {337} (\bibinfo {year} {2016})}\BibitemShut
  {NoStop}%
\bibitem [{\citenamefont {Kim}\ and\ \citenamefont
  {Murakami}(2016)}]{PhysRevB.93.195138}%
  \BibitemOpen
  \bibfield  {author} {\bibinfo {author} {\bibfnamefont {H.}~\bibnamefont
  {Kim}}\ and\ \bibinfo {author} {\bibfnamefont {S.}~\bibnamefont {Murakami}},\
  }\bibfield  {title} {\bibinfo {title} {{Emergent spinless Weyl semimetals
  between the topological crystalline insulator and normal insulator phases
  with glide symmetry}},\ }\href {https://doi.org/10.1103/PhysRevB.93.195138}
  {\bibfield  {journal} {\bibinfo  {journal} {Phys. Rev. B}\ }\textbf {\bibinfo
  {volume} {93}},\ \bibinfo {pages} {195138} (\bibinfo {year}
  {2016})}\BibitemShut {NoStop}%
\bibitem [{\citenamefont {Chen}\ \emph
  {et~al.}(2017{\natexlab{a}})\citenamefont {Chen}, \citenamefont {Zhang},
  \citenamefont {Yi}, \citenamefont {Song}, \citenamefont {Zhang},
  \citenamefont {Zhang}, \citenamefont {Shi}, \citenamefont {Weng},
  \citenamefont {Fang}, \citenamefont {Richard},\ and\ \citenamefont
  {Ding}}]{PhysRevB.96.064102}%
  \BibitemOpen
  \bibfield  {author} {\bibinfo {author} {\bibfnamefont {D.}~\bibnamefont
  {Chen}}, \bibinfo {author} {\bibfnamefont {T.-T.}\ \bibnamefont {Zhang}},
  \bibinfo {author} {\bibfnamefont {C.-J.}\ \bibnamefont {Yi}}, \bibinfo
  {author} {\bibfnamefont {Z.-D.}\ \bibnamefont {Song}}, \bibinfo {author}
  {\bibfnamefont {W.-L.}\ \bibnamefont {Zhang}}, \bibinfo {author}
  {\bibfnamefont {T.}~\bibnamefont {Zhang}}, \bibinfo {author} {\bibfnamefont
  {Y.-G.}\ \bibnamefont {Shi}}, \bibinfo {author} {\bibfnamefont {H.-M.}\
  \bibnamefont {Weng}}, \bibinfo {author} {\bibfnamefont {Z.}~\bibnamefont
  {Fang}}, \bibinfo {author} {\bibfnamefont {P.}~\bibnamefont {Richard}},\ and\
  \bibinfo {author} {\bibfnamefont {H.}~\bibnamefont {Ding}},\ }\bibfield
  {title} {\bibinfo {title} {{Robustness of topological states with respect to
  lattice instability in the nonsymmorphic topological insulator KHgSb}},\
  }\href {https://doi.org/10.1103/PhysRevB.96.064102} {\bibfield  {journal}
  {\bibinfo  {journal} {Phys. Rev. B}\ }\textbf {\bibinfo {volume} {96}},\
  \bibinfo {pages} {064102} (\bibinfo {year} {2017}{\natexlab{a}})}\BibitemShut
  {NoStop}%
\bibitem [{\citenamefont {Wieder}\ \emph {et~al.}(2018)\citenamefont {Wieder},
  \citenamefont {Bradlyn}, \citenamefont {Wang}, \citenamefont {Cano},
  \citenamefont {Kim}, \citenamefont {Kim}, \citenamefont {Rappe},
  \citenamefont {Kane},\ and\ \citenamefont {Bernevig}}]{wieder2018wallpaper}%
  \BibitemOpen
  \bibfield  {author} {\bibinfo {author} {\bibfnamefont {B.~J.}\ \bibnamefont
  {Wieder}}, \bibinfo {author} {\bibfnamefont {B.}~\bibnamefont {Bradlyn}},
  \bibinfo {author} {\bibfnamefont {Z.}~\bibnamefont {Wang}}, \bibinfo {author}
  {\bibfnamefont {J.}~\bibnamefont {Cano}}, \bibinfo {author} {\bibfnamefont
  {Y.}~\bibnamefont {Kim}}, \bibinfo {author} {\bibfnamefont {H.-S.~D.}\
  \bibnamefont {Kim}}, \bibinfo {author} {\bibfnamefont {A.~M.}\ \bibnamefont
  {Rappe}}, \bibinfo {author} {\bibfnamefont {C.}~\bibnamefont {Kane}},\ and\
  \bibinfo {author} {\bibfnamefont {B.~A.}\ \bibnamefont {Bernevig}},\
  }\bibfield  {title} {\bibinfo {title} {{Wallpaper fermions and the
  nonsymmorphic Dirac insulator}},\ }\href
  {https://science.sciencemag.org/content/361/6399/246} {\bibfield  {journal}
  {\bibinfo  {journal} {Science}\ }\textbf {\bibinfo {volume} {361}},\ \bibinfo
  {pages} {246} (\bibinfo {year} {2018})}\BibitemShut {NoStop}%
\bibitem [{\citenamefont {Kim}\ \emph {et~al.}(2019)\citenamefont {Kim},
  \citenamefont {Shiozaki},\ and\ \citenamefont
  {Murakami}}]{PhysRevB.100.165202}%
  \BibitemOpen
  \bibfield  {author} {\bibinfo {author} {\bibfnamefont {H.}~\bibnamefont
  {Kim}}, \bibinfo {author} {\bibfnamefont {K.}~\bibnamefont {Shiozaki}},\ and\
  \bibinfo {author} {\bibfnamefont {S.}~\bibnamefont {Murakami}},\ }\bibfield
  {title} {\bibinfo {title} {{Glide-symmetric magnetic topological crystalline
  insulators with inversion symmetry}},\ }\href
  {https://doi.org/10.1103/PhysRevB.100.165202} {\bibfield  {journal} {\bibinfo
   {journal} {Phys. Rev. B}\ }\textbf {\bibinfo {volume} {100}},\ \bibinfo
  {pages} {165202} (\bibinfo {year} {2019})}\BibitemShut {NoStop}%
\bibitem [{\citenamefont {Kim}\ and\ \citenamefont
  {Murakami}(2020)}]{kim2020glide}%
  \BibitemOpen
  \bibfield  {author} {\bibinfo {author} {\bibfnamefont {H.}~\bibnamefont
  {Kim}}\ and\ \bibinfo {author} {\bibfnamefont {S.}~\bibnamefont {Murakami}},\
  }\bibfield  {title} {\bibinfo {title} {Glide-symmetric topological
  crystalline insulator phase in a nonprimitive lattice},\ }\href
  {https://doi.org/10.1103/PhysRevB.102.195202} {\bibfield  {journal} {\bibinfo
   {journal} {Phys. Rev. B}\ }\textbf {\bibinfo {volume} {102}},\ \bibinfo
  {pages} {195202} (\bibinfo {year} {2020})}\BibitemShut {NoStop}%
\bibitem [{\citenamefont {Tanaka}\ \emph
  {et~al.}(2020{\natexlab{b}})\citenamefont {Tanaka}, \citenamefont
  {Takahashi}, \citenamefont {Zhang},\ and\ \citenamefont
  {Murakami}}]{PhysRevResearch.2.043274}%
  \BibitemOpen
  \bibfield  {author} {\bibinfo {author} {\bibfnamefont {Y.}~\bibnamefont
  {Tanaka}}, \bibinfo {author} {\bibfnamefont {R.}~\bibnamefont {Takahashi}},
  \bibinfo {author} {\bibfnamefont {T.}~\bibnamefont {Zhang}},\ and\ \bibinfo
  {author} {\bibfnamefont {S.}~\bibnamefont {Murakami}},\ }\bibfield  {title}
  {\bibinfo {title} {{Theory of inversion-${\mathbb{Z}}_{4}$ protected
  topological chiral hinge states and its applications to layered
  antiferromagnets}},\ }\href
  {https://doi.org/10.1103/PhysRevResearch.2.043274} {\bibfield  {journal}
  {\bibinfo  {journal} {Phys. Rev. Research}\ }\textbf {\bibinfo {volume}
  {2}},\ \bibinfo {pages} {043274} (\bibinfo {year}
  {2020}{\natexlab{b}})}\BibitemShut {NoStop}%
\bibitem [{\citenamefont {Liu}\ \emph {et~al.}(2014{\natexlab{b}})\citenamefont
  {Liu}, \citenamefont {Zhang},\ and\ \citenamefont
  {VanLeeuwen}}]{PhysRevB.90.085304}%
  \BibitemOpen
  \bibfield  {author} {\bibinfo {author} {\bibfnamefont {C.-X.}\ \bibnamefont
  {Liu}}, \bibinfo {author} {\bibfnamefont {R.-X.}\ \bibnamefont {Zhang}},\
  and\ \bibinfo {author} {\bibfnamefont {B.~K.}\ \bibnamefont {VanLeeuwen}},\
  }\bibfield  {title} {\bibinfo {title} {{Topological nonsymmorphic crystalline
  insulators}},\ }\href {https://doi.org/10.1103/PhysRevB.90.085304} {\bibfield
   {journal} {\bibinfo  {journal} {Phys. Rev. B}\ }\textbf {\bibinfo {volume}
  {90}},\ \bibinfo {pages} {085304} (\bibinfo {year}
  {2014}{\natexlab{b}})}\BibitemShut {NoStop}%
\bibitem [{\citenamefont {Shiozaki}\ \emph {et~al.}(2016)\citenamefont
  {Shiozaki}, \citenamefont {Sato},\ and\ \citenamefont
  {Gomi}}]{PhysRevB.93.195413}%
  \BibitemOpen
  \bibfield  {author} {\bibinfo {author} {\bibfnamefont {K.}~\bibnamefont
  {Shiozaki}}, \bibinfo {author} {\bibfnamefont {M.}~\bibnamefont {Sato}},\
  and\ \bibinfo {author} {\bibfnamefont {K.}~\bibnamefont {Gomi}},\ }\bibfield
  {title} {\bibinfo {title} {{Topology of nonsymmorphic crystalline insulators
  and superconductors}},\ }\href {https://doi.org/10.1103/PhysRevB.93.195413}
  {\bibfield  {journal} {\bibinfo  {journal} {Phys. Rev. B}\ }\textbf {\bibinfo
  {volume} {93}},\ \bibinfo {pages} {195413} (\bibinfo {year}
  {2016})}\BibitemShut {NoStop}%
\bibitem [{\citenamefont {Wang}\ \emph {et~al.}(2016)\citenamefont {Wang},
  \citenamefont {Alexandradinata}, \citenamefont {Cava},\ and\ \citenamefont
  {Bernevig}}]{wang2016hourglass}%
  \BibitemOpen
  \bibfield  {author} {\bibinfo {author} {\bibfnamefont {Z.}~\bibnamefont
  {Wang}}, \bibinfo {author} {\bibfnamefont {A.}~\bibnamefont
  {Alexandradinata}}, \bibinfo {author} {\bibfnamefont {R.~J.}\ \bibnamefont
  {Cava}},\ and\ \bibinfo {author} {\bibfnamefont {B.~A.}\ \bibnamefont
  {Bernevig}},\ }\bibfield  {title} {\bibinfo {title} {{Hourglass fermions}},\
  }\href {https://www.nature.com/articles/nature17410} {\bibfield  {journal}
  {\bibinfo  {journal} {Nature}\ }\textbf {\bibinfo {volume} {532}},\ \bibinfo
  {pages} {189} (\bibinfo {year} {2016})}\BibitemShut {NoStop}%
\bibitem [{\citenamefont {Fulga}\ \emph {et~al.}(2016)\citenamefont {Fulga},
  \citenamefont {Avraham}, \citenamefont {Beidenkopf},\ and\ \citenamefont
  {Stern}}]{PhysRevB.94.125405}%
  \BibitemOpen
  \bibfield  {author} {\bibinfo {author} {\bibfnamefont {I.~C.}\ \bibnamefont
  {Fulga}}, \bibinfo {author} {\bibfnamefont {N.}~\bibnamefont {Avraham}},
  \bibinfo {author} {\bibfnamefont {H.}~\bibnamefont {Beidenkopf}},\ and\
  \bibinfo {author} {\bibfnamefont {A.}~\bibnamefont {Stern}},\ }\bibfield
  {title} {\bibinfo {title} {{Coupled-layer description of topological
  crystalline insulators}},\ }\href
  {https://doi.org/10.1103/PhysRevB.94.125405} {\bibfield  {journal} {\bibinfo
  {journal} {Phys. Rev. B}\ }\textbf {\bibinfo {volume} {94}},\ \bibinfo
  {pages} {125405} (\bibinfo {year} {2016})}\BibitemShut {NoStop}%
\bibitem [{\citenamefont {Song}\ \emph {et~al.}(2017)\citenamefont {Song},
  \citenamefont {Huang}, \citenamefont {Fu},\ and\ \citenamefont
  {Hermele}}]{PhysRevX.7.011020}%
  \BibitemOpen
  \bibfield  {author} {\bibinfo {author} {\bibfnamefont {H.}~\bibnamefont
  {Song}}, \bibinfo {author} {\bibfnamefont {S.-J.}\ \bibnamefont {Huang}},
  \bibinfo {author} {\bibfnamefont {L.}~\bibnamefont {Fu}},\ and\ \bibinfo
  {author} {\bibfnamefont {M.}~\bibnamefont {Hermele}},\ }\bibfield  {title}
  {\bibinfo {title} {{Topological Phases Protected by Point Group Symmetry}},\
  }\href {https://doi.org/10.1103/PhysRevX.7.011020} {\bibfield  {journal}
  {\bibinfo  {journal} {Phys. Rev. X}\ }\textbf {\bibinfo {volume} {7}},\
  \bibinfo {pages} {011020} (\bibinfo {year} {2017})}\BibitemShut {NoStop}%
\bibitem [{\citenamefont {Huang}\ \emph {et~al.}(2017)\citenamefont {Huang},
  \citenamefont {Song}, \citenamefont {Huang},\ and\ \citenamefont
  {Hermele}}]{PhysRevB.96.205106}%
  \BibitemOpen
  \bibfield  {author} {\bibinfo {author} {\bibfnamefont {S.-J.}\ \bibnamefont
  {Huang}}, \bibinfo {author} {\bibfnamefont {H.}~\bibnamefont {Song}},
  \bibinfo {author} {\bibfnamefont {Y.-P.}\ \bibnamefont {Huang}},\ and\
  \bibinfo {author} {\bibfnamefont {M.}~\bibnamefont {Hermele}},\ }\bibfield
  {title} {\bibinfo {title} {{Building crystalline topological phases from
  lower-dimensional states}},\ }\href
  {https://doi.org/10.1103/PhysRevB.96.205106} {\bibfield  {journal} {\bibinfo
  {journal} {Phys. Rev. B}\ }\textbf {\bibinfo {volume} {96}},\ \bibinfo
  {pages} {205106} (\bibinfo {year} {2017})}\BibitemShut {NoStop}%
\bibitem [{sup()}]{sup}%
  \BibitemOpen
  \href@noop {} {}\bibinfo {note} {{See Supplemental Material at URL for
  further details on our effective surface theory of the TCI, the $(\alpha 0
  \gamma)$ surface with $\gamma \neq 1,2$, the tight-binding models, the TCI
  without $\mathcal{T}$ symmetry, the Wulff construction, the methods to
  calculate the surface energy $E_{\rm surf}^{(\alpha \beta \gamma)}$ and the
  SETS, the derivation of the function $F(\theta)$ based on the numbers of
  dangling bonds on the surface, and the methods of the first-principle
  calculations.}}\BibitemShut {Stop}%
\bibitem [{Pyt()}]{PythTB}%
  \BibitemOpen
  \href@noop {} {\bibinfo {title} {{PythTB}}},\ \bibinfo {howpublished}
  {\url{http://physics.rutgers.edu/pythtb/}}\BibitemShut {NoStop}%
\bibitem [{\citenamefont {Rahm}\ and\ \citenamefont
  {Erhart}(2020)}]{WulffPack}%
  \BibitemOpen
  \bibfield  {author} {\bibinfo {author} {\bibfnamefont {J.~M.}\ \bibnamefont
  {Rahm}}\ and\ \bibinfo {author} {\bibfnamefont {P.}~\bibnamefont {Erhart}},\
  }\bibfield  {title} {\bibinfo {title} {{WulffPack: A Python package for Wulff
  constructions}},\ }\href {https://doi.org/10.21105/joss.01944} {\bibfield
  {journal} {\bibinfo  {journal} {J. Open Source Softw.}\ }\textbf {\bibinfo
  {volume} {5}},\ \bibinfo {pages} {1944} (\bibinfo {year} {2020})}\BibitemShut
  {NoStop}%
\bibitem [{\citenamefont {Chen}\ \emph
  {et~al.}(2017{\natexlab{b}})\citenamefont {Chen}, \citenamefont {Zhang},
  \citenamefont {Yi}, \citenamefont {Song}, \citenamefont {Zhang},
  \citenamefont {Zhang}, \citenamefont {Shi}, \citenamefont {Weng},
  \citenamefont {Fang}, \citenamefont {Richard},\ and\ \citenamefont
  {Ding}}]{chen2017robustness}%
  \BibitemOpen
  \bibfield  {author} {\bibinfo {author} {\bibfnamefont {D.}~\bibnamefont
  {Chen}}, \bibinfo {author} {\bibfnamefont {T.-T.}\ \bibnamefont {Zhang}},
  \bibinfo {author} {\bibfnamefont {C.-J.}\ \bibnamefont {Yi}}, \bibinfo
  {author} {\bibfnamefont {Z.-D.}\ \bibnamefont {Song}}, \bibinfo {author}
  {\bibfnamefont {W.-L.}\ \bibnamefont {Zhang}}, \bibinfo {author}
  {\bibfnamefont {T.}~\bibnamefont {Zhang}}, \bibinfo {author} {\bibfnamefont
  {Y.-G.}\ \bibnamefont {Shi}}, \bibinfo {author} {\bibfnamefont {H.-M.}\
  \bibnamefont {Weng}}, \bibinfo {author} {\bibfnamefont {Z.}~\bibnamefont
  {Fang}}, \bibinfo {author} {\bibfnamefont {P.}~\bibnamefont {Richard}},\ and\
  \bibinfo {author} {\bibfnamefont {H.}~\bibnamefont {Ding}},\ }\bibfield
  {title} {\bibinfo {title} {{Robustness of topological states with respect to
  lattice instability in the nonsymmorphic topological insulator KHgSb}},\
  }\href {https://doi.org/10.1103/PhysRevB.96.064102} {\bibfield  {journal}
  {\bibinfo  {journal} {Phys. Rev. B}\ }\textbf {\bibinfo {volume} {96}},\
  \bibinfo {pages} {064102} (\bibinfo {year} {2017}{\natexlab{b}})}\BibitemShut
  {NoStop}%
\end{thebibliography}

%

\end{document}


\title{Supplemental Material for ``Anomalous crystal shapes of topological crystalline insulators''}
\author{Yutaro Tanaka}
\affiliation{
	Department of Physics, Tokyo Institute of Technology, 2-12-1 Ookayama, Meguro-ku, Tokyo 152-8551, Japan
}
\author{Tiantian Zhang}
\affiliation{
	Department of Physics, Tokyo Institute of Technology, 2-12-1 Ookayama, Meguro-ku, Tokyo 152-8551, Japan
}
\affiliation{
	TIES, Tokyo Institute of Technology, 2-12-1 Ookayama, Meguro-ku, Tokyo 152-8551, Japan
}
\author{Makio Uwaha}
\affiliation{
	Science Division, Center for General Education, Aichi Institute of Technology, 1247 Yachigusa, Yakusa-cho, Toyota, Aichi 470-0392, Japan
}
\author{Shuichi Murakami}
\affiliation{
	Department of Physics, Tokyo Institute of Technology, 2-12-1 Ookayama, Meguro-ku, Tokyo 152-8551, Japan
}
\affiliation{
	TIES, Tokyo Institute of Technology, 2-12-1 Ookayama, Meguro-ku, Tokyo 152-8551, Japan
}
\maketitle

\renewcommand{\theequation}{S.\arabic{equation}}
\setcounter{equation}{0}
\renewcommand{\tablename}{Table S}
\setcounter{table}{0}
\renewcommand{\figurename}{Supplementary Fig.}
\setcounter{figure}{0}

\renewcommand{\thesection}{Supplemental Material No.~\arabic{section}}
\setcounter{section}{0}

\onecolumngrid

\tableofcontents

\section{Methods}
 
\subsection{Effective surface model}
 Here we consider the model from the layer construction. We introduce effective surface Hamiltonians that includes  hybridizations of the helical edge modes between the adjacent layers. 
Because the layers $L_{A}$ and $L_{B}$ are 2D topological insulators, the Hamiltonians for the edges of the layers $L_{A}$ and $L_{B}$ along the $y$-axis  can be expressed as $H_{A}=vk_{y} \sigma_{1}$ and $H_{B}=-vk_{y} \sigma_{1}$, respectively, where $v$ is a real constant.   
Furthermore, we introduce the coupling term $H_{\delta}=\delta \sigma_{0}$ between adjacent layers $L_{A}$ and $L_{B}$, where $\delta$ is a real constant.
Therefore, in the case with $\gamma=2$, the effective surface Hamiltonian can be expressed as
\begin{align}\label{eq:surface_Hamiltonian_even}
\mathcal{H}^{(\gamma =2)}_{\alpha}(k_{y})
=&\sum_{z=0}^{\frac{\alpha-1}{2}}\bigl[c^{\dagger}_{z}H_{A}c_{z}\bigr]+\sum_{z=0}^{\frac{\alpha-3}{2}}\bigl[ c^{\dagger}_{z+\frac{1}{2}}H_{B}c_{z+\frac{1}{2}}\bigr]\nonumber \\
&+\sum_{z'=0, \frac{1}{2}, \cdots}^{\frac{\alpha-3}{2}}\bigl[ c^{\dagger}_{z'}H_{\delta}c_{z'+\frac{1}{2}} +{\rm h.c.}\bigr]. 
\end{align}
where $\alpha$ is an odd number, and $c_{z}=(c_{z\uparrow},c_{z\downarrow})^{T}$ ($2z\in \mathbb{Z}$) is an annihilation operator for electrons on the layer within the plane at $z$. 
On the other hand, when $\gamma=1$, the effective surface Hamiltonian is
\begin{align}\label{eq:surface_Hamiltonian_odd}
\mathcal{H}^{(\gamma =1)}_{\alpha}(k_y)=&\sum_{z=0}^{\alpha-1} \bigl[ c^{\dagger}_{z}H_{A}c_{z}+c^{\dagger}_{z+\frac{1}{2}}H_{B}c_{z+\frac{1}{2}} \bigr] \nonumber \\
&+\sum_{z'=0, \frac{1}{2}, \cdots}^{\alpha-1}\bigl[ c^{\dagger}_{z'}H_{\delta}c_{z'+\frac{1}{2}} +{\rm h.c.}\bigr].
\end{align}
For the details on how to obtain the eigenvalues of these Hamiltonians, see \ref{appendix:analysis_surface_hamiltonian}.

\subsection{Details of calculations of the surface energies}
To calculate the surface energy, we consider a slab geometry with the $(\alpha \beta \gamma)$ surface.  This slab has  finite thickness in the direction perpendicular to the $(\alpha \beta \gamma)$ surface and has periodic boundary conditions along the ($\alpha \beta \gamma$) directions. 
 Let us introduce hoppings between the top surface and the bottom surface of the slab, and then we replace this hopping parameter $t$ with $\lambda t$ with $\lambda$ being a real parameter. When $\lambda=0$, this system is just a slab geometry. On the other hand, when $\lambda=1$, this system is equal to a bulk crystal because it has the periodic boundary conditions along all directions. We introduce a Hamiltonian $H_{\rm slab}^{(\alpha \beta  \gamma)}(\lambda)$ with such geometry, and the surface energy of $H^{(\alpha \beta \gamma)}_{\rm slab}(\lambda)$  can be expressed as 
 \begin{equation}\label{Eq:definition_surface_energy}
E_{\rm surf}^{(\alpha \beta \gamma)}=\sum_{n=1}^{N} \frac{ E^{(\alpha \beta \gamma)}_{{\rm slab},n}|_{\lambda=0}-E^{(\alpha \beta \gamma)}_{{\rm slab},n}|_{\lambda=1} }{2 S_{\rm slab}},
\end{equation}
where $E^{(\alpha \beta \gamma)}_{{\rm slab},n}|_{\lambda}$ is the energy from a $n$-th band of $H^{(\alpha \beta \gamma)}_{\rm slab}(\lambda)$, $N$ is the total number of occupied bands, and $S_{\rm slab}$ represents the area of the slab surface of $H^{(\alpha \beta \gamma)}_{\rm slab}(\lambda)$.

Next, we discuss the details of the calculations of the SETS. 
Let us assume that $H_{\rm slab}^{(\alpha \beta  \gamma)}(0)$ has $N_{\rm surf}$ occupied bands forming the surface states, and the bands are labeled by $n= N-N_{\rm surf}+1, \cdots N$. 
 We define the surface energy from topological surface states (SETS) in terms of the difference between $E^{(\alpha \beta \gamma)}_{{\rm slab},n}|_{\lambda=0}$ and $E^{(\alpha \beta \gamma)}_{{\rm slab},n}|_{\lambda=1}$ with $n= N-N_{\rm surf}+1, \cdots N$:
\begin{align}\label{Eq:definition_SETS}
	E^{(\alpha \beta \gamma)}_{\rm SETS}=
	 \sum_{n= N-N_{\rm surf}+1}^{N}
 \frac{E^{(\alpha \beta \gamma)}_{{\rm slab},n}|_{\lambda=0}-E^{(\alpha \beta \gamma)}_{{\rm slab},n}|_{\lambda=1} }{2 S_{\rm slab}}.
\end{align}

\subsection{Wulff construction}
 According to the Wulff construction, the distance $h_{\alpha \beta \gamma}$ between a surface with a Miller index $(\alpha \beta \gamma)$ and the crystal center is proportional to  $E_{\rm surf}^{(\alpha \beta \gamma)}$:
$
E_{\rm surf}^{(\alpha \beta \gamma)}/h_{\alpha \beta \gamma} = l,
$
 where $l$ is a constant.
Thus, we can obtain the crystal shape through the following steps: (i) First, we calculate $h_{\alpha \beta  \gamma}= E_{\rm surf}^{(\alpha \beta \gamma)}$ for a Miller index $(\alpha \beta \gamma)$. (ii) We fix the origin $(0,0,0)^{T}$ and draw  the plane at a distance $E_{\rm surf}^{(\alpha \beta \gamma)}$ from the origin. This plane is along the ($\alpha \beta \gamma$) plane. (iii) By repeating (i) and (ii) for various Miller indices,  many planes will be obtained, and the innermost closed curve composed of these planes is the equilibrium crystal shape.

\subsection{Details of the first-principle calculations}
We performed first-principle calculations on KHgSb in the main text within the Perdew-Burke-Ernzerhof exchange-correlation \cite{perdew1996generalized} implemented in the Vienna $ab$ $initio$ simulation package \cite{kresse1996efficient}, combining with maximally localized Wannier functions \cite{marzari1997maximally} 
to obtain the surface local density of states by surface Green's function \cite{sancho1985highly,sancho1984quick,wu2018wanniertools}.

\section{Nonmagnetic topological crystalline insulator from the layer construction}\label{appendix:tight-binding_model}
\subsection{Layer constructions}
Here we construct the 3D tight-binding model of a topological crystalline insulator (TCI) protected by glide symmetry from the layer construction  \cite{song2018quantitative, PhysRevB.100.165202, kim2020glide, song2019topological}. This model is the first model $\mathcal{H}^{(1)}_{\rm TCI}(\boldsymbol{k})$ of the TCI in the main text. 
We start from a simple tight-binding model of a 2D topological insulator (TI) on a square lattice  written as
\begin{align}\label{Eq:2dTI}
	\mathcal{H}_{\rm TI}(\boldsymbol{k})=& - ( m-t_{1} \cos k_{x}-t_{1}\cos k_{y})\tau_{3}+t_{2}\sin k_y \tau_{1}\sigma_{1}+t_{3} \sin k_{x}\tau_{1}\sigma_{3},
\end{align} 
where $\tau_{j}$ and $\sigma_{j}\ (j=1, 2, 3)$ are Pauli matrices, $\tau_{0}$ and $\sigma_{0}$ are the $2\times 2$ identity matrices, and we set the lattice constants to 1.  In order to break the mirror $M_{x}$, $M_{y}$, and $M_{z}$ symmetries with $M_{x}=-i\tau_{3}\sigma_{1}$, $M_{y}=-i\tau_{3}\sigma_{2}$, and $M_{z}=-i\tau_{3}\sigma_{3}$, here, we introduce the second and third terms, $t_{2}\sin k_y \tau_{1}\sigma_{1}$ and $t_{3} \sin k_{x}\tau_{1}\sigma_{3}$. We set the Fermi energy $E_{\rm F}$ to be $E_{\rm F}=0$. 
This Hamiltonian has inversion (${I}$) symmetry and time-reversal ($\mathcal{T}$) symmetry:
\begin{align}
	{I} \mathcal{H}_{\rm TI}(\boldsymbol{k}) {I}^{-1}=\mathcal{H}_{\rm TI}(-\boldsymbol{k}),\nonumber \\
	\mathcal{T} \mathcal{H}_{\rm TI}(\boldsymbol{k}) \mathcal{T}^{-1}=\mathcal{H}_{\rm TI}(-\boldsymbol{k}),
\end{align}
where ${I}=\tau_{3} $, and $\mathcal{T}=-i\tau_{0} \sigma_{2}K$ with $K$ being the complex conjugation. 

Next, we place layers of the 2D TI of Eq.~(\ref{Eq:2dTI}) at the $z=n$ planes $(n \in \mathbb{Z})$ in real space. In this case, a Hamiltonian  $H_{A}$ consisting of the layers can be expressed as follows:
\begin{align}
	H_{A}=&\sum_{x,y,z \in \mathbb{Z}} \biggl[ c^{\dagger}(x+1,y,z) \biggr(\frac{t_{1}}{2} \tau_{3} + \frac{t_{3}}{2}i\tau_{1} \sigma_{3} \biggl) c(x,y,z) \nonumber \\
	&+c^{\dagger}(x,y+1,z) \biggr(\frac{t_{1}}{2} \tau_{3}+\frac{t_{2}}{2}i\tau_{1} \sigma_{1} \biggl) c(x,y,z) +{\rm h.c.}
	\biggr]
	-\sum_{x,y,z \in \mathbb{Z}} \biggl[ c^{\dagger}(x,y,z)m\tau_{3}  c(x,y,z)\biggr],
\end{align}
where $c^{\dagger} (x,y,z)$ is a creation operator at the $(x,y,z)$ site, and $x$, $y$ and $z$ run over integers. 
Now, we design the system to be invariant under the  glide $\hat{G}_{y}$ symmetry:
\begin{equation}
	\hat{G}_{y}=\{M_{y}|00\tfrac{1}{2}\},
\end{equation}
where $M_{y}$ represents the mirror reflection with respect to the $xz$ plane.
Under $\hat{G}_{y}$, a creation operation $c^{\dagger}_{\alpha}(x,y,z)$ transforms as
\begin{equation}
	\hat{G}_{y}:\ c_{\alpha}^{\dagger}(x,y,z)\ \rightarrow \  c_{\beta}^{\dagger}(x,-y,z+\tfrac{1}{2}) \bigl[ M_{y}\bigr]_{\beta \alpha},\nonumber 
\end{equation}
with
\begin{equation}
	M_{y}= -i\tau_{3} \sigma_{2},
\end{equation}
where $\alpha$ and $\beta$ run over the basis of Eq.~(\ref{Eq:2dTI}).  
In addition, by making copies of $H_{A}$ by the glide transformation $\hat{G}_{y}$ \cite{PhysRevB.100.165202, kim2020glide}, we obtain layers $B$ at $z'=\cdots,-1/2, 1/2, \cdots$, with the Hamiltonian of the layers $B$  expressed as
\begin{align}\label{Eq:sm_HArightarrowHB}
	\hat{G}_{y}:\ H_{A}\rightarrow \ H_{B}\equiv \sum_{x,y,z \in \mathbb{Z}} \biggl[ &c^{\dagger}(x+1,-y,z+\tfrac{1}{2}) \biggr( \frac{t_{1}}{2} \tau_{3} + \frac{t_{3}}{2}i\tau_{1} \sigma_{3} \biggl) c(x,-y, z+\tfrac{1}{2}) \nonumber \\
	&+c^{\dagger}(x,-y-1,z+\tfrac{1}{2}) \biggr(\frac{t_{1}}{2} \tau_{3}+\frac{t_{2}}{2}i\tau_{1} \sigma_{1} \biggl) c(x,-y,z+\tfrac{1}{2}) +{\rm h.c.}
	\biggr]\nonumber \\
	-\sum_{x,y,z \in \mathbb{Z}} \biggl[ &c^{\dagger}(x,-y,z+\tfrac{1}{2})m\tau_{3}  c(x,-y,z+\tfrac{1}{2})\biggr]\nonumber \\
	= \sum_{x,y,z \in \mathbb{Z}} \biggl[ &c^{\dagger}(x+1,y,z+\tfrac{1}{2}) \biggr( \frac{t_{1}}{2} \tau_{3} + \frac{t_{3}}{2}i\tau_{1} \sigma_{3} \biggl) c(x,y, z+\tfrac{1}{2}) \nonumber \\
	&+c^{\dagger}(x,y+1,z+\tfrac{1}{2}) \biggr(\frac{t_{1}}{2} \tau_{3}-\frac{t_{2}}{2}i\tau_{1} \sigma_{1} \biggl) c(x,y,z+\tfrac{1}{2}) +{\rm h.c.}
	\biggr]\nonumber \\
	-\sum_{x,y,z \in \mathbb{Z}} \biggl[ &c^{\dagger}(x,y,z+\tfrac{1}{2})m\tau_{3}  c(x,y,z+\tfrac{1}{2})\biggr].
\end{align}

\begin{figure}
	\includegraphics[width=0.7\columnwidth]{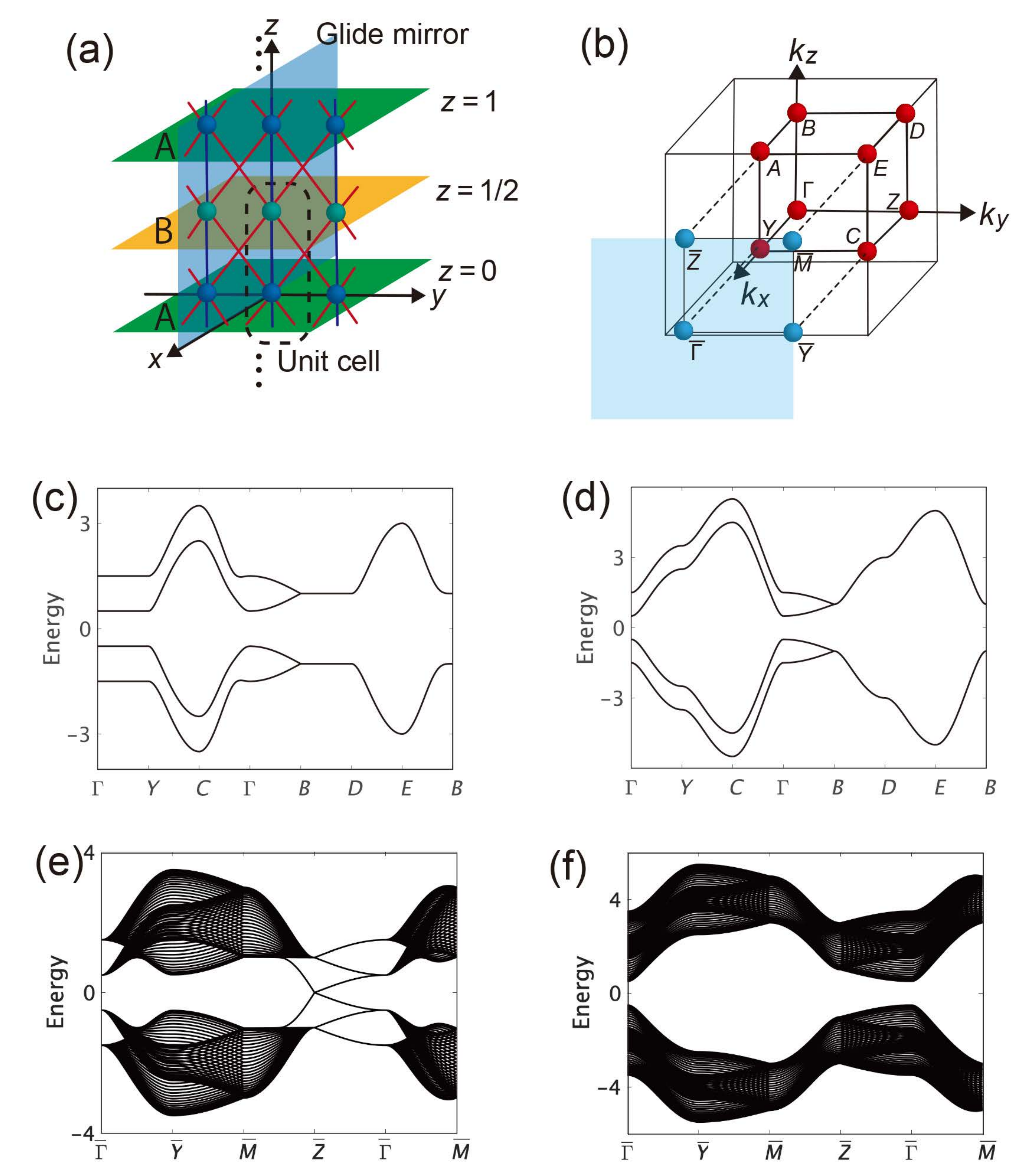}
	\caption{Tight-binding model $\mathcal{H}^{(1)}_{\rm TCI} (\boldsymbol{k})$ of a TCI with $G_y$ symmetry. (a) The 2D TI layer A at $z=0$ and the 2D TI layer B at $z=1/2$ in the unit cell. The layer construction from these layers has the $\hat{G}_y$ symmetry. 
		(b) The bulk Brillouin zone and the (100) surface Brillouin zone. (c,d) The band structures of the bulk Hamiltonian $\mathcal{H}^{(1)}_{\rm TCI}(\boldsymbol{k})$. (e,f) The band structures in the (100) slab with the finite system size along the $x$-direction $L=30$. The parameters are $t_{1}=t_{2}=t_{3}=m=1$, $t_{AB}=0.5$, $t'_{AB}=0$ in (c) and (e), and $t_{1}=t_{2}=t_{3}=1$, $t_{AB}=0.5$, $t'_{AB}=0$, $m=3$ in (d) and (f).} 
	\label{fig:SM_tight-binding_model}
\end{figure}

By combining the layers $B$ with the layers $A$, we obtain the layer constructions with the glide $\hat{G}_{y}$ symmetry. 
Furthermore, we introduce interlayer hoppings between the $A$ and $B$ layers. 
Then we obtain the overall Hamiltonian of the TCI written as 
\begin{equation}
	H^{(1)}_{\rm TCI}=H_{A}+H_{B}+H_{AB}, 
\end{equation}
where $H_{AB}$ denotes the Hamiltonian of the interlayer couplings
\begin{align}
	H_{AB}=\sum_{x,y,z \in \mathbb{Z}}\biggl[ \frac{t_{AB}}{2}\Bigl( c_{A}^{\dagger} &(x,y,z)  c_{B} (x,y,z+\tfrac{1}{2}) 
	+c_{B}^{\dagger} (x,y,z+\tfrac{1}{2})  c_{A} (x,y,z+1)  +{\rm h.c.} \Bigr)\nonumber \\
	+\frac{t'_{AB}}{2}\Bigl( &c_{A}^{\dagger} (x,y,z)  c_{B} (x+1,y,z+\tfrac{1}{2})+c_{A}^{\dagger} (x,y,z)  c_{B} (x-1,y,z+\tfrac{1}{2})\nonumber \\ 
	&+c_{B}^{\dagger} (x,y,z+\tfrac{1}{2})  c_{A} (x+1,y,z+1)+ c_{B}^{\dagger} (x,y,z+\tfrac{1}{2})  c_{A} (x-1,y,z+1) +{\rm h.c.} \Bigr)\biggr],
\end{align}
with $c_{A}^{\dagger}(x,y,z)$ and $c_{B}^{\dagger}(x,y,z)$  being the creation operators at the layers $A$ and $B$, respectively  [see Supplementary Fig.~\ref{fig:SM_tight-binding_model}(a)]. 
In this way, through the layer construction with 2D TIs, we have obtained the Hamiltonian preserving $G_y$ symmetry because this interlayer coupling does not break the glide symmetry.

\subsection{Bulk Hamiltonian and symmetry}
Let us introduce the Fourier transformations
\begin{equation}
	c^{\dagger}_{\alpha}(\boldsymbol{k})=\frac{1}{\sqrt{N}}\sum_{\boldsymbol{R}}e^{i\boldsymbol{k}\cdot (\boldsymbol{R}+\boldsymbol{r}_{\alpha})}c^{\dagger}_{\alpha}( \boldsymbol{R}+\boldsymbol{r}_{\alpha}),
\end{equation} 
where $N$ is the number of unit cells, $\alpha$ runs over eight basis states spanned by $\tau_{3}=\pm$, $\sigma_{3}=\pm$, and A/B sublattices, $\boldsymbol{r}_{\alpha}$ denotes the positions of the sites in the unit cell, and  $\boldsymbol{R}=(x,y,z)$ is the position of the unit cell with $x,y,z \in \mathbb{Z}$. By using the Fourier transformations, the Bloch Hamiltonian of $H^{(1)}_{\rm TCI}$ is 
written in terms of the eight basis states. We obtain the following Bloch Hamiltonian of $H^{(1)}_{\rm TCI}$:
\begin{align}\label{SM:eq:3dbloch_TCI}
	\mathcal{H}^{(1)}_{\rm TCI}(\boldsymbol{k})=&- ( m-t_{1} \cos k_{x}-t_{1}\cos k_{y})\mu_{0}\tau_{3}
	+t_{2}\sin k_y \mu_{3}\tau_{1}\sigma_{1}\nonumber \\
	&+t_{3} \sin k_{x} \mu_{0} \tau_{1}\sigma_{3}+(t_{AB}+2t'_{AB} \cos k_{x})\cos \tfrac{k_{z}}{2}\mu_{1}\tau_{0},
\end{align}
where $\mu_{i}$ $(i=1,2,3)$ are the Pauli matrices corresponding to the A/B sublattices, and $\mu_{0}$ is the $2\times 2$ identity matrix.
Under $\hat{G}_{y}$, a creation operator $c^{\dagger}_{\alpha}(\boldsymbol{k})$ transforms as
\begin{align}
	\hat{G}_{y}&:\ c_{\alpha}^{\dagger}(k_{x},k_{y},k_{z})\ \rightarrow \ c_{\beta}^{\dagger}(k_{x},-k_{y},k_{z}) \bigl[ G_{y} (k_{z}) \bigr]_{\beta \alpha},
\end{align}
with
\begin{align}
	G_{y}(k_{z})= -ie^{-i\frac{k_{z}}{2}}\mu_{1}\tau_{3} \sigma_{2},
\end{align}
where $\alpha$ and $\beta$ run over the basis of Eq.~(\ref{SM:eq:3dbloch_TCI}).  
The Hamiltonian $\mathcal{H}^{(1)}_{\rm TCI}(\boldsymbol{k})$ has the glide $\hat{G}_{y}$ symmetry:
\begin{equation}
	G_{y}(k_{z}) \mathcal{H}^{(1)}_{\rm TCI}(k_x, k_y, k_z)G^{-1}_{y}(k_{z})=\mathcal{H}^{(1)}_{\rm TCI}(k_x, -k_y, k_z).
\end{equation} 

Supplementary Figure \ref{fig:SM_tight-binding_model}(b) shows the bulk Brillouin zone and the (100) surface Brillouin zone. 
From the Bloch Hamiltonian $\mathcal{H}^{(1)}_{\rm TCI}(\boldsymbol{k})$, we calculate the bulk band structures [Supplementary Fig.~\ref{fig:SM_tight-binding_model}(c,d)]. Furthermore, we calculate the band structures for the (100) slab when  the system size along the $x$-direction is $L_{x}=30$ [Supplementary Fig.~\ref{fig:SM_tight-binding_model}(e,f)].

\begin{figure}
	\includegraphics[width=0.8\columnwidth]{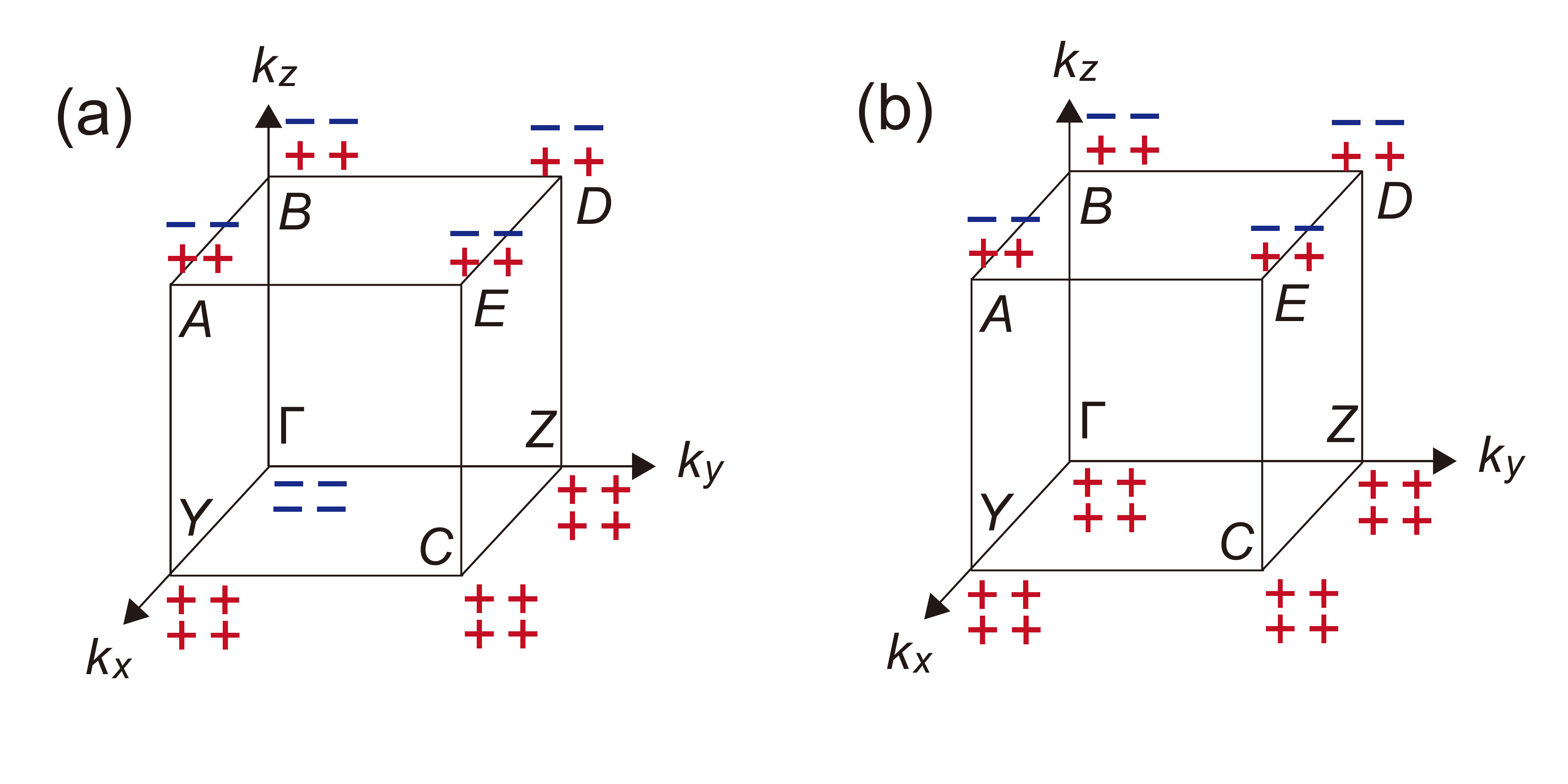}
	\caption{The parity eigenvalues of the occupied states of the bulk Hamiltonian [Eq.~(\ref{SM:eq:3dbloch_TCI})] at the high-symmetry points in the $k$-space. The parameters are $t_{1}=t_{2}=t_{3}=1$, $t_{AB}=0.5$, $t'_{AB}=0$, (a) $m=1$, and (b)  $m=3$. The parity eigenvalues at the high-symmetry points satisfy $(\mathbb{Z}_2, \mathbb{Z}_2, \mathbb{Z}_2, \mathbb{Z}_4)=(0,0,0,2)$ in (a), and $(\mathbb{Z}_2, \mathbb{Z}_2, \mathbb{Z}_2, \mathbb{Z}_4)=(0,0,0,0)$ in (b), respectively. } 
	\label{fig:SM_parity_eigenvalues}
\end{figure}

Under $\hat{I}$, a creation operator $c^{\dagger}_{\alpha}(\boldsymbol{k})$ transforms as
\begin{align}
	\hat{I}&:\ c_{\alpha}^{\dagger}(k_{x},k_{y},k_{z})\ \rightarrow \ c_{\beta}^{\dagger}(-k_{x},-k_{y},-k_{z}) \bigl[ I  \bigr]_{\beta \alpha},
\end{align}
with
$
I=  \tau_{3}.
$
Thus, the Hamiltonian $\mathcal{H}^{(1)}_{\rm TCI}(\boldsymbol{k})$ has ${I}$ symmetry:
\begin{equation}
I \mathcal{H}^{(1)}_{\rm TCI}(k_x, k_y, k_z)I^{-1}=\mathcal{H}^{(1)}_{\rm TCI}(-k_x, -k_y, -k_z).
\end{equation}
Because of $I$ symmetry, we can diagnose the topological invariant in terms of the parity eigenvalues of the occupied states at high-symmetry points in the $k$-space. 
The symmetry-based indicator \cite{bradlyn2017topological, po2017symmetry, PhysRevX.7.041069} in the space group No.~13 ($P2/c$) can be expressed as $(\mathbb{Z}_{2},  \mathbb{Z}_{2}, \mathbb{Z}_{2}, \mathbb{Z}_{4})$ with the strong $\mathbb{Z}_4$ indicator being
\begin{equation}
	z_{4}=\frac{1}{4}\sum_{\Gamma_{j}}\Bigl( n_{+}(\Gamma_{j})-n_{-}(\Gamma_{j}) \Bigr), 
\end{equation}
where $n_{+}(\Gamma_{j})$ ($n_{-}(\Gamma_{j})$) is the number of occupied states with even (odd) parity eigenvalues at the high-symmetry points $\Gamma_{j}$ $(=\Gamma$, $Y$, $Z$, $C$, $B$, $A$, $D$, $E)$ in the Brillouin zone. When the three weak $\mathbb{Z}_{2}$ indices are trivial ($(\mathbb{Z}_2, \mathbb{Z}_2, \mathbb{Z}_2 )=(0,0,0)$), $z_{4}=2$ indicates that a given insulator is a TCI protected by the glide symmetry \cite{song2018quantitative, PhysRevB.100.165202}. We obtain the parity eigenvalues of the occupied states from the bulk Hamiltonian [Eq.~(\ref{SM:eq:3dbloch_TCI})] with the parameters being  $t_{1}=t_{2}=t_{3}=1$, $t_{AB}=0.5$, $m=1$ [Supplementary Fig.~\ref{fig:SM_parity_eigenvalues}(a)], and  $m=3$ [Supplementary Fig.~\ref{fig:SM_parity_eigenvalues}(b)]. Supplementary Fig.~\ref{fig:SM_parity_eigenvalues}(a) ($m=1$) shows parity eigenvalues for a topologically nontrivial case with the symmetry-based indicator $(\mathbb{Z}_2, \mathbb{Z}_2, \mathbb{Z}_2, \mathbb{Z}_4)=(0,0,0,2)$.  In the band structures in the slab geometry with these parameters, the gapless states appear on the (100) surface as shown in Supplementary Fig.~\ref{fig:SM_tight-binding_model}(e). On the other hand, Supplementary Fig.~\ref{fig:SM_parity_eigenvalues}(b) ($m=3$) shows a topologically trivial case with the symmetry-based indicator $(\mathbb{Z}_2, \mathbb{Z}_2, \mathbb{Z}_2, \mathbb{Z}_4)=(0,0,0,0)$, which results in the absence of the gapless states on the (100) surface [Supplementary Fig.~\ref{fig:SM_tight-binding_model}(f)]. 

\begin{figure*}
	\includegraphics[width=0.8\columnwidth]{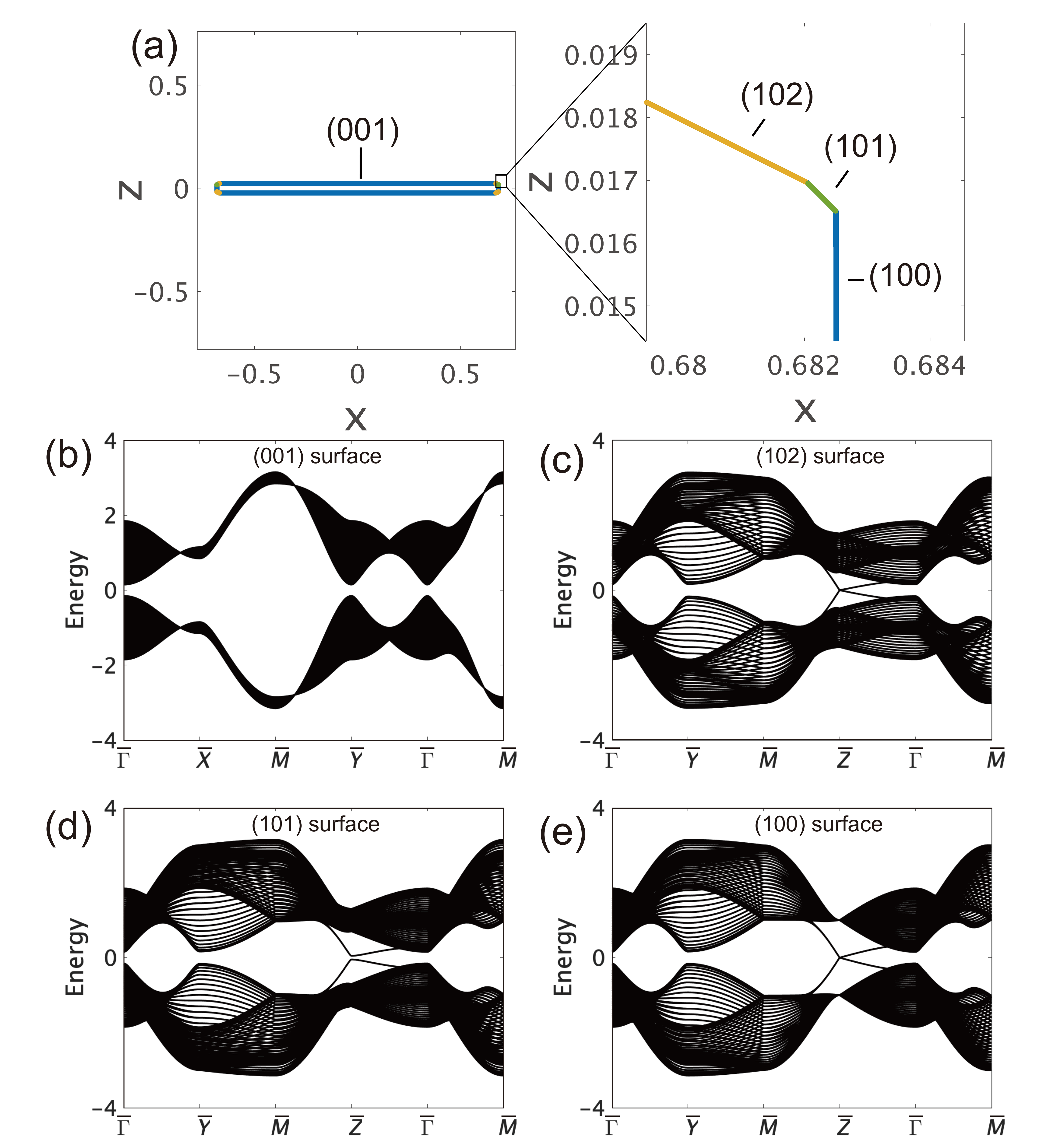}
	\caption{Equilibrium shape of the crystals for the Hamiltonian $\mathcal{H}^{(1)}_{\rm TCI}(\boldsymbol{k})$ from the Wulff construction. The parameters of the tight-binding model [Eq.~(\ref{SM:eq:3dbloch_TCI})] are set to be  $t_{1}=t_2=t_3=m=1$, $t_{AB}=0.35$, $t'_{AB}=0.25$. The slab thickness is $L=30$. $E_{\rm surf}^{\alpha \gamma }$ is obtained from Eq.~(\ref{SM:eq:3dbloch_TCI}). The maximum absolute Miller indices are $5$.  (a) The crystal shapes obtained from the results of the surface energy $E_{\rm surf}^{(\alpha 0 \gamma)}$.  (b-e) Band structures calculated in the slab geometry, where the slabs have the (001) surface in (b), the (102) surface in (c), the (101) surface in (d), and the $(100)$ surface in (e). } 
	\label{fig:surface_energy_Miller}
\end{figure*}

\subsection{Crystal shape of the nonsymmorphic topological crystalline insulator}
Now we discuss an equilibrium crystal shape for the TCI phase protected by $G_y$ symmetry [Eq.~(\ref{SM:eq:3dbloch_TCI})], which is determined to minimize the surface energy of the crystal. 
Here, we study a crystal with a cylindrical shape along the $y$-axis, which limits the crystal surfaces to be along $(\alpha 0 \gamma)$.
By using the Wulff construction \cite{wulffconst}, we can obtain an equilibrium crystal shape from the surface energy $E_{\rm surf}^{(\alpha 0 \gamma)}$. For the details of the calculations of the surface energies, see ``Methods'' in the main text. 
The surface energy are calculated up to a maximum absolute value of the Miller index  of 5. 
From these surface energies, we can obtain an equilibrium shape of the crystal [Supplementary Fig.~\ref{fig:surface_energy_Miller}(a)]. 
The equilibrium crystal shape of the TCI has the (001) surface over a wide range because the (001) surface energy is much lower than those of the other surfaces. 
This property results from two factors. First, the absence of topological surface states on the (001) surface results in the preference of the (001) surface over other surfaces with the topological gapless states. 
Second,  in this calculation of the tight-binding model,  the hopping amplitudes in the $z$ direction, $t_{AB}$ and $t'_{AB}$ are smaller than those in the $x$ and $y$ directions,  $t_{1}$, $t_2$, and $t_3$. It lowers the (001) surface energy over the (100) surface because the surface energy increases as the strengths of the chemical bondings along  the direction perpendicular to the surface increase. 

We find that the crystal shape of the TCI has the (001), (102), (101), and (100) facets in Supplementary  Fig.~\ref{fig:surface_energy_Miller}(a). 
In addition, we calculate the band structures in the slab geometry with the (001), (102), (101), and (100) surfaces [Supplementary Fig.~\ref{fig:surface_energy_Miller}(b-e)].
Because the (001) and the (101)  surfaces are not glide symmetric, the surface states are gapped [Supplementary Figs.~\ref{fig:surface_energy_Miller}(b) and  \ref{fig:surface_energy_Miller}(d)], whereas the glide-symmetric (102) and (100)  surfaces are gapless [Supplementary Figs.~\ref{fig:surface_energy_Miller}(c) and \ref{fig:surface_energy_Miller}(e)].
Thus, the crystal of the TCI is formed by the (001), (102), (101), and (100) facets with the (100) and  (102) facets being gapless and the (001) and (101) facets being gapped. Our calculations of surface energies automatically include contributions both from surface states and from bulk occupied states. The latter one can be physically regarded as a contribution from chemical bonding between atoms. 

\subsection{Surface energy of $\mathcal{H}^{(1)}_{\rm TCI}(\boldsymbol{k})$}\label{appendix_surface_energy_first_model}
In this section, we discuss the surface energies of our model $\mathcal{H}^{(1)}_{\rm TCI}(\boldsymbol{k})$. Henceforth, we choose the parameters  $t_{1}=t_{2}=t_{3}=1$.  As we show in Fig. 1(d-f) in the main text, the flat bands appear along $\bar{\Gamma}$-$\bar{Z}$ on the (101), (102), and (201) surfaces when $t'_{AB}=0$. On the other hand, when $t'_{Ab}\neq 0$, they are no longer flat as shown in Supplementary Fig.~\ref{fig:surface_energy_Miller}.  In the following, we discuss the cases of $t'_{AB}\neq 0$ and  $t'_{AB}=0$. 

\begin{figure}[t]
	\includegraphics[width=1.\columnwidth]{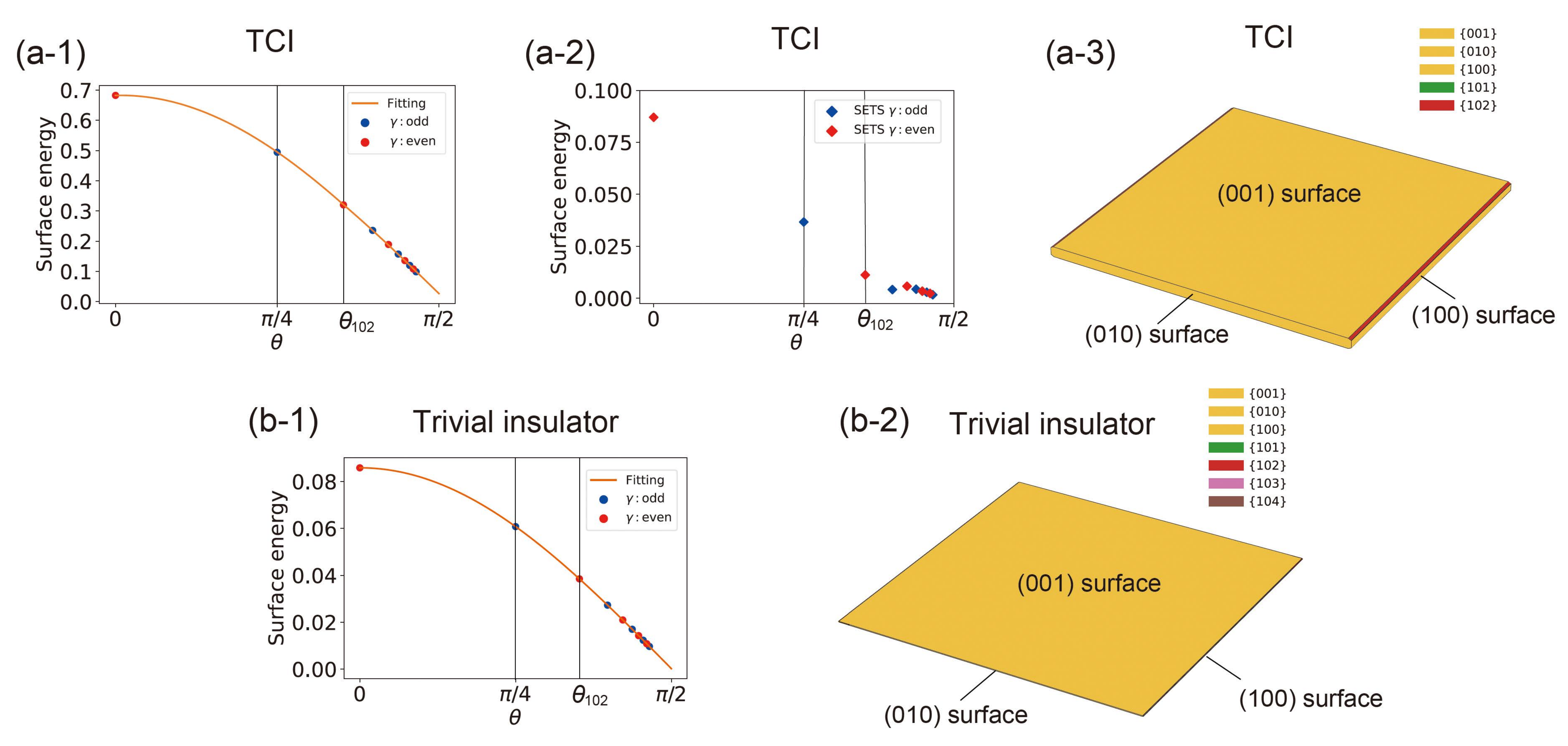}
	\caption{Surface energy and crystal shapes of $\mathcal{H}^{(1)}_{\rm TCI}(\boldsymbol{k})$ when $t'_{AB}\neq 0$. We choose the parameters $t_{1}=t_{2}=t_{3}=1$, $t_{AB}=0.35$, $t'_{AB}=0.25$, (a) $m=1$ (TCI), and  (b) $m=6$ (trivial insulator). (a-1,b-1) The ($10\gamma$) surface energy from $\gamma=0$ to $\gamma=9$. The dots are determined by  $E^{(\alpha \beta \gamma)}_{\rm surf}$. We fit the surface energy with $F(\theta)$, where the fitting parameters are $\Delta_{x}=0.6642$, $\Delta_{z}=0.01764$, $\Delta_{xz}=0.01018$  in (a-1) and $\Delta_{x}=0.08562$, $\Delta_{z}=0.0001106$, $\Delta_{xz}=0.0001134$ in (b-1). (a-2) The SETS $E^{(\alpha \beta \gamma)}_{\rm SETS}$ of the TCI. (a-3) The equilibrium crystal shape of the TCI. (b-2) The equilibrium crystal shape of the trivial insulator. } \label{fig:surface_energy_first_model}
\end{figure}

\subsubsection{$\mathcal{H}^{(1)}_{\rm TCI}(\boldsymbol{k})$ with $t'_{AB}\neq 0$.}
To begin with, we discuss the case of the TCI phase by setting the parameters $m=1$, $t_{AB}=0.35$, and $t'_{AB}=0.25$.  Supplementary Fig.~\ref{fig:surface_energy_first_model}(a-1) shows the surface energy $E^{(\alpha \beta \gamma)}_{\rm surf}$ of $\mathcal{H}^{(1)}_{\rm TCI}(\boldsymbol{k})$ as a function of the angle $\theta$ between the surface and the (100) plane, defined by $\tan \theta=\gamma / \alpha$. 
Then Supplementary Fig.~\ref{fig:surface_energy_first_model}(a-2) shows the $E^{(\alpha \beta \gamma)}_{\rm SETS}$. The SETS does not have a peak at $\theta=\theta_{102}$, unlike the $\mathcal{H}^{(2)}_{\rm TCI}(\boldsymbol{k})$ we discuss in the main text. 
Next, as in the main text,  we introduce the following trial function:
\begin{align}\label{eq:fitting_secb}
	F(\theta)=\Delta_{x}\cos|\theta|+\Delta_{z}\sin|\theta|
	+\Delta_{xz}\sin|\theta-\theta_{102}|+\Delta_{xz}\sin|\theta+\theta_{102}|,
\end{align}
where $\theta_{102}$ is the angle satisfying $\cos \theta_{102}=1/\sqrt{5}$ and $\sin \theta_{102}=2/\sqrt{5}$.  
This function can be obtained by the analysis in terms of the numbers of dangling  bonds on the surface in \ref{Sup:surface_ene_chem_bond}.
We fit $E^{(\alpha \beta \gamma)}_{\rm surf}$ in Supplementary Fig.~\ref{fig:surface_energy_first_model}(a-1) with $F(\theta)$. The function does not have the cusp at $\theta= \theta_{102}$ unlike the case of  $\mathcal{H}^{(2)}_{\rm TCI}(\boldsymbol{k})$ in the main text. In addition, the surface energy at $\theta=\pi/2$ is small compared to the other surface energies, leading to the (001) surface occupying the majority of the total surface of the crystal [Fig.~\ref{fig:surface_energy_first_model}(a-3)]. 

Next, we discuss the case of the trivial insulator phase by setting $m=6$, $t_{AB}=0.35$, and $t'_{AB}=0.25$. 
 Supplementary Fig.~\ref{fig:surface_energy_first_model}(b) shows the dependence of $E^{(\alpha \beta \gamma)}_{\rm surf}$ on the surface orientations as a function of the angle $\theta$, and we fit $E^{(\alpha \beta \gamma)}_{\rm surf}$ with $F(\theta)$. This fitting and that for TCI [Supplementary Fig.~\ref{fig:surface_energy_first_model}(a-1)] have a similar dependence on $\theta$. 
Furthermore, Supplementary Fig.~\ref{fig:surface_energy_first_model}(b-2) shows the equilibrium crystal shape of the trivial insulator. This crystal shape is similar to that of the TCI [Supplementary Fig.~\ref{fig:surface_energy_first_model}(a-3)] in that the (001) surface occupies the majority of the total surface of the crystals.  
These results indicate that it is difficult to see the effect of topological surface states on the crystal shape in $\mathcal{H}^{(1)}_{\rm TCI}(\boldsymbol{k})$, unlike $\mathcal{H}^{(2)}_{\rm TCI}(\boldsymbol{k})$ in the main text. 
In this model, because the interlayer couplings are weak,  the (001) surface energy becomes significantly smaller than other surfaces, and the SETS cannot affect the crystal shape. 

\begin{figure}[t]
	\includegraphics[width=1.\columnwidth]{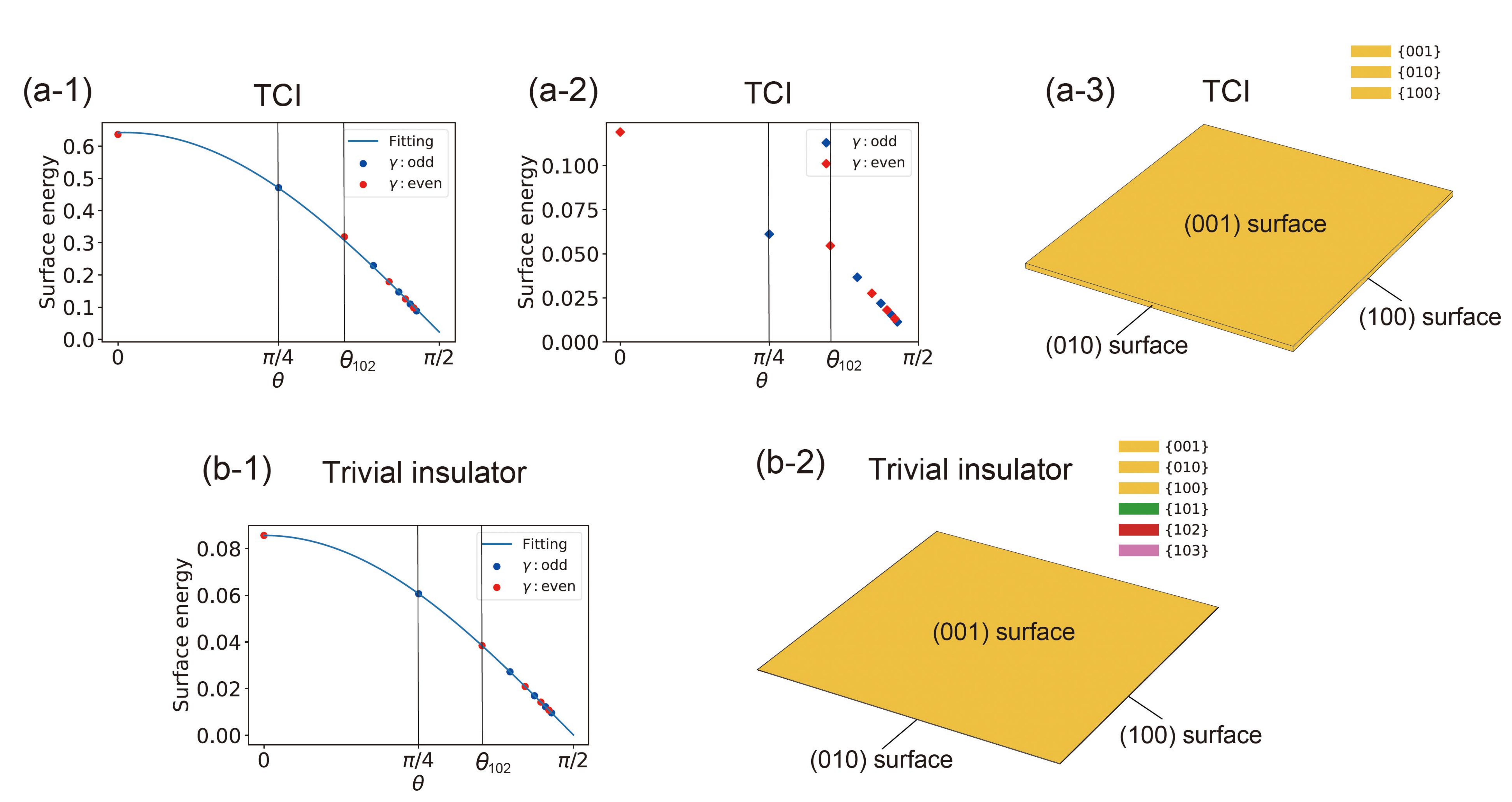}
	\caption{Surface energy and crystal shapes of $\mathcal{H}^{(1)}_{\rm TCI}(\boldsymbol{k})$ when $t'_{AB}= 0$. We choose the parameters $t_{1}=t_{2}=t_{3}=1$, $t_{AB}=0.5$, $t'_{AB}=0$, (a) $m=1$ (TCI), and  (b) $m=6$ (trivial insulator). (a-1,b-1) The ($10\gamma$) surface energy from $\gamma=0$ to $\gamma=9$. The dots are determined by  $E^{(\alpha \beta \gamma)}_{\rm surf}$. We fit the surface energy with $F(\theta)$, where the fitting parameters are $\Delta_{x}=0.6422$, $\Delta_{z}=0.02270$  in (a-1) and $\Delta_{x}=0.08560$, $\Delta_{z}=0.0001406$ in (b-1). (a-2) The SETS $E^{(\alpha \beta \gamma)}_{\rm SETS}$ of the TCI. (a-3) The equilibrium crystal shape of the TCI. (b-2) The equilibrium crystal shape of the trivial insulator. }\label{fig:surface_energy_first_model_flat}
\end{figure}

\subsubsection{$\mathcal{H}^{(1)}_{\rm TCI}(\boldsymbol{k})$ with $t'_{AB}= 0$.}
Here, we show that the SETS cannot affect the crystal shape also when $t'_{AB}= 0$. We begin with the case of the TCI phase with $m=1$, $t_{AB}=0.5$, and  $t'_{AB}=0$. 
Supplementary Fig.~\ref{fig:surface_energy_first_model_flat}(a-1) shows the dependence of the surface energies $E^{(\alpha \beta \gamma)}_{\rm surf}$ on the surface orientations as a function of the angle $\theta$. In addition, Supplementary Fig.~\ref{fig:surface_energy_first_model_flat}(a-2) indicates the $E^{(\alpha \beta \gamma)}_{\rm SETS}$. Because $t'_{AB}= 0$,  the second-nearest neighbor hopping vanishes in this model, which means that the third and fourth terms in $F(\theta)$ are not necessary [Eq.~(\ref{eq:fitting_secb})].  Therefore, we introduce the fitting function 
\begin{align}\label{eq:fitting_secb_second}
	F'(\theta)=\Delta_{x}\cos|\theta|+\Delta_{z}\sin|\theta|.
\end{align}
We fit $E^{(\alpha \beta \gamma)}_{\rm surf}$ in Supplementary Fig.~\ref{fig:surface_energy_first_model_flat}(a-1) with $F'(\theta)$. As in the case with $t_{AB}\neq 0$, the surface energy at $\theta=\pi/2$ is so small compared to the other surface energies also in this case. It leads to the (001) surface occupying the majority of the total surface of the crystal [Fig.~\ref{fig:surface_energy_first_model_flat}(a-3)]. 

In addition, the case of the trivial insulator with $m=6$, $t_{AB}=0.5$, $t'_{AB}=0$ shows the similar behaviors as shown in Supplementary Figs.~\ref{fig:surface_energy_first_model_flat}(b-1) and \ref{fig:surface_energy_first_model_flat}(b-2). 
Thus, it is difficult to see the effects of the topological surface states on the crystal shape, as in the case with $t'_{AB}\neq 0$.


\subsection{Surface energy without topological surface states}\label{appendix_surface_energy_from_chemical}
Here we estimate the values of the surface energy due to chemical bonding by calculating the energy of the slab of the TCI protected by $G_y$ symmetry. 
For this purpose, we consider the simplest slab of the facet with the Miller index $(001)$ because topological surface states do not appear on this surface. 

\begin{figure}[t]
	\includegraphics[width=1.\columnwidth]{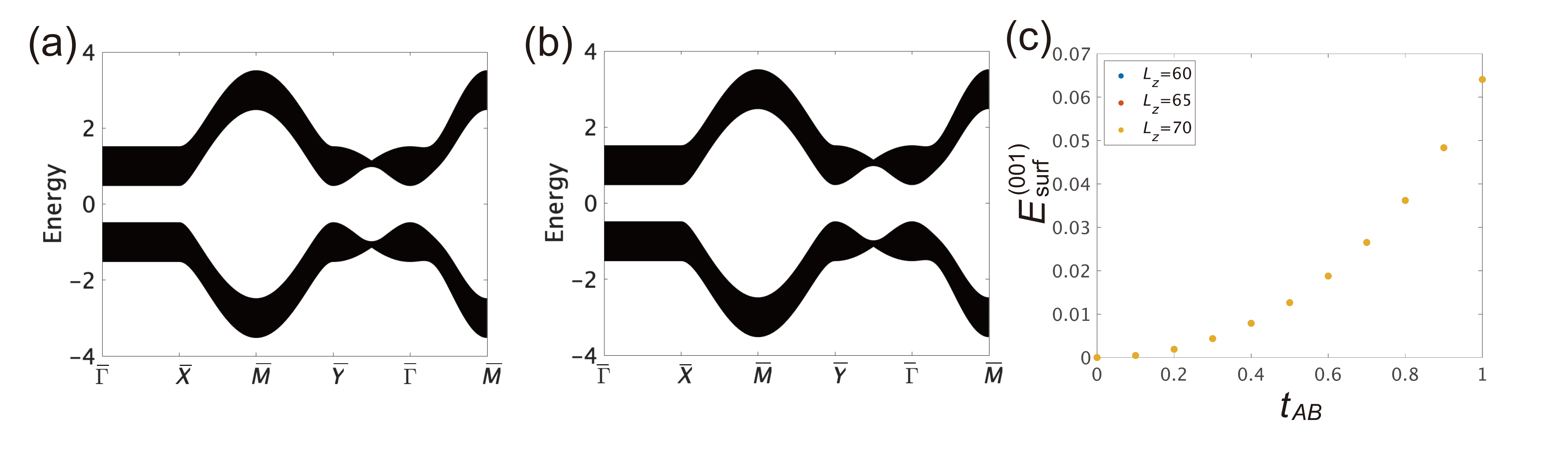}
	\caption{Surface energy without topological surface states with the parameters being $t_{1}=t_{2}=t_{3}=m=1$, $t'_{AB}=0$.  (a,b) The band structures of the slab with $t_{AB}=0.5$. The system size is $L_{z}=60$. The boundary conditions in the $z$-direction are (a) periodic boundary condition and (b) open boundary condition. (c) The surface energy $E^{(001)}_{\rm surf}$ with various values of $t_{AB}$ and  a mesh in the $k$-space being $30\times 30$. } 
	\label{fig:the surface energy from the chemical bondings}
\end{figure}

We use the simple tight-binding model [Eq.~(\ref{SM:eq:3dbloch_TCI})]. 
This slab has a finite thickness in the $z$-direction and has periodic boundary conditions in the $x$- and $y$-directions.
Let us introduce hoppings between the top surface and the bottom surface of the slab, and then we replace this hopping parameter $t_{AB}$ as $\lambda t_{AB}$ with $\lambda$ being a real parameter. When $\lambda=0$, this system is just a slab geometry. On the other hand, when $\lambda=1$, this system is equal to a bulk crystal because it has the periodic boundary condition in the $x$, $y$, and $z$-directions.  Let $H^{(001)}_{\rm slab}(\lambda)$ denote the Hamiltonian with such geometry.
Supplementary Figs.~\ref{fig:the surface energy from the chemical bondings} (a) and \ref{fig:the surface energy from the chemical bondings}(b) show  band structures of $H^{(001)}_{\rm slab}(\lambda)$ with $\lambda=1$ and $\lambda=0$, respectively. 
From this result, we confirm that  the topological gapless states do not appear on the (001) surface.  
The surface energy of the Hamiltonian $H^{(001)}_{\rm slab}(\lambda)$  can be expressed as 
\begin{equation}\label{Seq:definition_surface_energy}
	E_{\rm surf}^{(001)}=\sum^{N}_{n=1}\frac{E^{(001)}_{{\rm slab}, n}|_{\lambda=0}-E^{(001)}_{{\rm slab},n}|_{\lambda=1}}{2S_{\rm slab}},
\end{equation}
where $E^{(\alpha \beta \gamma)}_{{\rm slab},n}|_{\lambda}$ is the energy from the $n$-th lowest band of $H^{(\alpha \beta \gamma)}_{\rm slab}(\lambda)$, $N$ is the total number of occupied bands, and $S_{\rm slab}$ represents the area of the slab surface of $H^{(\alpha \beta \gamma)}_{\rm slab}(\lambda)$.
Dependence of the surface energies, $E^{(001)}_{\rm surf}$ on $t_{AB}$ with the parameters $t_{1}=t_{2}=t_{3}=m=1$  is shown in  Supplementary Fig.~\ref{fig:the surface energy from the chemical bondings}(c), and it shows that $E_{\rm surf}^{(001)}$ increases as $t_{AB}$ increases.  This result supports that the surface energies without the topological surface states is largely related to the strength of the chemical bonds. 

\section{Wilson loop approach for diagnosing topological invariants}\label{AppendixC:Wilsonloops_diagnosing_topological}
\subsection{Wilson loop matrix and its eigenvalues}
Here we review the Wilson loop approach for calculations of the topological invariant based on Refs.~\cite{PhysRevB.84.075119, PhysRevB.89.155114, wang2016hourglass, wieder2018wallpaper} and calculate the Wilson loop for our model.
First of all, we introduce the notations of tight-binding model for calculations of the Wilson loop eigenvalues and diagnosing topological invariants.  We start from the tight-binding orbitals $\phi_{\boldsymbol{R} \alpha}(\boldsymbol{r})=\varphi_{\alpha}(\boldsymbol{r}-\boldsymbol{R}-\boldsymbol{r}_{\alpha})$  in the unit cell indexed by the lattice vector $\boldsymbol{R}$, where $\boldsymbol{r}_{\alpha}$ denotes the positions of the sites in the unit cell.  By the Fourier transformations, the wave function is given by
$
	\ket{\chi_{\alpha}(\boldsymbol{k})} =\sum_{\boldsymbol{R}}e^{i\boldsymbol{k}\cdot (\boldsymbol{R}+\boldsymbol{r}_{\alpha})} \ket{\phi_{\boldsymbol{R} \alpha}}/\sqrt{N_{\rm u}},
$
 where $N_{\rm u}$ is the number of unit cells. 
A Hamiltonian $\hat{H}$ can be expressed  as the Hamiltonian matrix in terms of these wave functions,  
$
	[\hat{\mathcal{H}}(\boldsymbol{k})]_{\alpha \beta}= \bra{\chi_{\alpha}(\boldsymbol{k})}\hat{H}\ket{\chi_{\beta}(\boldsymbol{k})},
$
where this Hamiltonian has eigenstates defined as $\ket{u_{n}(\boldsymbol{k})}$. 
Under a translation of a reciprocal lattice vector $\boldsymbol{G}$ in the $k$-space,  this Hamiltonian matrix transforms as
\begin{equation}
	\hat{\mathcal{H}}(\boldsymbol{k}+\boldsymbol{G})=\hat{V}^{-1}(\boldsymbol{G})\hat{\mathcal{H}}(\boldsymbol{k})\hat{V}(\boldsymbol{G}),
\end{equation}
where the matrix representation of $\hat{V}(\boldsymbol{G})$ with the basis of  $[\hat{\mathcal{H}}(\boldsymbol{k})]_{\alpha \beta}$ is given by 
$
	[V ({\boldsymbol{G}})]_{\alpha \beta}=\delta_{\alpha \beta}e^{i\boldsymbol{G}\cdot \boldsymbol{r}_{\alpha}}.
$
Therefore,  the eigenstates of the Hamiltonian at the $\boldsymbol{k}+\boldsymbol{G}$ point in the $k$-{space} can be written as 
\begin{equation}
	\ket{u_{n}(\boldsymbol{k}+\boldsymbol{G})}=\hat{V}(\boldsymbol{G})^{-1}\ket{u_{n}(\boldsymbol{k})}.
\end{equation}

Next, we discuss the methods of  calculations of the Wilson loops. 
We can diagnose the topological invariant in a TCI with nonsymmorphic crystal symmetries \cite{PhysRevB.90.085304, wang2016hourglass, wieder2018wallpaper} by tracking the evolution of Wannier function centers.
Here we consider a 3D crystal with mutually orthogonal crystal axes. 
The spectrum of the Wannier function centers along the $x$-direction is obtained by diagonalizing  a $x$-directed Wilson loop matrix \cite{PhysRevB.84.075119, PhysRevB.89.155114, wieder2018wallpaper} defined by
\begin{align}
	&[\mathcal{W}(k_{x0},k_{y},k_z)]_{nm} \nonumber \\
	\equiv & \biggl[ \mathcal{P} \exp \Bigl( i\int_{k_{x0}}^{k_{x0}+2\pi} dk_{x} \mathcal{A}_x (k_{x},k_y, k_z) \Bigr)  \biggr]_{nm},
\end{align}
where  $ \mathcal{A}_x (\boldsymbol{k}) $  is a non-Abelian Berry connection
$
[\mathcal{A}_x (\boldsymbol{k}) ]_{nm}=i\bra{u_{n}(\boldsymbol{k})}\partial_{x} \ket{u_{m}(\boldsymbol{k})}
$,  $\ket{u_{n}(\boldsymbol{k})}$ is the periodic part  of the Bloch wave function, and $\mathcal{P}$ denotes the path-ordered operation.
By using a discrete form of $\mathcal{A}_x ( k_{x0}+\tfrac{2\pi j}{N}, k_y, k_z )$,  we obtain the following equation:
\begin{align}
	& \biggl[ \exp \biggl( i \frac{2\pi}{N}\mathcal{A}_x ( k_{x0}+\tfrac{2\pi j}{N}, k_y, k_z ) \biggl)  \biggl]_{nm}\nonumber \\
	\approx &  \delta_{nm} + i\frac{2\pi}{N}\biggl[ \mathcal{A}_{x} (k_{x0}+\tfrac{2\pi j}{N}, k_{y},k_{z})      \biggl]_{nm}\nonumber \\
	\approx & \delta_{nm}-\bra{u_n (k_{x0}+\tfrac{2\pi j}{N},k_{y},k_{z})}  \Bigl( \ket{u_{m}(k_{x0}+\tfrac{2\pi j}{N}, k_{y}, k_{z})}-\ket{u_{m}(k_{x0}+\tfrac{2\pi  (j-1)}{N}, k_{y}, k_{z})}   \Bigr)\nonumber \\
	=& \braket{u_n \bigl( k_{x0}+\tfrac{2\pi j}{N}, k_{y}, k_{z} \bigr) | u_{m}(k_{x0}+\tfrac{2\pi (j-1)}{N}, k_{y}, k_{z})}.\label{eq:discrete_berry_con_kffe}
\end{align}
By rewriting $dk_x$ as $dk_{x} = 2\pi /N$ with $dk_{x}\ll 2\pi$ ($N$ is a large integer) and using Eq.~(\ref{eq:discrete_berry_con_kffe}),  the discrete form of the Wilson loop matrix can be expressed as 
\begin{align}\label{appendix:willsonloop_def}
	[\mathcal{W}(k_{x0},k_{y},k_z)]_{nm} 
	&\approx  \biggl[ \mathcal{P}  \prod_{j=1}^{N} \exp \Bigl( i\frac{2\pi}{N}  \mathcal{A}_x (k_{x0}+\tfrac{2\pi j}{N},k_y, k_z) \Bigr)  \biggr]_{nm}\nonumber \\
	&\approx \bra{u_{n}(k_{x0} +2\pi,k_y, k_z)} \biggl( \mathcal{P} \prod_{j=1}^{N-1} \hat{P} (k_{x0}+\tfrac{2\pi j}{N}, k_{y}, k_{z}) \biggl) \ket{u_{m}(k_{x0},k_y,k_z)}\nonumber \\
	& = \bra{u_{n}(k_{x0},k_y, k_z)} \hat{V}(2\pi \hat{x}) \biggl( \mathcal{P}  \prod_{j=1}^{N-1} \hat{P} (k_{x0}+\tfrac{2\pi j}{N}, k_{y}, k_{z}) \biggl)  \ket{u_{m}(k_{x0},k_y,k_z)},
\end{align}
where $\hat{P}(\boldsymbol{k})$ is the projection operator onto the occupied states defined by
\begin{equation}
	\hat{P}(\boldsymbol{k})=\sum_{n}^{n_{\rm occ}}\ket{u_n (\boldsymbol{k})} \bra{u_n (\boldsymbol{k})}.
\end{equation}
The Wilson loop matrix of Eq.~(\ref{appendix:willsonloop_def}) is a $U(n_{\rm occ})$ unitary matrix, and the eigenvalues of Eq.~(\ref{appendix:willsonloop_def}) are gauge invariant.
We label the eigenvalues of the  Wilson loop matrix $\mathcal{W}(k_{x0}, k_y, k_z)$ by $\exp [i \theta_n (k_{y}, k_{z})]$ with $n=1,\cdots, n_{\rm occ}$. The phase $\theta_n (k_{y}, k_{z})$ gives us Wannier centers of the occupied bands. 

\begin{figure}
	\includegraphics[width=0.8\columnwidth]{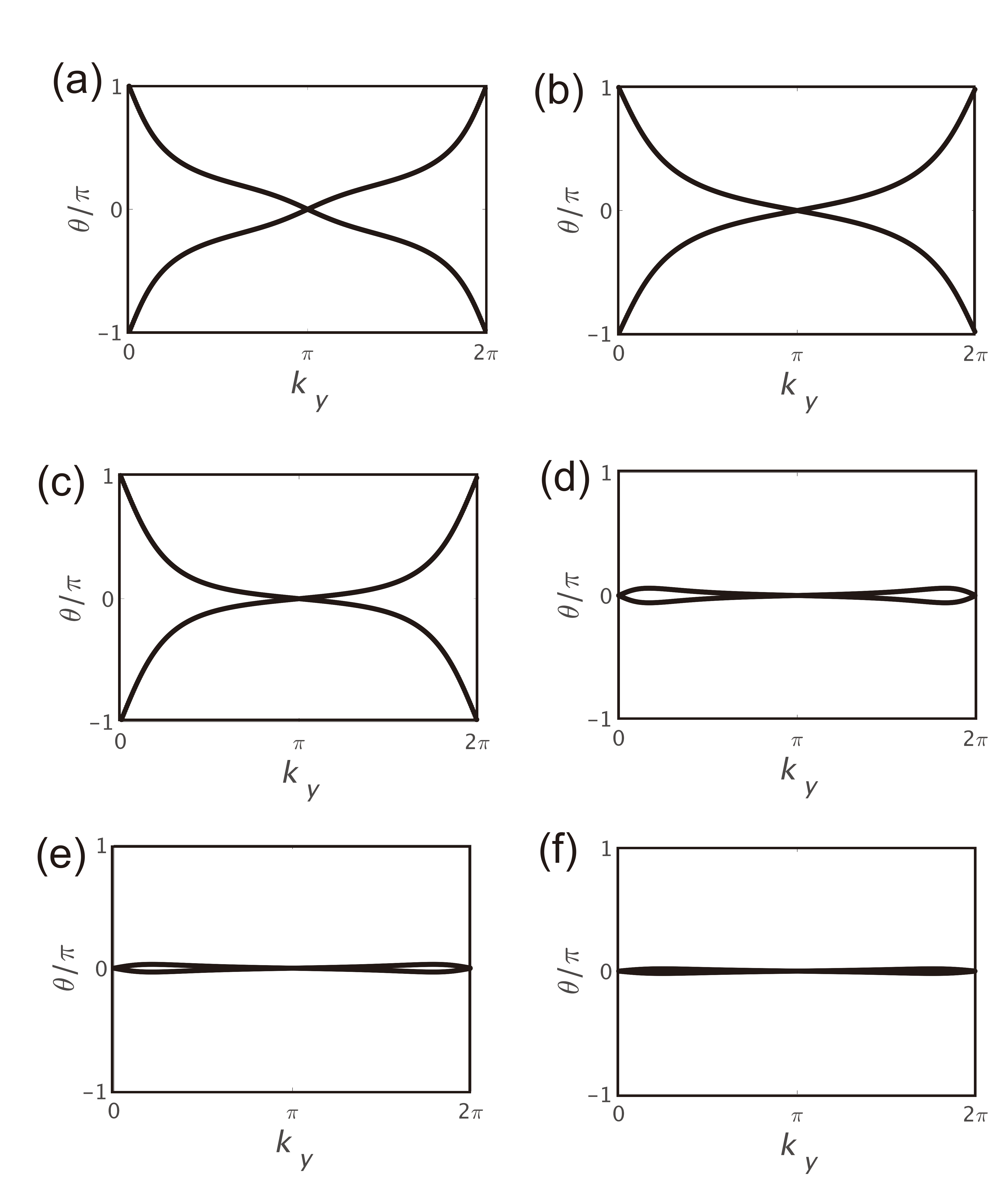}
	\caption{The spectra of the $x$-directed Wilson loop matrix with $k_z=0$ of Eq.~(\ref{SM:eq:3dbloch_TCI}). The parameters are set to be $t_{1}=t_{2}=t_{3}=1$, $t_{AB}=0.5$, and (a) $m=0.6$, (b) $m=1$, (c) $m=1.4$, (d) $m=2.6$, (e) $m=3$, (f) $m=3.4$. (a-c) Topologically nontrivial phases. (d-f) Trivial phases.} 
	\label{fig:nonmagnetic_wilsonloop}
\end{figure}

We calculate the spectra of the $x$-directed Wilson loop matrix in the tight-binding model of Eq.~(\ref{SM:eq:3dbloch_TCI}) [Supplementary Fig.~\ref{fig:nonmagnetic_wilsonloop}]. 
If the change of the phase $\theta_{n}$ through the change of $k_y$ is nonzero, the system is topological. 
The bulk closes when with $t_{1}=t_{2}=t_{3}=1$, $t_{AB}=0.5$, and $1.5\leq m \leq 2.5$. When $0.5 \leq m < 1.5$, the Wilson loop spectra wind in the change of $k_y$ from zero to $2\pi$ [Supplementary Fig.~\ref{fig:nonmagnetic_wilsonloop}(a-c)]. On the other hand, the Wilson loop spectra with $m >2.5 $ are trivial [Supplementary Fig.~\ref{fig:nonmagnetic_wilsonloop}(e-f)]. From these results, we conclude that the topological phase transition occurs in changing the parameter $m$.

\subsection{Topological invariant for glide-symmetric TCI}
Here we review a topological invariant for the 3D TCI protected by glide $G_y = \{M_y | 00 \frac{1}{2} \}$ symmetry introduced in the previous work \cite{PhysRevB.93.195413}.
In the following, we show that this topological invariant can be calculated from the Wilson loop spectra based on the discussions in this previous work. 
Let us consider a spinful 3D insulator with time-reversal ($\mathcal{T}$) symmetry
\begin{equation}
	\mathcal{T} \mathcal{H} (\boldsymbol{k}) \mathcal{T}^{-1}=\mathcal{H}(-\boldsymbol{k}), \ \mathcal{T}^2=-1.
\end{equation}
Next, we define the Berry connections as 
 \begin{gather}
 A_{j}(\boldsymbol{k})=i\sum_{n}^{n_{\rm occ}}\bra{u_n (\boldsymbol{k})}\partial_{k_j} \ket{u_n (\boldsymbol{k})},\label{SM:eq_Berry_connection}\\ 
 A_{\pm,j}(\boldsymbol{k})=i\sum_{n}^{n_{\rm occ}} \bra{u^{(\pm)}_n (\boldsymbol{k})}\partial_{k_j} \ket{u_{n}^{(\pm)} (\boldsymbol{k})},\label{SM:eq_Berry_connection_glide_sector}
 \end{gather}
with $j=x,y,z$, where the subscript $\pm$ represents the glide sector of the glide eigenvalues $\pm ie^{-ik_z/2}$, and we define the Berry curvatures as   
\begin{gather}
	F_{ij}(\boldsymbol{k})=\partial_{k_i}A_{j}(\boldsymbol{k})-\partial_{k_j}A_{i}(\boldsymbol{k}),\label{SM:eq_Berry_curvature}\\
	F_{\pm, ij}(\boldsymbol{k})=\partial_{k_i}A_{\pm,j}(\boldsymbol{k})-\partial_{k_j}A_{\pm,i}(\boldsymbol{k}).\label{SM:eq_Berry_curvature_glide_sector}
\end{gather}
According to Ref.~\cite{PhysRevB.93.195413}, the $\mathbb{Z}_{4}$ topological invariant for the $\mathcal{T}$-symmetric TCI is given by 
\begin{align}\label{eq:supplement_topo_inv_3DTRSTCI}
	\nu \equiv & \frac{2}{\pi} \int^{\pi}_{-\pi} d k_{x} \biggl( A^{\rm I}_{+,x}(k_x, \pi,\pi)- A^{\rm I}_{+,x}(k_x, 0, \pi) \biggl) \nonumber \\
	&+\frac{1}{\pi} \int^{\pi}_{0}dk_z \int^{\pi}_{-\pi}dk_x \biggl( F_{+,xz}(k_{x},\pi, k_z)- F_{+,xz}(k_{x}, 0, k_{z})  \biggr) \nonumber \\
	&+\frac{1}{2\pi}\int^{\pi}_{0}dk_y \int^{\pi}_{-\pi}dk_x F_{xy}(k_x,k_y,0)\ \ {\rm mod}\ 4,
\end{align}
where the subscript ${\rm I}$ of $A^{\rm I}_{+,j}(\boldsymbol{k})$ indicates the first occupied states of Kramers pairs $[\ket{u^{\rm I}_{n}(\boldsymbol{k})}, \ket{u^{\rm II}_{n}(\boldsymbol{k})}]$. 
$\nu=0$ indicates that a given insulator is trivial,  $\nu=1,3$ indicates that the insulator is a  strong $\mathbb{Z}_2$ TI, and $\nu=2$ indicates that the insulator is the TCI protected by $G_y$ symmetry. 

In the following, we discuss how to calculate this topological invariant from the spectra of the Wilson loop matrix.
The third term in the right-hand side of Eq.~(\ref{eq:supplement_topo_inv_3DTRSTCI}) can be rewritten as 
\begin{align}\label{eq:first_term_topo_inv}
	&\frac{1}{\pi} \int^{\pi}_{0}dk_z \int^{\pi}_{-\pi}dk_x F_{+,xz}(k_{x},\pi, k_z) \nonumber \\
	=& -\frac{1}{\pi} \int^{\pi}_{-\pi} dk_x A_{+,x}(k_x, \pi, \pi)+\frac{1}{\pi} \int^{\pi}_{-\pi} dk_x A_{+,x}(k_x, \pi, 0)+2M^{+}_{\bar{M} \bar{Y}}, 
\end{align}
where $M^{+}_{\bar{M}\bar{Y}}$ is the winding number of the Wilson loop spectra along the $\bar{M}$-$\bar{Y}$ line in the $+$ glide sector. 
In addition, on the $(k_{y}, k_{z})=(\pi, \pi)$ line,  $\mathcal{T}$ maps the occupied states in the $+$ glide sector to itself. The first and second occupied states of a Kramers pair have the $+$ glide sector: $[\ket{u^{(+)\rm I}_{n}(k_x, \pi, \pi)}, \ket{u^{(+)\rm II}_{n}(k_x, \pi, \pi)}]$. Thus, we obtain the following equation:
\begin{equation}\label{eq:kramers_kakikae_k_z=pi}
	-\frac{1}{\pi} \int^{\pi}_{-\pi} dk_x A_{+,x}(k_x, \pi, \pi)=-\frac{2}{\pi} \int^{\pi}_{-\pi} dk_x A^{\rm I}_{+,x}(k_x, \pi, \pi).
\end{equation}
On the other hand, the glide eigenvalue of occupied  states is given by $\pm i$ on the $(k_{y}, k_{z})=(\pi, 0)$ line, and the glide sectors exchange under $\mathcal{T}$. Thus, under $\mathcal{T}$, an occupied state  $\ket{u^{(+)}_{n}(k_x, \pi,0)}$ in the $+$ glide sector transforms into a Kramers pair $\ket{u^{(-)}_{n}(k_x, \pi,0)}$ in the $-$ glide sector.  We can rewrite a Kramers pair as $[\ket{u^{\rm I}_{n}(k_x, \pi,0)}, \ket{u^{\rm II}_{n}(k_x, \pi,0)}]=[\ket{u^{(+)}_{n}(k_x, \pi,0)}, \ket{u^{(-)}_{n}(k_x, \pi,0)}]$. Thus, the second term in the last line in Eq.~(\ref{eq:first_term_topo_inv}) can be written as
\begin{equation}\label{eq:kramers_kakikae_k_z=zero}
	\frac{1}{\pi} \int^{\pi}_{-\pi} dk_x A_{+,x}(k_x, \pi, 0)=	\frac{1}{\pi} \int^{\pi}_{-\pi} dk_x A^{\rm I}_{+,x}(k_x, \pi, 0).
\end{equation} 
By combining Eqs.~(\ref{eq:first_term_topo_inv}), (\ref{eq:kramers_kakikae_k_z=pi}), and (\ref{eq:kramers_kakikae_k_z=zero}), we obtain
\begin{align}\label{eq:first_part_topo_TCI}
	\frac{1}{\pi} \int^{\pi}_{0}dk_z \int^{\pi}_{-\pi}dk_x F_{+,xz}(k_{x},\pi, k_z)
	= -\frac{2}{\pi} \int^{\pi}_{-\pi} dk_x A^{\rm I}_{+,x}(k_x, \pi, \pi)
	+\frac{1}{\pi} \int^{\pi}_{-\pi} dk_x A^{\rm I}_{+,x}(k_x, \pi, 0)+2M^{+}_{\bar{M} \bar{Y}}.
\end{align}

Similarly, we can rewrite the fourth and the fifth terms of the right-hand side in Eq.~(\ref{eq:supplement_topo_inv_3DTRSTCI}) as 
\begin{align}
	&-\frac{1}{\pi} \int^{\pi}_{0}dk_z \int^{\pi}_{-\pi}dk_x  F_{+,xz}(k_{x}, 0, k_{z})
	=\frac{2}{\pi}\int^{\pi}_{-\pi}dk_x A^{\rm I}_{x}(k_x, 0, \pi)-\frac{1}{\pi} \int^{\pi}_{-\pi} dk_x A^{\rm I}_{x}(k_x,0,0)+2M^{+}_{\bar{\Gamma}\bar{Z}}\label{eq:second_part_topo_TCI},\\
	&\frac{1}{2\pi}\int^{\pi}_{0}dk_y \int^{\pi}_{-\pi}dk_x F_{xy}(k_x,k_y,0)=-\frac{1}{2\pi}\int^{\pi}_{-\pi}dk_x A_{x}(k_x,\pi,0)+\frac{1}{2\pi}\int^{\pi}_{-\pi}dk_x A_x (k_x,0,0)+M_{\bar{Y}\bar{\Gamma}},\label{eq:third_part_topo_TCI}
\end{align}
respectively, where $M_{\bar{\Gamma}\bar{Z}}$ is the winding number of the Wilson loop spectra along the $\bar{\Gamma}$-$\bar{Z}$ line in the $+$ glide sector. Furthermore, $M_{\bar{Y}\bar{\Gamma}}$ is the winding number of the Wilson loop spectra of all occupied bands along the $\bar{Y}$-$\bar{\Gamma}$ line. By substituting Eqs.~(\ref{eq:first_part_topo_TCI}), (\ref{eq:second_part_topo_TCI}), and (\ref{eq:third_part_topo_TCI}) into Eq.~(\ref{eq:supplement_topo_inv_3DTRSTCI}), we get
\begin{equation}\label{eq:z4_topological_invariant_wilson_loop}
 \nu \equiv 2M^{+}_{\bar{M} \bar{Y}}+2M^{+}_{\bar{\Gamma}\bar{Z}}+M_{\bar{Y}\bar{\Gamma}}\ \ {\rm mod}\ 4.
\end{equation}
Thus, we can calculate the $\mathbb{Z}_{4}$ topological invariant $\nu$ from the Wilson loop spectra along the $\bar{M}$-$\bar{Y}$,  $\bar{\Gamma}$-$\bar{Z}$, and $\bar{Y}$-$\bar{\Gamma}$ lines.
This result agrees with the method introduced in Ref.~\cite{wieder2018wallpaper} for calculating the topological invariant for the TCI from the Wilson loop spectra. 

From the spectra of the $x$-directed Wilson loop in Supplementary Fig.~\ref{fig:nonmagnetic_wilsonloop}, we can calculate the $\mathbb{Z}_{4}$  topological invariant $\nu$ of the tight-binding model of Eq.~(\ref{SM:eq:3dbloch_TCI}). We confirmed that by calculating the spectra of the $x$-directed Wilson loops, they do not wind along the $\bar{M}$-$\bar{Y}$ and  $\bar{\Gamma}$-$\bar{Z}$ lines, which leads to $M^{+}_{\bar{M} \bar{Y}}=0$ and $M^{+}_{\bar{\Gamma}\bar{Z}}=0$ in this model. Thus, the Wilson loop spectra along the $\bar{Y}$-$\bar{\Gamma}$ line determine $\nu$. Supplementary Figs.~\ref{fig:nonmagnetic_wilsonloop}(a-c) show that the doubly degenerate spectra wind by $2\pi$, and therefore it follows that $M_{\bar{Y}\bar{\Gamma}}\equiv 2$ mod 4, namely, $\nu \equiv 2$ mod 4 in these cases. These results show that our model with these parameters is the TCI protected by $G_y$ symmetry.

\section{Magnetic topological crystalline insulator protected by nonsymmorphic symmetries}
\subsection{Tight-binding model}

Here, we discuss surface states of a magnetic TCI protected by glide symmetry to confirm that our theory of the non-magnetic TCI in the main text can be applied to the magnetic TCI.
For this purpose, we construct a 3D tight-binding model of the TCI in the space group No.~13 ($P2/c$). We start from a 2D tight-binding model of a Chern insulator \cite{PhysRevB.78.195424} on a square lattice with two sublattices at $y=\tfrac{1}{4}$ and $y=\tfrac{3}{4}$ in the unit cell [see Supplementary Fig.~\ref{fig:SM_MTCI}(a)]: 
\begin{align}\label{apendix:chern ins model}
	H_{{\rm Chern}; A}=&-\sum_{j=1,2}\sum_{x,y,z \in \mathbb{Z}} t \biggl[ 
	c^{\dagger}_{Aj}(x+1,y+\tfrac{2j-1}{4},z) \Bigl(\frac{\sigma_{3}+i\sigma_{1}}{2} \Bigr)c_{Aj}(x,y+\tfrac{2j-1}{4},z) +{\rm h.c.}\biggr]
	\nonumber \\
	&-\sum_{x,y,z \in \mathbb{Z}} t \biggl[
	c^{\dagger}_{A2}(x,y+\tfrac{3}{4},z) \Bigl(\frac{\sigma_{3}+i\sigma_{2}}{2} \Bigr)c_{A1}(x,y+\tfrac{1}{4},z) \nonumber \\
	&\ \ \ \ \ \ \ \ \ \ \ \ \ \ \ +c^{\dagger}_{A1}(x,y+\tfrac{5}{4},z) \Bigl(\frac{\sigma_{3}+i\sigma_{2}}{2} \Bigr)c_{A2}(x,y+\tfrac{3}{4},z)+{\rm h.c.}
	\biggr]\nonumber \\
	&+\sum_{j=1,2}\sum_{x,y,z \in \mathbb{Z}}\biggl[ m c_{Aj}^{\dagger}(x,y+\tfrac{2j-1}{4},z) \sigma_{3} c_{Aj}(x,y+\tfrac{2j-1}{4},z)  \biggr], 
\end{align} 
where $c_{Aj}^{\dagger}(x,y,z)$ is a creation operators at the ($x,y,z$) site within the layers $A$ with $j$-sublattice ($j=1,2$).
This Hamiltonian has four basis states spanned by $\sigma_3 =\pm$ and sublattices $A_1/A_2$. 
The coordinate  $(x,y,z)$ runs over the square lattice sites, and we set the lattice constant to 1. 

We construct the 3D tight-binding model from this 2D Chern insulator \cite{PhysRevB.100.165202, kim2020glide}. 
This Chern insulator is located at $z=n$ plane ($n \in \mathbb{Z}$) in real space. 
In the space group No.~13, the symmetry generators are given by
\begin{equation}
	\hat{G}_{y}=\{ M_{y}|00\tfrac{1}{2} \},\ \ \hat{I}=\{ I|000 \},
\end{equation}
where $M_{y}$ represents the mirror reflection with respect to the $xz$ plane. 
Under $\hat{G}_{y}$ and $\hat{I}$, the creation operator $c^{\dagger}(x,y,z)$ transforms as 
\begin{align}
	&\hat{I} : c^{\dagger}_{\alpha} (x,y,z) \rightarrow c^{\dagger}_{\beta}(-x,-y,-z)[I]_{\beta \alpha} \ \ {\rm with}\ I=\tau_1  \sigma_{3}, \\
	&\hat{G}_{y} : c^{\dagger}_{\alpha} (x,y,z) \rightarrow c^{\dagger}_{\beta}(x,-y,z+\tfrac{1}{2})[M_{y}]_{\beta \alpha} \ \ {\rm with}\ M_{y}=\tau_0,
\end{align}
where $\alpha$ and $\beta$ run over the basis of Eq.~(\ref{apendix:chern ins model}), $\tau_i$ ($i=1,2,3$) is Pauli matrix  corresponding to $A_1/ A_2$ sublattices, and $\tau_0$ is the $2\times 2$  identity matrix. 

Now, we design the system to be invariant under the glide $\hat{G}_{y}$ symmetry. 
By making copies of the layers of $H_{{\rm Chern}; A}$ by the $\hat{G}_{y}$ transformation, we obtain the layers $B$:
\begin{align}
		\hat{G}_{y} : H_{{\rm Chern}; A} \rightarrow  H_{{\rm Chern}; B} \equiv &-\sum_{j=1,2}\sum_{x,y,z \in \mathbb{Z}} t \biggl[ 
	c^{\dagger}_{Bj}(x+1,y+\tfrac{2j-1}{4},z) \Bigl(\frac{\sigma_{3}+i\sigma_{1}}{2} \Bigr)c_{Bj}(x,y+\tfrac{2j-1}{4},z) +{\rm h.c.}\biggr]
	\nonumber \\
	&-\sum_{x,y,z \in \mathbb{Z}} t \biggl[
	c^{\dagger}_{B2}(x,y+\tfrac{3}{4},z+\tfrac{1}{2}) \Bigl(\frac{\sigma_{3}-i\sigma_{2}}{2} \Bigr)c_{B1}(x,y+\tfrac{1}{4},z+\tfrac{1}{2})\nonumber \\
	&\ \ \ \ \ \ \ \ \ \ \ \ \ \ \ +c^{\dagger}_{B1}(x,y+\tfrac{5}{4},z+\tfrac{1}{2}) \Bigl(\frac{\sigma_{3}-i\sigma_{2}}{2} \Bigr)c_{B2}(x,y+\tfrac{3}{4},z+\tfrac{1}{2})+{\rm h.c.}
	\biggr]\nonumber \\
	&+\sum_{j=1,2}\sum_{x,y,z \in \mathbb{Z}}\biggl[ m c_{Bj}^{\dagger}(x,y+\tfrac{2j-1}{4},z) \sigma_{3} c_{Bj}(x,y+\tfrac{2j-1}{4},z)  \biggr],
\end{align}
where  $c_{Bj}^{\dagger}(x,y,z)$ is the creation operators at the layer $B$.
Next, we introduce an interlayer hopping between the $A$ and $B$ layers. Thus, we obtain the overall Hamiltonian of the magnetic TCI 
\begin{equation}
	H_{\rm MTCI}=H_{{\rm Chern}; A}+H_{{\rm Chern}; B}+H'_{AB},
\end{equation}
where $H'_{AB}$ denotes the Hamiltonian of the interlayer couplings given by
\begin{align}
	H'_{AB}=\sum_{x,y,z \in \mathbb{Z}} \frac{t_{AB}}{2} \Biggr[ 
	c^{\dagger}_{A1}(x,y+\tfrac{1}{4},z)c_{B1}(x,y+\tfrac{1}{4},z+\tfrac{1}{2}) + c^{\dagger}_{A2}(x,y+\tfrac{3}{4},z+1)c_{B2}(x,y+\tfrac{3}{4},z+\tfrac{1}{2}) +{\rm h.c.}
	\Biggl].
\end{align}
This interlayer coupling $H'_{AB}$ breaks the mirror $\hat{M}_{z}$ symmetry. 
In this way, through the layer construction with 2D Chern insulators, we obtain the Hamiltonian of the magnetic TCI protected by  $G_{y}$ symmetry.

\begin{figure}
	\includegraphics[width=0.8\columnwidth]{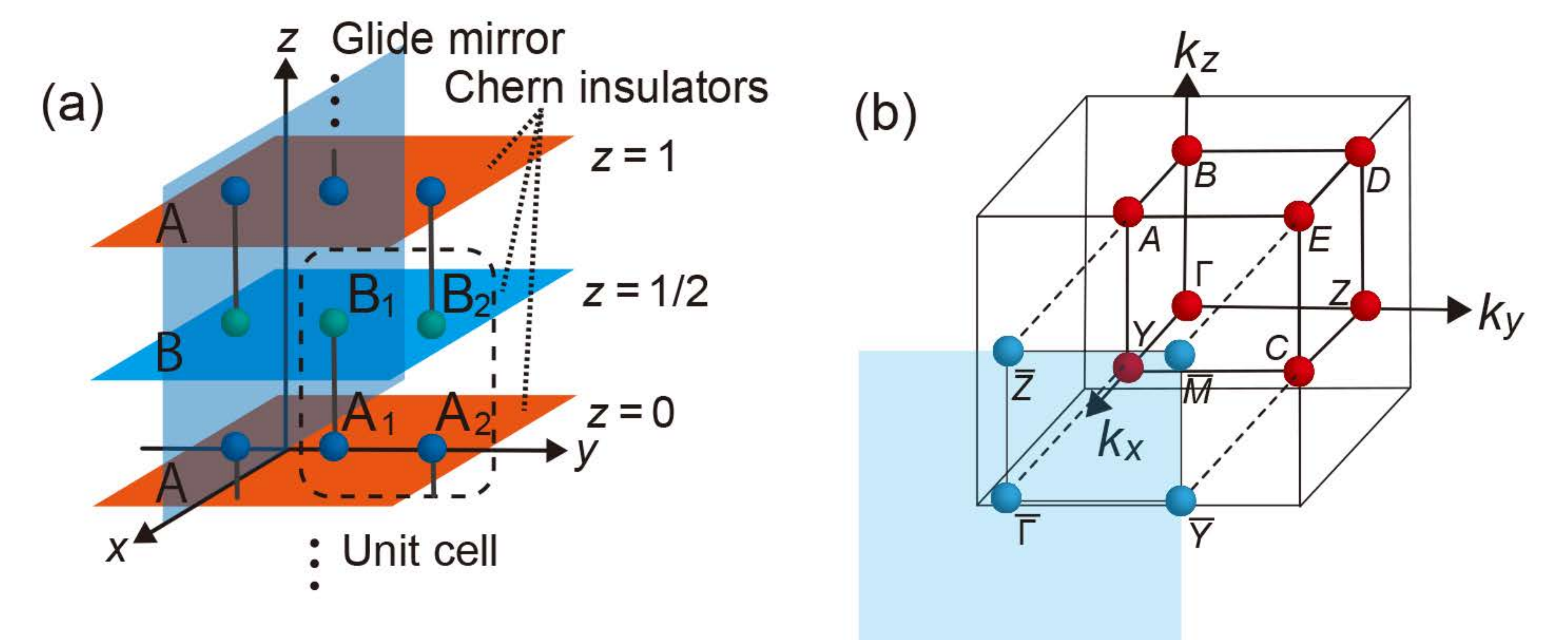}
	\caption{(a) 2D Chern insulator layer A at $z=0$ and 2D Chern insulator layer B at $z=1/2$ in the unit cell. The layer construction from these layers has the ${G}_y$ symmetry.  (b) The bulk Brillouin zone and the (100) surface Brillouin zone.} 
	\label{fig:SM_MTCI}
\end{figure}

Next, to obtain the Bloch Hamiltonian, we  introduce the Fourier transformations $c^{\dagger}_{\alpha}(\boldsymbol{k})=\sum_{\boldsymbol{R}} e^{i \boldsymbol{k}\cdot (\boldsymbol{R}+\boldsymbol{r}_{\alpha})}c_{\alpha}^{\dagger}(\boldsymbol{R}+\boldsymbol{r}_{\alpha})/\sqrt{N}$, where $N$ is the number of unit cells, $\alpha$ runs over eight basis states spanned by $\sigma_{3}=\pm$, A/B sublattices, and $X_{1}/X_{2}$ ($X=A,B$),   $\boldsymbol{r}_{\alpha}$ denotes the positions of the sites in the unit cell, and $\boldsymbol{R}=(x,y,z)$ is the position of the unit cell with $x,y,z \in \mathbb{Z}$. 
Thus,  we obtain the Bloch Hamiltonian of $H_{\rm MTCI}$ 
\begin{align}\label{appendix:MTCI_bloch}
	\mathcal{H}_{\rm MTCI}(\boldsymbol{k})=&(m-t\cos k_{x})\sigma_{3}-t \sin k_{x}  \sigma_{1}-t\cos \tfrac{k_{y}}{2}\tau_{1}\sigma_{3}-t \sin \tfrac{k_{y}}{2}\mu_{3}\tau_{1}\sigma_{2}+\mathcal{H}'_{AB}(k_{z})
\end{align}
where $\tau_{0}$ and $\mu_{0}$ are the 2$\times$2 identity matrices, $\mu_{i}$ ($i=1,2,3$) are the Puli matrices corresponding to the $A/B$ sublattices,  
and $\tau_{i}$ ($i=1,2,3$) are the Pauli matrices corresponding to the $X_{1}$/$X_{2}$ ($X=A,B$) sublattices. 
Here, the interlayer Hamiltonian $\mathcal{H}'_{AB}(k_{z})$ can be expressed as  
\begin{align}
	\mathcal{H}'_{AB}(k_{z})= \frac{t_{AB}}{2} \begin{pmatrix}
		0 & 0 &e^{i\tfrac{k_{z}}{2}} & 0   \\
		0 & 0 & 0 & e^{-i\tfrac{k_{z}}{2}} \\
		e^{-i\tfrac{k_{z}}{2}} & 0 &0 & 0  \\
		0 & e^{i\tfrac{k_{z}}{2}} &0 & 0  \\
	\end{pmatrix}\otimes \sigma_{0}.
\end{align}
Under $\hat{G}_{y}$ and $\hat{I}$, the creation operator  $c^{\dagger}_{\alpha}$ transforms as
\begin{align}
	\hat{G}_{y}: c_{\alpha}^{\dagger}(k_{x},k_{y},k_{z}) \rightarrow \  c^{\dagger}_{\beta}(k_{x},-k_{y},k_{z}) [G_{y}(k_{z})]_{\beta \alpha}, \\
	\hat{I}:  c_{\alpha}^{\dagger}(k_{x},k_{y},k_{z}) \rightarrow \  c^{\dagger}_{\beta}(-k_{x},-k_{y},-k_{z}) [I]_{\beta \alpha},
\end{align}
with 
\begin{align}
	G_{y}(k_{z})=e^{-i\tfrac{k_{z}}{2}} \mu_{1} \tau_{0},\ \ I=\mu_0 \tau_1 \sigma_3
\end{align}
where $\alpha$ and $\beta$ run over the basis of Eq.~(\ref{appendix:MTCI_bloch}). 
The Hamiltonian $\mathcal{H_{\rm MTCI}}(\boldsymbol{k})$ has $G_y$ symmetry and $I$ symmetry:
\begin{gather}
	G_{y}(k_{z}) \mathcal{H}_{\rm MTCI}(k_{x},k_{y},k_{z}) G_{y}^{-1}(k_{z}) =\mathcal{H}_{\rm MTCI} (k_{x},-k_{y},k_{z}), \\
	I  \mathcal{H}_{\rm MTCI}(\boldsymbol{k}) I^{-1}= \mathcal{H}_{\rm MTCI} (-\boldsymbol{k}).
\end{gather}
The $G_y$ matrix has the glide eigenvalues $g_y(k_z)=\pm  \exp \left(-i\tfrac{k_z}{2} \right)$ because $G^2_y(k_z)=\exp \left( -i k_z\right)$. 

\begin{figure}
	\includegraphics[width=1.\columnwidth]{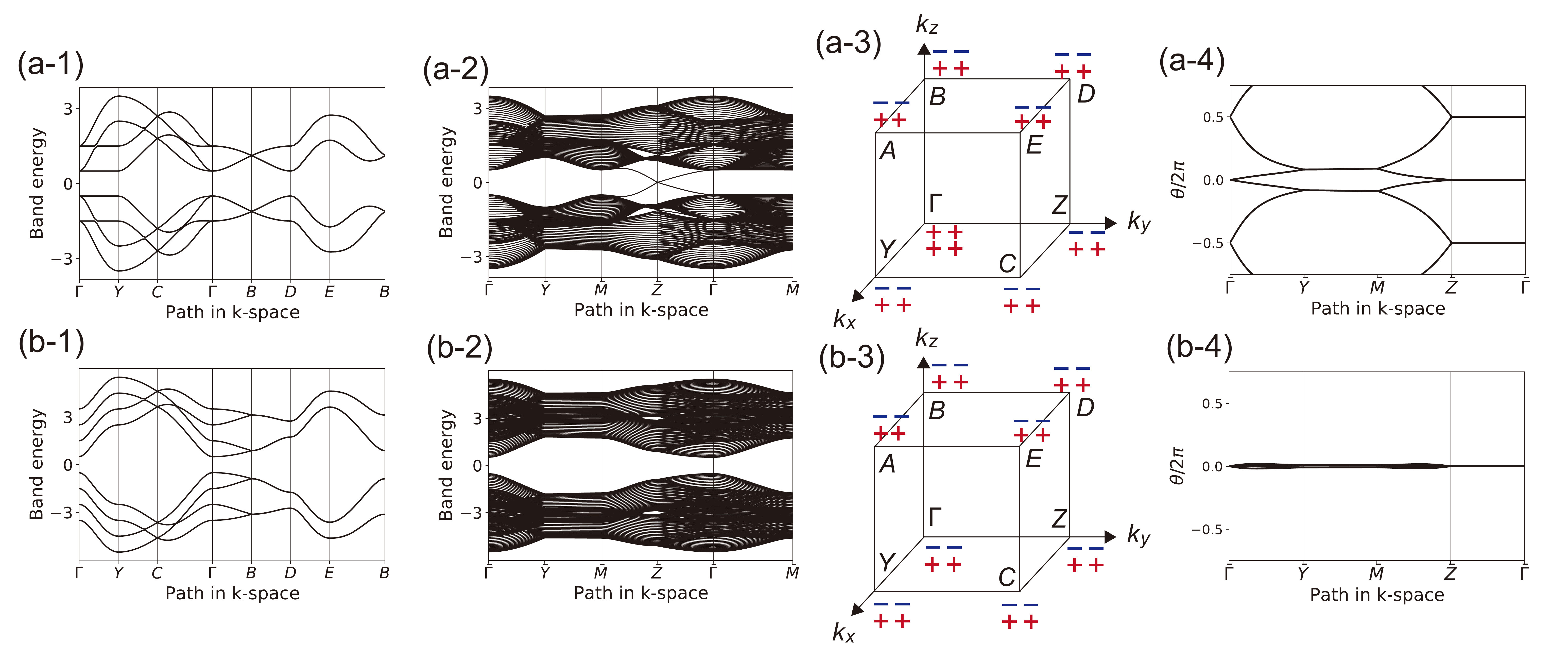}
	\caption{Band structures and topological invariants for (a) The TCI protected by $G_y$ symmetry and (b) the trivial insulator. The parameters of $\mathcal{H}_{\rm MTCI}(\boldsymbol{k})$ are $t=m=t_{AB}=1$ in (a), and the parameters are $t=t_{AB}=1$, and $m=3$ in (b). (a-1,b-1) The band structures of the bulk Hamiltonian $\mathcal{H}_{\rm MTCI}(\boldsymbol{k})$. (a-2,b-2) The band structures in the (100) slab with the finite system size along the $x$-direction $L_x=30$. (a-3,b-3) The parity eigenvalues of the occupied states of the bulk Hamiltonian $\mathcal{H}_{\rm MTCI}(\boldsymbol{k})$ at the high-symmetry points in the $k$-space. (a-4,b-4) The spectra of the $x$-directed Wilson loop matrix. } 
	\label{fig:SM_MTCI_topo_inv}
\end{figure}

\begin{figure}
	\includegraphics[width=0.8\columnwidth]{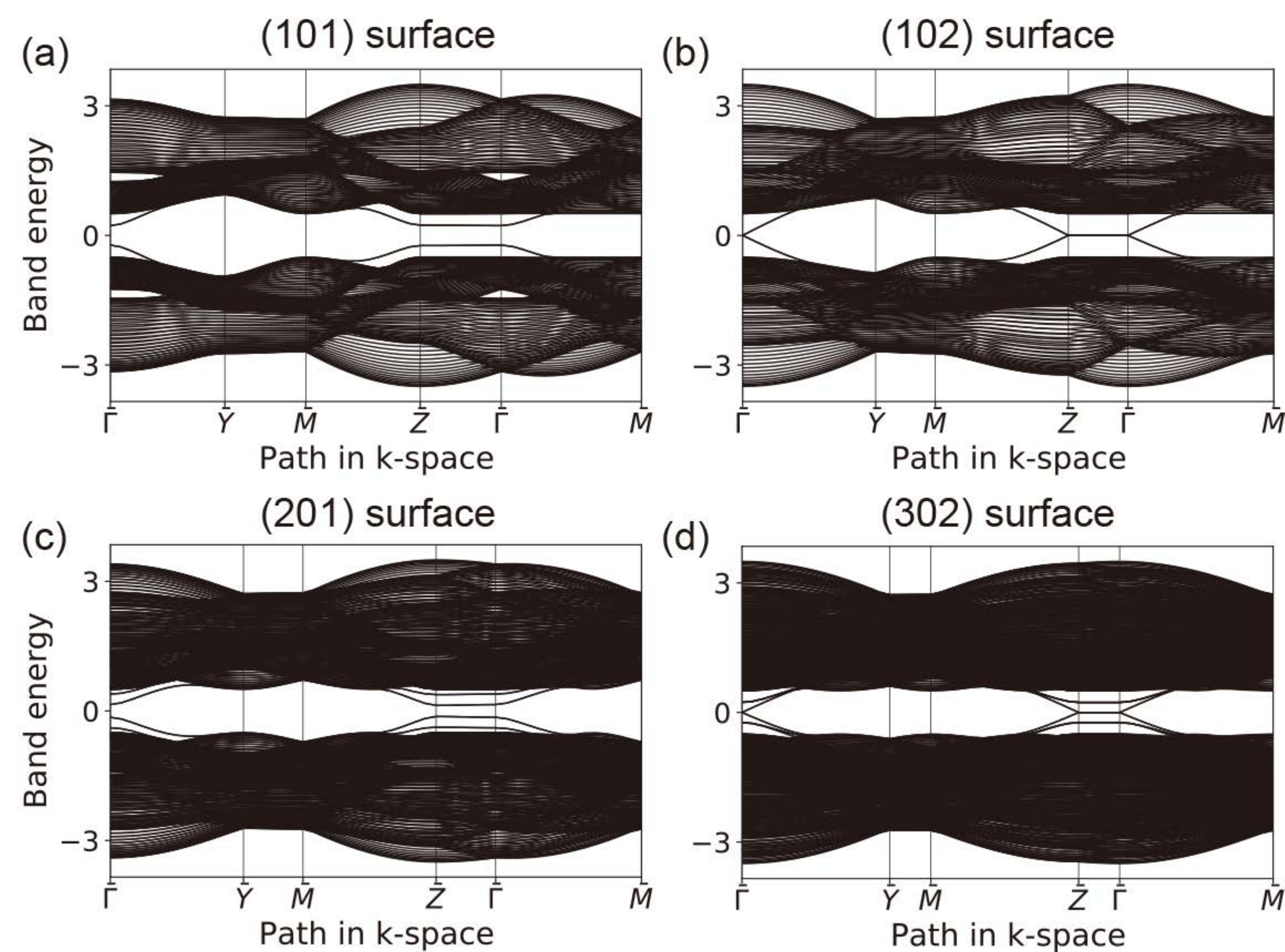}
	\caption{The band structures of $\mathcal{H}_{\rm MTCI}(\boldsymbol{k})$ in the slab geometry with surfaces having various Miller indices ($\alpha 0 \gamma$). They are calculated in a slab geometry with the system size along the $x$-direction $L_{x}=30$. The parameters are $t=m=t_{AB}=1$. Here we set the Fermi energy to be zero. (a,c) When $\gamma$ is odd, gapless states do not appear. (b,d) When $\gamma$ is even, gapless states appear.} 
	\label{fig:SM_MTCI_various_surface}
\end{figure} 

Here, the high-symmetry points in the Brillouin zone in the bulk are given by $\{ \Gamma=(0,0,0),\ Y=(\pi, 0, 0),\  Z=(0,\pi, 0),\  C=(\pi,\pi, 0),\  B=(0,0,\pi),\  A=(\pi,0,\pi),\  D=(0,\pi,\pi),\  E=(\pi, \pi, \pi) \}$, and the high-symmetry  points in the (100) surface Brillouin zone are $\{\bar{\Gamma}=(0,0),\ \bar{Y}=(\pi, 0),\  \bar{Z}=(0,\pi),\  \bar{M}=(\pi,\pi) \}$ [Figs.~\ref{fig:SM_MTCI}(b)].
The band structures of $\mathcal{H}_{\rm MTCI}(\boldsymbol{k})$ in the bulk  are shown in Supplementary Figs.~\ref{fig:SM_MTCI_topo_inv}(a-1) and \ref{fig:SM_MTCI_topo_inv}(b-1) for $(t,t_{AB}, m)=(1,1,1)$ and $(1,1,3)$,  respectively, and those in the slab geometry with the (100) surface are shown in Supplementary Figs.~\ref{fig:SM_MTCI_topo_inv}(a-2) and \ref{fig:SM_MTCI_topo_inv}(b-2), respectively. In Supplementary Fig.~\ref{fig:SM_MTCI_topo_inv}(a-2), the gapless surface states appear, while those do not appear in Supplementary Fig.~\ref{fig:SM_MTCI_topo_inv}(b-2). As we show in the following, the former case corresponds to the magnetic TCI phase protected by $G_y$ symmetry, and the latter case corresponds to the trivial insulator phase.  

\subsection{Topological invariant for magnetic TCI protected by glide symmetry}
Here, we introduce a topological invariant for the magnetic TCI protected by $G_y$ symmetry. According to  Refs.~\cite{PhysRevB.100.165202, PhysRevB.91.155120, PhysRevB.91.161105}, the $\mathbb{Z}_2$ topological invariant is given by
\begin{align}\label{SM:magnetic_TCI_Z2_inv}
	\nu' =& \frac{1}{\pi}\int^{\pi}_{-\pi}dk_x \biggl( A_{+,x}(k_x,0,\pi)-A_{-,x}(k_y, \pi, \pi) \biggr) \nonumber \\
	&-\frac{1}{2\pi}  \int^{\pi}_{-\pi} dk_x \int^{\pi}_{-\pi} dk_z  \biggl( F_{-,xz}(k_x, 0, k_z)- F_{+,xz}(k_x,\pi,k_z) \biggr) \nonumber \\
	&-\frac{1}{2\pi}\int^{\pi}_{0}dk_y \int^{\pi}_{-\pi}dk_x F_{xy}(k_x,k_y,\pi)\ \ {\rm mod}\ 2,
\end{align}
where $A_j (\boldsymbol{k})$ and $A_{\pm, j} (\boldsymbol{k})$ ($j=x,y,z$) are the Berry connections defined as in  Eqs.~(\ref{SM:eq_Berry_connection}) and (\ref{SM:eq_Berry_connection_glide_sector}), respectively, and  $F_{ij}(\boldsymbol{k})$ and  $F_{\pm, ij} (\boldsymbol{k})$ ($i,j =x,y,z$) are the Berry curvatures given by Eqs.~(\ref{SM:eq_Berry_curvature}) and (\ref{SM:eq_Berry_curvature_glide_sector}), respectively. 
We note that the formula (\ref{SM:magnetic_TCI_Z2_inv}) follows the definition in Ref.~\cite{PhysRevB.100.165202}, which is different from the one in Refs. \cite{ PhysRevB.91.155120, PhysRevB.91.161105} by the Chern number within the $k_x$-$k_z$ plane. The two definitions are equal when we restrict ourselves to the cases with the Chern number being zero. 
By using the result in the previous works~\cite{PhysRevB.100.165202, kim2020glide}, when $\mathcal{I}$ symmetry is preserved and the inversion center is in the glide plane,  the $\mathbb{Z}_{2}$ topological invariant can be expressed as
\begin{equation}
	\nu ' \equiv \frac{1}{2}\sum_{\Gamma_{j}}\biggl( n_{+}(\Gamma_{j})-n_{-}(\Gamma_{j})\biggr) \ {\rm mod}\ 2, 
\end{equation}
where $n_{+}(\Gamma_{j})$ ($n_{-}(\Gamma_{j})$) is the number of occupied states with even (odd) parity eigenvalues at the high-symmetry points $\Gamma_{j}$ (=$Y$, $Z$, $C$, $B$, $A$, $D$, $E$) in the Brillouin zone. 
By applying this formula to our model $\mathcal{H}_{\rm MTCI}(\boldsymbol{k})$ having $\mathcal{I}$ symmetry, 
we obtain $\nu ' \equiv 1$ mod 2 and $\nu ' \equiv 0$ mod 2 for $(t,t_{AB},m)=(1,1,1)$ and $(1,1,3)$, respectively, from the parity eigenvalues shown in Supplementary Figs.~\ref{fig:SM_MTCI_topo_inv}(a-3) and \ref{fig:SM_MTCI_topo_inv}(b-3). Therefore, the former is the magnetic TCI protected by $G_y$ symmetry, and  the later is the trivial insulator. 

We also calculate the spectra of the $x$-directed Wilson loop matrix [Supplementary Figs.~\ref{fig:SM_MTCI_topo_inv}(a-4) and \ref{fig:SM_MTCI_topo_inv}(b-4)].  In Supplementary Fig.~\ref{fig:SM_MTCI_topo_inv}(a-4), the Wilson loop spectra along the $\bar{\Gamma}$-$\bar{Y}$ lines in the $k$ space wind by $2\pi$.  On the other hand, in Supplementary Fig.~\ref{fig:SM_MTCI_topo_inv}(b-4), the Wilson loop spectra do not wind. These results also show the above conclusion. 

\subsection{Surface states of the MTCI}
Next, to see dependence of the topological surface states on the surface orientation, we calculate band structures in the slab geometry with various $(\alpha 0 \gamma)$ surfaces [Supplementary Fig.~\ref{fig:SM_MTCI_various_surface}]. In Supplementary Figs.~\ref{fig:SM_MTCI_various_surface}(a), \ref{fig:SM_MTCI_various_surface}(c), the gapless surface states do not appear because the ($\alpha 0 \gamma$) surface is not glide symmetric when $\gamma \equiv 1$ mod 2. On the other hand, Supplementary Figs.~\ref{fig:SM_MTCI_various_surface}(b) and \ref{fig:SM_MTCI_various_surface}(d) show that the gapless surface states appear because the $(\alpha 0 \gamma)$ surface is glide symmetric when $\gamma \equiv 0$ mod 2. From these results, we conclude that our theory on the dependence of the topological surface states on the surface orientations can also be applied to the MTCI protected by $G_y$ symmetry. 

\section{Second model of  nonmagnetic topological crystalline insulator }\label{appendix:second_tight-binding_model}
\subsection{Tight-binding model}
Here we show details of the second model of the TCI $\mathcal{H}^{(2)}_{\rm TCI}(\boldsymbol{k})$  in the main text. The method to construct this model is similar to that in \ref{appendix:tight-binding_model}.  We start from the following simple tight-binding model of a 3D TI on a primitive cubic lattice: 
\begin{align}
	\mathcal{H}^{(A)}_{\rm 3DTI}(\boldsymbol{k})=\biggl(-m + \sum_{j=x,y,z}t_{j} \cos k_j \biggr) \tau_3 \sigma_0 +v_x \sin k_x \tau_{1}\sigma_{3} +v_y \sin k_y \tau_1 \sigma_1 +v_z \sin k_z \tau_1 \sigma_{2}.
\end{align}
This model has ${I}$ symmetry and $\mathcal{T}$ symmetry: ${I} \mathcal{H}^{(A)}_{\rm 3DTI} {I}^{-1}=\mathcal{H}^{(A)}_{\rm 3DTI}(-\boldsymbol{k})$, $\mathcal{T} \mathcal{H}^{(A)}_{\rm 3DTI} \mathcal{T}^{-1}=\mathcal{H}^{(A)}_{\rm 3DTI}(-\boldsymbol{k})$, where ${I}=\tau_{3} $, and $\mathcal{T}=-i\tau_{0} \sigma_{2}K$ with $K$ being the complex conjugation. 
The Hamiltonian has four basis states spanned by $\tau_{3}=\pm$, $\sigma_{3}=\pm$. 

\begin{figure}
\includegraphics[width=1.\columnwidth]{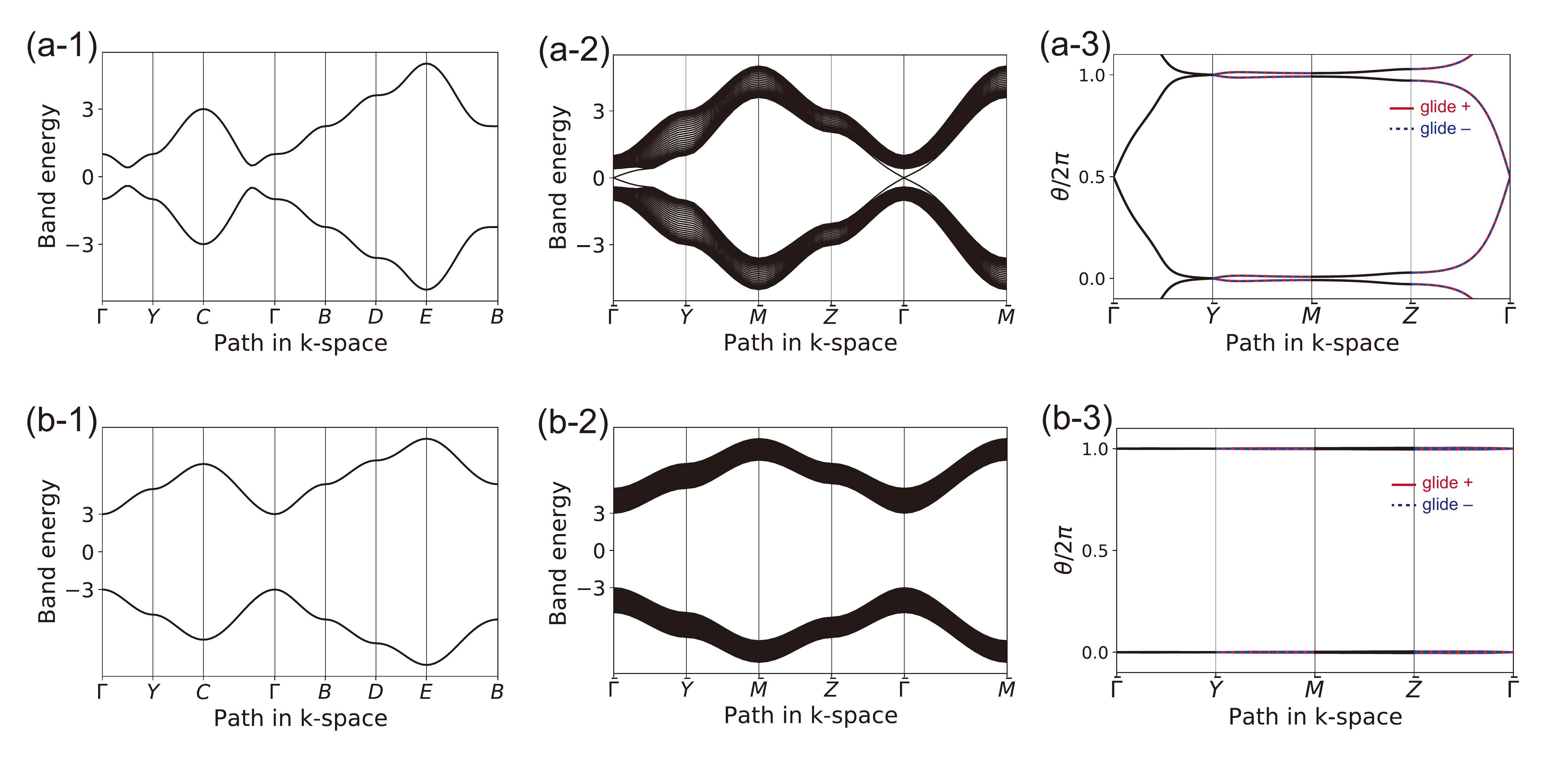}
\caption{The band structures and the Wilson loop spectra of $\mathcal{H}^{(2)}_{\rm TCI}(\boldsymbol{k})$ for (a) the TCI protected by $G_y$ symmetry and (b) the trivial insulator. (a-1,b-1) The band structures in the bulk. (a-2,b-2) The band structures in the (100) slab with the finite system size along the $x$-direction $L_x=30$. (a-3,b-3) The spectra of the $x$-directed Wilson loop matrix. The parameters are $m=2$, $t_x=t_y=t_z=1$, $v_x=v_y=v_z=0.4$, $v_{ab1}=0.8$, $v_{ab2}=1.2$ in (a) and $m=6$, $t_x=t_y=t_z=1$, $v_x=v_z=0.4$, $v_z=0.6$, $v_{ab1}=0.8$, $v_{ab2}=1.2$ in (b).} \label{fig:SM_second_TCI}
\end{figure}

Next, we consider a glide $\hat{G}_{y}=\{M_{y}|00\tfrac{1}{2}\}$ symmetry,  where $M_{y}$ represents the mirror reflection with respect to the $xz$ plane. Under $\hat{G}_{y}$, the Hamiltonian $\mathcal{H}^{(A)}_{\rm 3DTI}(\boldsymbol{k})$ can be transformed into $\mathcal{H}^{(B)}_{\rm 3DTI}(\boldsymbol{k})$:
\begin{equation}
	\hat{G}_{y}:\ \mathcal{H}^{(A)}_{\rm 3DTI}(k_x, k_y, k_z) \rightarrow \mathcal{H}^{(B)}_{\rm 3DTI}(k_x, k_y, k_z)=\mathcal{H}^{(A)}_{\rm 3DTI}(k_x, -k_y, k_z),
\end{equation}
in a similar way to the discussion in \ref{appendix:tight-binding_model}.
By combining $\mathcal{H}^{(A)}_{\rm 3DTI}(\boldsymbol{k})$ and $\mathcal{H}^{(B)}_{\rm 3DTI}(\boldsymbol{k})$, the Hamiltonian as the direct sum of these Hamiltonians is written as
\begin{align}
\mathcal{H}^{(A)}_{\rm 3DTI}(\boldsymbol{k})\oplus  \mathcal{H}^{(B)}_{\rm 3DTI}(\boldsymbol{k}) 
 =& \biggl(-m + \sum_{j=x,y,z}t_{j} \cos k_j \biggr)\mu_{0} \tau_3 \nonumber \\
  &+v_x \sin k_x \mu_{0} \tau_{1}\sigma_{3} +v_y \sin k_y \mu_{3} \tau_1 \sigma_1 +v_z \sin k_z \mu_0 \tau_1 \sigma_{2},
 \end{align}
where $\mu_i$ ($i=1,2,3$) are the Pauli matrices corresponding to the A/B sublattices, $\mu_0$ is the $2\times 2$ identity matrix, the site of the A sublattice is $(x,y,z)$ with $x,y,z \in \mathbb{Z}$, and the site of the B sublattice is $(x,y,z+\tfrac{1}{2})$. 
We add the coupling term between $A$ and $B$
\begin{align}
	\mathcal{H}^{(AB)}(\boldsymbol{k})=\Bigl( v_{ab1}  +v_{ab2}\cos k_x \Bigr)  \mu_{1}  \tau_1 \sigma_1 \sin \tfrac{k_z}{2}
\end{align}
 to $\mathcal{H}^{(A)}_{\rm 3DTI}(\boldsymbol{k})\oplus  \mathcal{H}^{(B)}_{\rm 3DTI}(\boldsymbol{k})$. Thus, we obtain the following overall Hamiltonian:
 \begin{equation}
 	\mathcal{H}^{(2)}_{\rm TCI}(\boldsymbol{k})=\mathcal{H}^{(A)}_{\rm 3DTI}(\boldsymbol{k})\oplus  \mathcal{H}^{(B)}_{\rm 3DTI}(\boldsymbol{k}) +\mathcal{H}^{(AB)}(\boldsymbol{k}).
 \end{equation} 
Here, the $G_y$ glide is given by $G_y(k_z)=-i \exp \left(-i \tfrac{k_z}{2} \right)\mu_1 \tau_3 \sigma_2$, and $\mathcal{H}^{(2)}_{\rm TCI}(\boldsymbol{k})$ has $G_y$ symmetry: 
\begin{equation}
G_y(k_z) \mathcal{H}^{(2)}_{\rm TCI}(k_x, k_y, k_z)G^{-1}_y(k_z)= \mathcal{H}^{(2)}_{\rm TCI}(k_x, -k_y, k_z).
\end{equation}  
The $G_y$ matrix has glide eigenvalues $g_y(k_z)=\pm i \exp \left(-i\tfrac{k_z}{2} \right)$. By using a unitary matrix 
\begin{align}
	U_{g_y}=\frac{1}{\sqrt{2}}\begin{pmatrix}
		1&0&0&0&1&0&0&0 \\
		0 &1&0&0&0&1&0&0\\
		0&0&1&0&0&0&1&0\\
		0&0&0&1&0&0&0&1\\
		0&i&0&0&0&-i&0&0\\
		-i&0&0&0&i&0&0&0\\
		0&0&0&-i&0&0&0&i\\
		0&0&i&0&0&0&-i&0
		\end{pmatrix},
\end{align}
 $\mathcal{H}_{\rm TCI}^{(2)}(\boldsymbol{k})$ can be transformed into the following form where the glide symmetry is manifest. 
\begin{equation}
	U_{g_y} \mathcal{H}_{\rm TCI}^{(2)}(\boldsymbol{k})U_{g_y}^{\dagger}=\begin{pmatrix} \mathcal{H}_{+}^{(2)}(\boldsymbol{k}) & M_{g_y} (k_y) \\
		M_{g_y}^{\dagger}(k_y) &  \mathcal{H}_{-}^{(2)}(\boldsymbol{k})
		\end{pmatrix},
\end{equation}
with
\begin{align}
	\mathcal{H}^{(2)}_{+}(\boldsymbol{k}) =&\biggl(-m + \sum_{j=x,y,z}t_{j} \cos k_j \biggr) \tau_3 \sigma_0 +v_x \sin k_x \tau_{1}\sigma_{3} \nonumber \\
	&+v_z \sin k_z \tau_1 \sigma_{2}+(v_{ab1}+v_{ab2}\cos k_x)\sin \tfrac{k_z}{2} \tau_1 \sigma_1,\\
	\mathcal{H}^{(2)}_{-}(\boldsymbol{k}) =&\biggl(-m + \sum_{j=x,y,z}t_{j} \cos k_j \biggr) \tau_3 \sigma_0 +v_x \sin k_x \tau_{1}\sigma_{3} \nonumber \\
	&-v_z \sin k_z \tau_1 \sigma_{2}-(v_{ab1}+v_{ab2}\cos k_x)\sin \tfrac{k_z}{2} \tau_1 \sigma_1,\\
	M_{g_y}(k_y)=&-v_y \sin k_y \tau_2 \sigma_3.
 \end{align}
On the glide-invariant planes $k_y=0, \pi$, $M_{g_y} (k_y)$ vanishes. Within these planes,  $\mathcal{H}_{+}^{(2)}(\boldsymbol{k})$ and $\mathcal{H}_{-}^{(2)}(\boldsymbol{k})$ are the Hamiltonians in the $+$ and the $-$  sectors  labeled by the glide  eigenvalues $g_{y,+}(k_z)$ and $g_{y,-}(k_z)$, respectively. Thus, the  Hamiltonian with the $k_y=0$ or $k_z = \pi$ can be expressed as the direct sum of the Hamiltonians in the $+$ and $-$ glide sectors. 

\subsection{Band structures and Wilson loops}

\begin{figure}
	\includegraphics[width=1.\columnwidth]{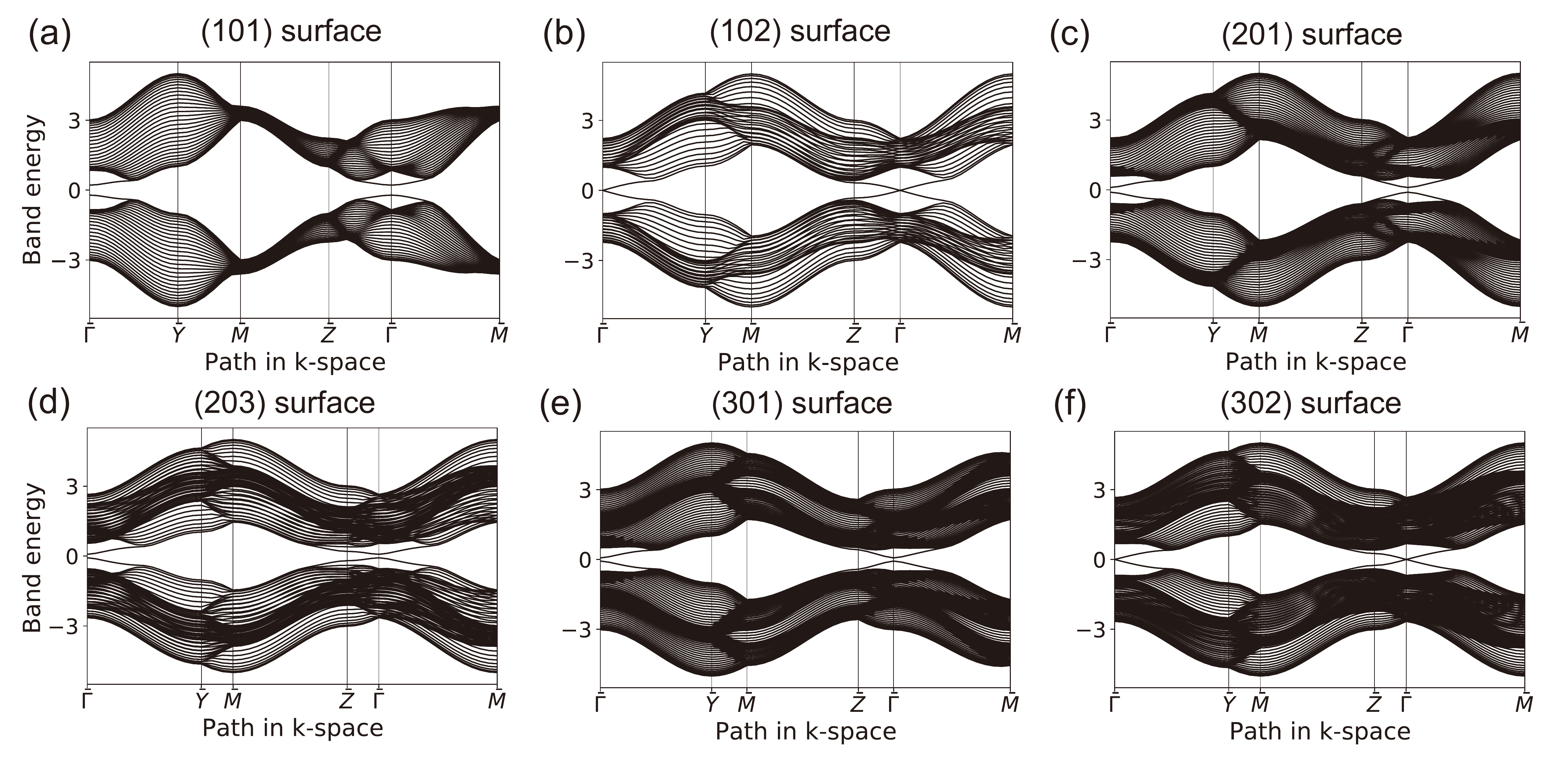}
	\caption{The band structures of $\mathcal{H}^{(2)}_{\rm TCI}(\boldsymbol{k})$ in the slab geometry with surfaces having various Miller indices ($\alpha 0 \gamma$). They are calculated in a slab geometry with the system size along the $x$-direction $L_{x}=30$.  The parameters are $m=2$, $t_x=t_y=t_z=1$, $v_x=v_y=v_z=0.4$, $v_{ab1}=0.8$, and $v_{ab2}=1.2$. Here we set the Fermi energy to be zero. (a,c,d,e) When $\gamma$ is odd, gapless states do not appear. (b,f) When $\gamma$ is even, gapless states appear.} 
	\label{fig:SM_second_model_various_surface}
\end{figure}

 Here we study band structures and topological invariants of our second model $\mathcal{H}_{\rm TCI}^{(2)}(\boldsymbol{k})$. 
 The high-symmetry points in the Brillouin zone in the bulk are given by $\{ \Gamma=(0,0,0),\ Y=(\pi, 0, 0),\  Z=(0,\pi, 0),\  C=(\pi,\pi, 0),\  B=(0,0,\pi),\  A=(\pi,0,\pi),\  D=(0,\pi,\pi),\  E=(\pi, \pi, \pi) \}$, and the high-symmetry  points in the (100) surface Brillouin zone are $\{\bar{\Gamma}=(0,0),\ \bar{Y}=(\pi, 0),\  \bar{Z}=(0,\pi),\  \bar{M}=(\pi,\pi) \}$.
 In the calculations, we use two sets of parameter values 
$
 (m, t_x, t_y, t_z, v_x, v_y, v_z, v_{ab1}, v_{ab2})=(2,1,1,1,0.4,0.4,0.4,0.8,1.2)$,  $(6,1,1,1,0.4,0.4,0.6,0.8,1.2)
$
 in  Supplementary Figs.~\ref{fig:SM_second_TCI}(a) and ~\ref{fig:SM_second_TCI}(b), respectively. 
 We calculate band structures of $\mathcal{H}_{\rm TCI}^{(2)}(\boldsymbol{k})$ in the bulk [Supplementary Figs.~\ref{fig:SM_second_TCI}(a-1) and \ref{fig:SM_second_TCI}(b-1)] and in the slab geometry with the(100) surface [Supplementary Figs.~\ref{fig:SM_second_TCI}(a-2) and \ref{fig:SM_second_TCI}(b-2)]. In addition, Supplementary Figs.~\ref{fig:SM_second_TCI}(a-3) and \ref{fig:SM_second_TCI}(b-3) show the Wilson loop spectra of the occupied states of $\mathcal{H}_{\rm TCI}^{(2)}(\boldsymbol{k})$. The former shows that the doubly degenerate spectra wind along the $\bar{\Gamma}$-$\bar{Y}$ line, and the spectra in the $+$ glide sector along the $\bar{M}$-$\bar{Y}$ and $\bar{\Gamma}$-$\bar{Z}$ lines are not winding. From these results, we conclude that the $\mathbb{Z}_{4}$ topological invariant $\nu$ [see Eq.~(\ref{eq:z4_topological_invariant_wilson_loop})] is 2, and therefore Supplementary Fig.~\ref{fig:SM_second_TCI}(a) corresponds to the TCI phase protected by $G_y$ symmetry.  On the other hand, Supplementary Fig.~\ref{fig:SM_second_TCI}(b-3) shows that the Wilson loop spectra do not wind, namely, $\nu =0$, and therefore Supplementary Fig.~\ref{fig:SM_second_TCI}(b) corresponds to the trivial insulator.

Next, to see the dependence of the topological surface states on the surface orientation, we calculate band structures in the slab geometry with various $(\alpha 0 \gamma)$ surfaces [Supplementary Fig.~\ref{fig:SM_second_model_various_surface}]. As we discuss in the main text, the nonsymmorphic property of $G_y$ symmetry leads to the singular dependence of the surface states on the parity of $\gamma$.
On the (101), (201), (203) and (301) surfaces [Supplementary Figs.~\ref{fig:SM_second_model_various_surface}(a), \ref{fig:SM_second_model_various_surface}(c), \ref{fig:SM_second_model_various_surface}(d), and \ref{fig:SM_second_model_various_surface}(e)], the gapless surface states do not appear because the ($\alpha 0 \gamma$) surface is not glide symmetric when $\gamma \equiv 1$ mod 2. On the other hand, on the (102) and (302) surfaces  [Supplementary Figs.~\ref{fig:SM_second_model_various_surface}(b) and \ref{fig:SM_second_model_various_surface}(f)], the gapless surface states appear because the $(\alpha 0 \gamma)$ surface is glide symmetric when $\gamma \equiv 0$ mod 2. From these results, we conclude that the topological surface states in the second model of the TCI $\mathcal{H}_{\rm TCI}^{(2)}(\boldsymbol{k})$ have a similar dependence on the surface orientation to the first model of the TCI $\mathcal{H}_{\rm TCI}^{(1)}(\boldsymbol{k})$ in the main text. 

\subsection{Correspondence between hopping and fitting parameters}
In this section, we discuss correspondence between the hoppings of $\mathcal{H}_{\rm TCI}^{(2)}(\boldsymbol{k})$ and the fitting parameters $\Delta_{x}$, $\Delta_{z}$, $\Delta_{xz}$ of the trial function $F(\theta)$ to fit the surface energy in the main text. 
$\Delta_{x}$, $\Delta_{z}$, and $\Delta_{xz}$ correspond to the strength of the hopping parameters in the [100], [001], and [201] directions, respectively, as we discuss in \ref{Sup:surface_ene_chem_bond}. 
In our calculation on the two cases of the trivial insulator and the TCI, we used the same hopping parameters $t_{i}$, $v_{i}$ and $v_{abj}$ of $\mathcal{H}^{(2)}_{\rm TCI}(\boldsymbol{k})$ in the [100], [001], and [201] directions, as illustrated in Supplementary Fig.~\ref{fig:bond}(a). 
These hopping parameters along various directions are of the same order of magnitude.
Therefore, the fitting parameters $\Delta_{x}$, $\Delta_{z}$ and $\Delta_{xz}$, which physically correspond to the strengths of bondings along various directions, are expected to have the similar size. It is indeed the case in the trivial insulator [Supplementary Fig.~\ref{fig:bond}(b)].  
 On the other hand, in the case of the TCI,  such a feature does not appear [Supplementary Fig.~\ref{fig:bond}(c)]  
although we do not change the parameters of the hopping in the $x$ and $z$ directions. 
$\Delta_{xz}$ compared to $\Delta_{x}$  and $\Delta_{z}$ becomes much smaller compared to that of the trivial insulator. This unphysical behavior disappears in the case of the TCI without the SETS, and  the ratios between the values of $\Delta_{x}$, $\Delta_{z}$, and $\Delta_{xz}$ are close to the trivial insulator, compared to the TCI [Supplementary Fig.~\ref{fig:bond}(d)].  Thus, we conclude that the unique features of the surface energies of the TCI  are due to the SETS. 
The above analyses show that the crystal shape of the TCI is determined by the interplay between the SECB and the SETS. 

\begin{figure}[b]
\includegraphics[width=0.8\columnwidth]{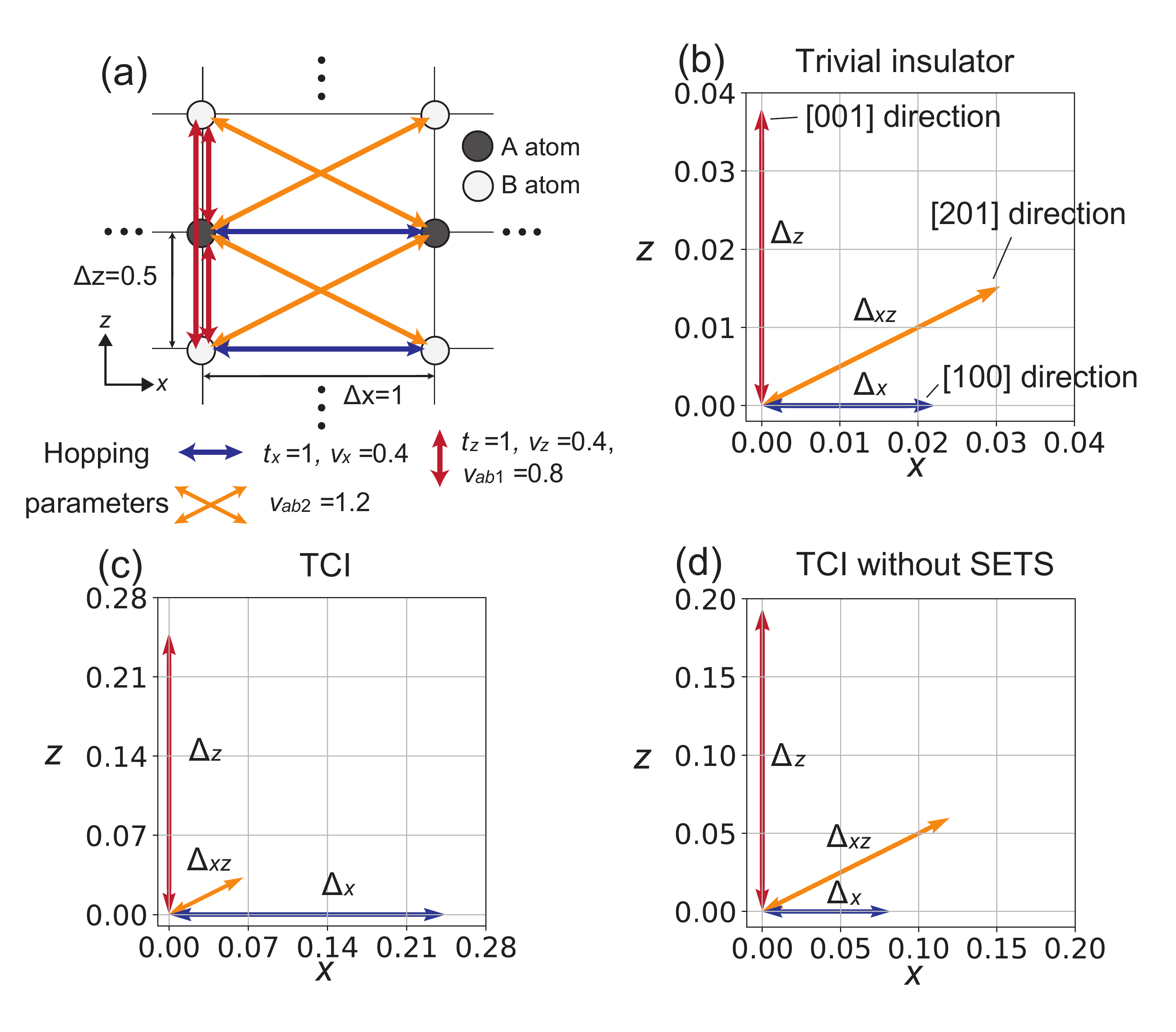}
\caption{Correspondence between hopping parameters and fitting parameters $\Delta_{x}$, $\Delta_{z}$, and $\Delta_{xz}$. (a) The bondings between the atoms of $\mathcal{H}_{\rm TCI}^{(2)}(\boldsymbol{k})$. (b-d) $\Delta_{x}$, $\Delta_{z}$, and $\Delta_{xz}$ in the [100], [001], and [201] directions, respectively, for (b) the trivial insulator, (c) the TCI, and (d) the TCI without SETS.}
\label{fig:bond}
\end{figure}

\clearpage

\section{Surface energy from chemical bonds}\label{Sup:surface_ene_chem_bond}
\begin{figure}[b]
	\includegraphics[width=0.8\columnwidth]{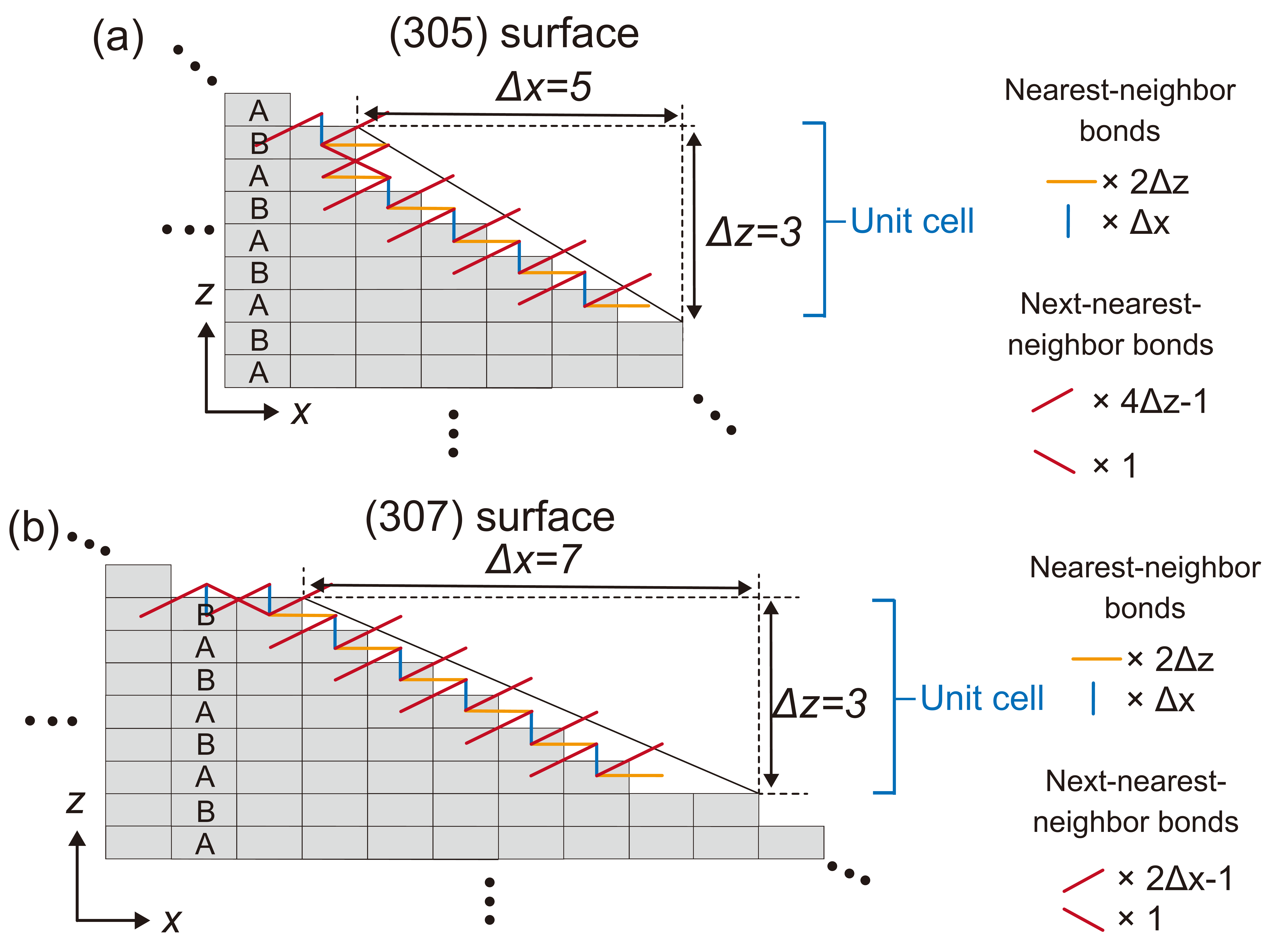}
	\caption{Examples of the steps for the ($\alpha 0 \gamma$) surfaces. The sizes of the unit cell in the $x$- and the $z$-direction are $\Delta x =\gamma$ and $\Delta z =\alpha$, respectively. (a) The configuration of the steps for the $(305)$ surface with $\gamma \leq 2 \alpha$.  (b) The configuration of the steps for the $(307)$ surface with $ 2 \alpha < \gamma$. In both cases, the numbers of nearest-neighbor bonds along the $x$- and the $z$-directions cut by the surface  are $2\Delta z$ and $\Delta x$, respectively. The total numbers of next-nearest-neighbor bonds cut by the surface are given by $4\Delta z$ in (a) and $2\Delta x$ in (b), respectively. } 
	\label{fig:SM_chemical_steps}
\end{figure}

In this section, we derive the trial function $F(\theta)$ for the surface energy  from the analysis in terms of the numbers of broken bonds between the atoms  when creating a surface from a bulk crystal. 
For this purpose, we consider a 3D network of atoms connected by bonds, which corresponds to the tight-binding models in this paper. 
The atoms in the netwrok are  on an orthorhombic lattice with $G_y =\{ M_y|00\tfrac{1}{2} \}$ symmetry, where we set the lattice constants to 1. 
The atoms are located at $(x,y,z)$ ($A$ sublattice) and at $(x,y,z+1/2)$ ($B$ sublattice) with $x,y,z \in \mathbb{Z}$.
We call the atomic layers at $z=n$ and at $z=n+1/2$ ($n\in \mathbb{Z}$) $A$ and $B$ layers, respectively. 
 The system has nearest-neighbor bonds along the $x$-, the $y$-, and the $z$-directions and next-nearest-neighbor bonds  along the $\vec{r}=(1,0,\pm 1/2), (-1,0,\pm 1/2)$ directions between the atoms. 
In the following, we discuss the surface energy of the ($\alpha 0 \gamma$) surface to derive the function $F(\theta)$ of the surface energy in the main text.
On the ($\alpha 0 \gamma$) surface,  the size of the unit cell along the $x$-direction is $\Delta x = \gamma$ and that along the $z$-direction is $\Delta z=\alpha$ [Supplementary Fig.~\ref{fig:SM_chemical_steps}]. 

First, we consider the nearest-neighbor bonds. When making the ($\alpha 0 \gamma$) surface from a bulk crystal, $2\alpha$ bonds between the atoms along the $x$-directions are broken, and $\gamma$ bonds along the $z$-directions are broken [Supplementary Figs.~\ref{fig:SM_chemical_steps}(a) and \ref{fig:SM_chemical_steps}(b)].  
Let the energy per a chemical bond along the $x$-direction and that along the $z$-direction be $\delta_x$ and $\delta_z$, respectively. 
Thus, the surface energy from the nearest-neighbor bonds for the $(\alpha 0 \gamma)$ surface can be expressed as
\begin{align}
	F^{(1)}(\theta)=&\frac{\delta_{x}}{2} \times \frac{2 \alpha}{S_{\rm cell}}+\frac{\delta_{z}}{2} \times \frac{\gamma}{S_{\rm cell}}\nonumber \\
	=&\delta_{x} \cos \theta + \frac{\delta_{z}}{2} \sin \theta,
\end{align}
where $\theta$ is an angle satisfying $\tan \theta =\gamma/ \alpha$, and $S_{\rm cell}=\sqrt{\alpha^2+\gamma^2}$ is the area of the unit cell. The reason $\delta_{i}$ ($i=x,z$) is multiplied by $1/2$ is that the process of cutting the bonds  from the bulk  result in appearance of two surfaces.  

Next, we discuss the surface energy from the next-nearest-neighbor bonds. Let us define the energy per  a  next-nearest-neighbor bond as $\delta_{xz}$.  In this case, we need to consider the following two cases: (i) $\gamma \leq 2 \alpha$ and (ii) $2\alpha < \gamma$. In the case of (i), when making the ($\alpha 0 \gamma$) surface from a bulk crystal,  the number of broken next-nearest-neighbor bonds is $4\Delta z = 4\alpha$ [Supplementary Fig.~\ref{fig:SM_chemical_steps}(a)].  Thus, the surface energy from the next-nearest-neighbor bonds can be expressed as
\begin{align}\label{SM:eq_F2_part1}
	F^{(2)}(\theta) &= \frac{\delta_{xz}}{2} \times  \frac{4 \alpha}{S_{\rm cell}} =2 \delta_{xz} \cos \theta \ \ \ (\gamma \leq 2 \alpha ).
\end{align}
Meanwhile, in the case of (ii) $2\alpha < \gamma$, when making the ($\alpha 0 \gamma$) surface from a bulk crystal,  the number of broken next-nearest-neighbor bonds  is $2\Delta x = 2\gamma$ [Supplementary Fig.~\ref{fig:SM_chemical_steps}(b)]. 
In this case, the surface energy from the next-nearest-neighbor bonds is given by
\begin{align}\label{SM:eq_F2_part2}
	F^{(2)}(\theta) &= \frac{\delta_{xz}}{2} \times  \frac{2 \gamma }{S_{\rm cell}} = \delta_{xz} \sin \theta \ \ \ (2\alpha  < \gamma ).
\end{align}
Furthermore, Eqs.~(\ref{SM:eq_F2_part1}) and (\ref{SM:eq_F2_part2}) can be rewritten in a unified from as follows:
\begin{equation} 
	F^{(2)} (\theta)=\frac{\sqrt{5}}{2} \delta_{xz} \Bigl( \sin |\theta-\theta_{102}|+  \sin |\theta+\theta_{102}| \Bigr),
\end{equation}
where $\theta_{102}$ is the angle satisfying $\cos \theta_{102}=1/\sqrt{5}$ and $\sin \theta_{102} =2/\sqrt{5}$. 
Thus, the total surface energy from the chemical bonds is given by
\begin{equation}
	F(\theta)=F^{(1)}(\theta)+F^{(2)}(\theta),
\end{equation}
and this function is the trial function $F(\theta)$ for the surface energy from chemical bonds in the main text. 

\section{Surface theory of the topological crystalline insulators protected by nonsymmorphic symmetry}\label{appendix:analysis_surface_hamiltonian}
\subsection{Effective surface theory from layer construction}
\begin{figure}
	\includegraphics[width=0.8\columnwidth]{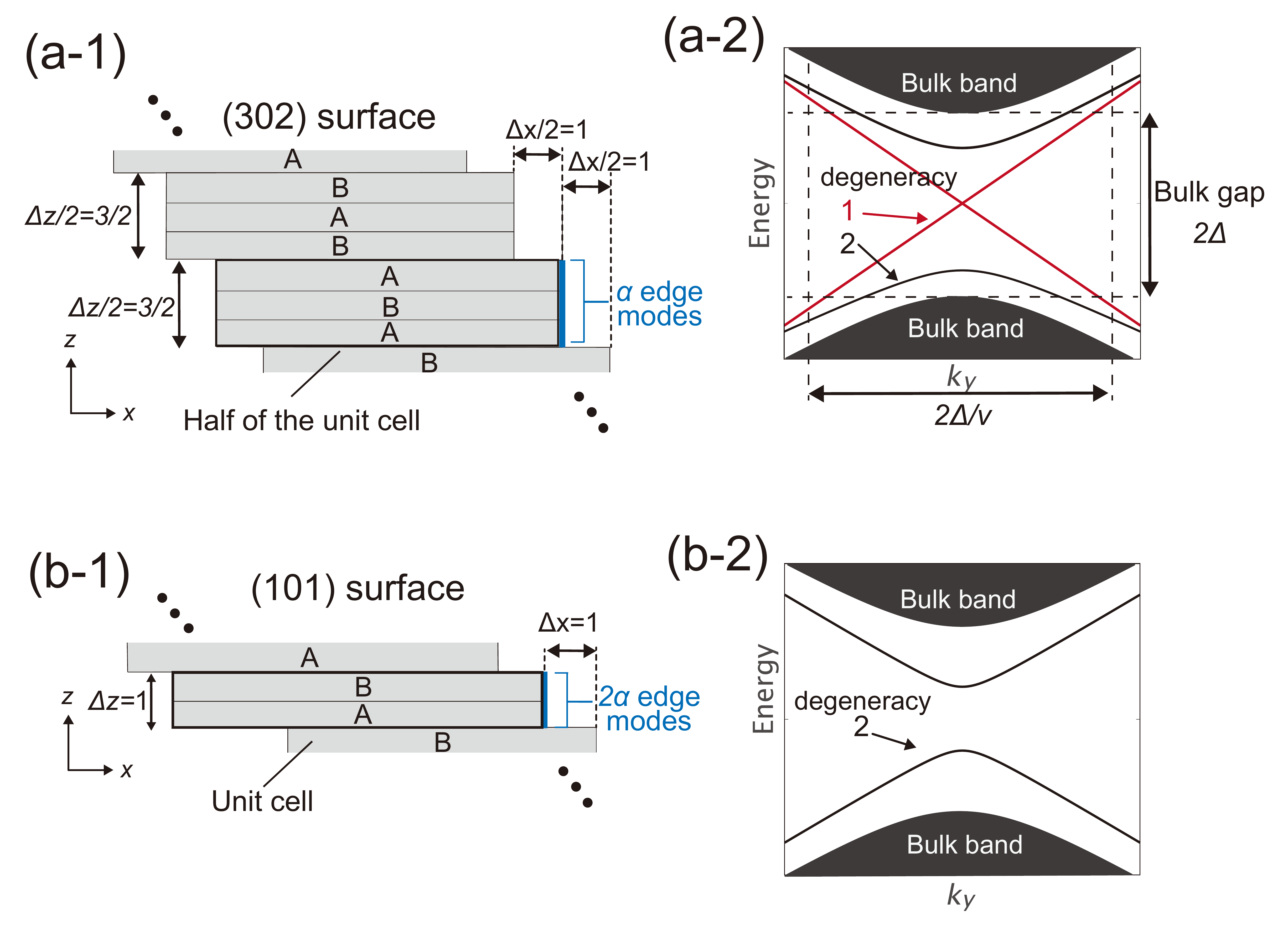}
	\caption{Examples of the steps and surface band structures (a) on the ($302$) surface and (b) on the (101) surface. (a-1,b-1) The steps of the surface consisting of the layers $L_{A}$ and $L_{B}$ when (a-1) $\gamma=2$ and (b-1) $\gamma=1$, respectively. (a-2) The surface bands of $\mathcal{H}^{(\gamma =1)}_{\alpha}(k_y)$ with $\alpha =3$. The degeneracy of the red bands is $1$ at $k_{y}\neq 0$, and the degeneracy of the other bands is $2$. (b-2) The surface bands of $\mathcal{H}^{(\gamma=2)}_{\alpha}(k_y)$ with $\alpha=1$.} 
	\label{fig:SM_LC_glide}
\end{figure}

To construct a surface theory of the TCI, we use  layer constructions \cite{song2018quantitative, PhysRevB.100.165202, kim2020glide, song2019topological}.
Here we introduce 2D topological insulator layers, which are periodically located and decoupled. This layer construction consists of two kinds of layers $L_{A}$ and $L_{B}$ [Supplementary Fig.~\ref{fig:SM_LC_glide}]. We introduce weak inter-layer couplings without closing gap, where the inter-layer couplings preserve the glide $G_{y}$ symmetry.
In the following, in order to study surface states as the ($\alpha 0 \gamma$) surface, we consider the helical edge modes along the $y$-axis from the layers $L_{A}$ and $L_{B}$ and their hybridization.  
Then the effective Hamiltonians describing the helical edge states for the layers $L_A$ and $L_B$ can be written as   $H_{A}=vk_{y}\sigma_{1}$, $H_{B}=-H_{A}$, and $\sigma_{j}\ (j=1, 2, 3)$  is Pauli matrices. 
We consider ($\alpha 0 \gamma$) surfaces with $\gamma=1, 2$ in order to see the difference between a $G_{y}$-symmetric ($\gamma=2$) case [Supplementary Fig.~\ref{fig:SM_LC_glide}(a)] and a $G_{y}$-asymmetric ($\gamma=1$) case [Supplementary Fig.~\ref{fig:SM_LC_glide}(b)]. 
In the case with $\gamma =2$, the effective Hamiltonian in the matrix representation can be expressed as
\begin{gather}
	\mathcal{H}^{(\gamma=2)}_{\alpha}(k_{y})=
	\underbrace{
		\begin{pmatrix}
			H_A & H_{\delta} &  & \\
			H_{\delta}^{\dagger} & H_B & \ddots  &  \\
			&  \ddots & \ddots & H_\delta \\
			& & H_{\delta}^{\dagger} & H_A
		\end{pmatrix}
	}_{\alpha \ {\rm blocks},\ {\rm matrix\ size}:\ 2\alpha}\nonumber \\
=\sum_{z=0}^{\frac{\alpha-1}{2}}\bigl[c^{\dagger}_{z}H_{A}c_{z}\bigr]+\sum_{z=0}^{\frac{\alpha-3}{2}}\bigl[ c^{\dagger}_{z+\frac{1}{2}}H_{B}c_{z+\frac{1}{2}}\bigr] +\sum_{z'=0, \frac{1}{2}, \cdots}^{\frac{\alpha-3}{2}}\bigl[ c^{\dagger}_{z'}H_{\delta}c_{z'+\frac{1}{2}} +{\rm h.c.}\bigr], 
\end{gather}
where $H_{\delta}=\delta \sigma_0$ ($\sigma_0$ is $2\times 2$ identity matrix), $\alpha$ $(>1)$ is an odd number, and $c_{z}=(c_{z \uparrow}, c_{z \downarrow})$ ($2z\in \mathbb{Z}$) is an annihilation operator for electrons on that layer within the plane at $z$. When $\alpha=1$ and $\gamma=2$, the surface Hamiltonian is simply written as $H_{A}$ or $H_{B}$, and therefore the helical edge modes of each layer do not hybridize in this case. 
Meanwhile, when $\gamma=1$, the effective surface Hamiltonian is
\begin{gather}
 \mathcal{H}^{(\gamma =1)}_{\alpha}(k_{y})=
	\underbrace{
		\begin{pmatrix}
			H_A & H_{\delta} &  & \\
			H_{\delta}^{\dagger} & H_B & \ddots  &  \\
			&  \ddots & \ddots & H_\delta \\
			& & H_{\delta}^{\dagger} & H_B
		\end{pmatrix}
	}_{2\alpha \ {\rm blocks},\ {\rm matrix\ size}:\ 4\alpha} \nonumber \\
	=\sum_{z=0}^{\alpha-1} \bigl[ c^{\dagger}_{z}H_{A}c_{z}+c^{\dagger}_{z+\frac{1}{2}}H_{B}c_{z+\frac{1}{2}} \bigr] +\sum_{z'=0, \frac{1}{2}, \cdots}^{\alpha-1}\bigl[ c^{\dagger}_{z'}H_{\delta}c_{z'+\frac{1}{2}} +{\rm h.c.}\bigr].
\end{gather}

First, we consider the case with $\gamma=2$. Let $\psi_{\alpha}$ denote an eigenstate satisfying
\begin{equation}\label{appeq:schodinger_gamma_even}
	\mathcal{H}^{(\gamma =2)}_{\alpha} (k_y) \psi_{\alpha}(k_y) = E^{(\gamma=2)}_{\alpha}  \psi_{\alpha}(k_y),
\end{equation}
with $
\psi_{\alpha}(k_y)=(\psi_{A1 \uparrow}, \psi_{A1 \downarrow}, \psi_{B2 \uparrow}, \psi_{B2 \downarrow},  \cdots, \psi_{A2j-1 \uparrow}, \psi_{A2j-1 \downarrow}, \psi_{B2j \uparrow}, \psi_{B2j \downarrow}  ,\cdots, \psi_{A \alpha \uparrow}, \psi_{A \alpha \downarrow})^{T}.$
We write the eigenstate as 
$
\psi_{A2j-1 \uparrow}=\phi_{A \uparrow} \sin \bigl( (2j-1)k \bigr),  \psi_{A2j-1 \downarrow}=\phi_{A \downarrow} \sin \bigl( (2j-1)k \bigr), \psi_{B2j\uparrow}=\phi_{B \uparrow} \sin (2jk),  \psi_{B2j \downarrow}=\phi_{B\downarrow} \sin (2jk)$. 
By substituting this eigenstate into Eq.~(\ref{appeq:schodinger_gamma_even}), we can obtain the following equations:
\begin{align}
	&vk_{y}\phi_{A\downarrow}+ 2\delta \phi_{B\uparrow} \cos k=E^{(\gamma=2)}_{\alpha} \phi_{A \uparrow}, \nonumber \\
	&vk_{y}\phi_{A\uparrow}+ 2\delta \phi_{B \downarrow} \cos k=E^{(\gamma=2)}_{\alpha} \phi_{A\downarrow},  \nonumber \\
	&2\delta \phi_{A\uparrow} \cos k -vk_y \phi_{B \downarrow}=E^{(\gamma=2)}_{\alpha} \phi_{B\uparrow}, \nonumber \\
	&2\delta \phi_{A\downarrow} \cos k -vk_y \phi_{B \uparrow}=E^{(\gamma=2)}_{\alpha} \phi_{B \downarrow}.
\end{align}
Therefore, by solving these coupled equations, we can obtain the eigenvalue 
\begin{equation}
E^{(\gamma=2)}_{\alpha}=\pm \sqrt{v^2 k^{2}_{y}+4\delta^2 \cos^2 k}.
\end{equation}
From the boundary condition, we obtain
\begin{equation}
	k=\frac{\pi n}{\alpha +1},\ n=1,2,\cdots (\alpha+1)/2.
\end{equation}
Therefore, the eigenvalue is written as
$
E^{(\gamma=2)}_{\alpha}=\pm \sqrt{v^{2}k^{2}_{y}+(m^{(\gamma=2)}_{\alpha}(n))^2},
$
where $m^{(\gamma=2)}_{\alpha}(n)$ is a Dirac mass term given by 
\begin{equation}\label{appendix_Diracmass_equaiton_gamma_even}
	m^{(\gamma=2)}_{\alpha}(n)=2\delta \cos \left( \frac{\pi n}{\alpha +1}\right),
\end{equation}
and  $n$ runs over $n=1,2,\cdots  (\alpha+1)/2$.
When $\gamma=1$, in a similar way, we can obtain the eigenvalue $E^{(\gamma=1)}_{\alpha}$ of the Hamiltonian $\mathcal{H}_{\alpha}^{(\gamma=1)}(\boldsymbol{k})$  as
$
E^{(\gamma=1)}_{\alpha}=\pm \sqrt{v^{2}k^{2}_{y}+(m^{(\gamma=1)}_{\alpha}(n))^2},
$
where 
\begin{equation}\label{appendix_Diracmass_equaiton_gamma_odd}
	m^{(\gamma=1)}_{\alpha}(n)=2\delta \cos \left( \frac{\pi n}{2\alpha +1}\right),
\end{equation}
with $n=1,2,\cdots , \alpha$.  The difference between Eq.~(\ref{appendix_Diracmass_equaiton_gamma_even}) and Eq.~(\ref{appendix_Diracmass_equaiton_gamma_odd}) comes from the different boundary conditions.  

For the Hamiltonian $\mathcal{H}^{(\gamma =2)}_{\alpha}(\boldsymbol{k})$, the bands with $n=1,2,\cdots (\alpha-1)/2$ are doubly degenerate. On the other hand, the band with $n=(\alpha+1)/2$ is not degenerate at $k_{y}\neq 0$, and these bands are gapless. 
Let 2$\Delta$ denote the bulk gap, and we assume that the surface energy can be determined by the bands in the $k$-space with $-\Delta /v \leq k_{y} < \Delta /v$ and $-\pi \leq k_{z} < \pi$.
Here, the energy is independent of $k_{z}$ because the gapless states are due to the modes along the edges of the $xy$ layers. 
Therefore, the total energy  with $\gamma=2$ can be expressed as 
\begin{align}\label{E_dirac_gamma_even}
	E^{(\gamma =2)}_{\rm Dirac, \alpha}=\frac{v}{ 2\Delta }  \int^{\frac{\Delta}{v}}_{-\frac{\Delta}{v}}dk_{y} 
	\biggl[  -v|k_{y}|+ \Delta 
	+  2\sum_{n=1}^{\frac{\alpha-1}{2}}  \Bigl( -\sqrt{v^{2}k^{2}_{y}+(m^{(\gamma =2)}_{\alpha }(n))^2}+ \Delta  \Bigr) \biggr],
\end{align}
where the energy is measured from $-\Delta$. 
On the other hand, when $\gamma=1$, all the bands are doubly degenerate, and gapless states do not appear.  Therefore, the surface energy in the unit cell can be written as 
\begin{align}\label{E_dirac_gamma_odd}
	E^{(\gamma=1)}_{\rm Dirac, \alpha}=\frac{v}{ \Delta } \sum_{n=1}^{\alpha}  \int^{\frac{\Delta}{v}}_{-\frac{\Delta}{v}}dk_{y}  \biggl[  -\sqrt{v^{2}k^{2}_{y}+(m^{(\gamma=1)}_{\alpha }(n))^2}+ \Delta  \biggr]. 
\end{align} 
From these surface energies $E_{\rm Dirac, \alpha}^{(\gamma=1)}$ and $E_{\rm Dirac, \alpha}^{(\gamma=2)}$, we can obtain  the surface energy for the $(\alpha 0 \gamma)$ surface with $\gamma=1,2 $ per unit area as follows:
\begin{align}\label{Dirac_eq_alpha_gamma} 
	E^{(\alpha 0 \gamma)}_{{\rm Dirac}}=
	\begin{cases}
		2E^{(\gamma =2)}_{\rm Dirac}/S_{\rm cell}& (\gamma =2) \\
		E^{(\gamma=1)}_{\rm Dirac}/S_{\rm cell}& (\gamma =1), 
	\end{cases}
\end{align}
where $S_{\rm cell}$ is the area of the surface in the unit cell.

\subsection{Effective surface theory with $\gamma \neq 1,2$}

In the above discussion, we study the effective surface theory on the ($\alpha 0 \gamma$) surfaces with $\gamma=1,2$. Next, we consider the cases with $\gamma \neq 1,2$. We construct the surface theory on the ($\alpha 0 \gamma$) surfaces with $\gamma \neq 1,2$ by combining the cases with $\gamma =1,2$.  Here atoms occupy all the positions $(x,y,z)$ and $(x,y,z+1/2)$ with $x,y,z \in \mathbb{Z}$  below the plane $\boldsymbol{n}\cdot \boldsymbol{r}=0$, where $\boldsymbol{n}=(\alpha /\sqrt{\alpha^2+\gamma^2},0,\gamma/\sqrt{\alpha^2+\gamma^2})^{T}$ and $\boldsymbol{r}=(x,y,z)^{T}$. Here, we set  the origin being on this plane. To obtain the surface energy $E^{(\alpha 0 \gamma)}_{\rm Dirac}$, we need to consider the following two cases: (i) $\gamma \leq 2 \alpha$ and (ii) $ 2 \alpha < \gamma$.

\begin{figure}
	\includegraphics[width=0.9\columnwidth]{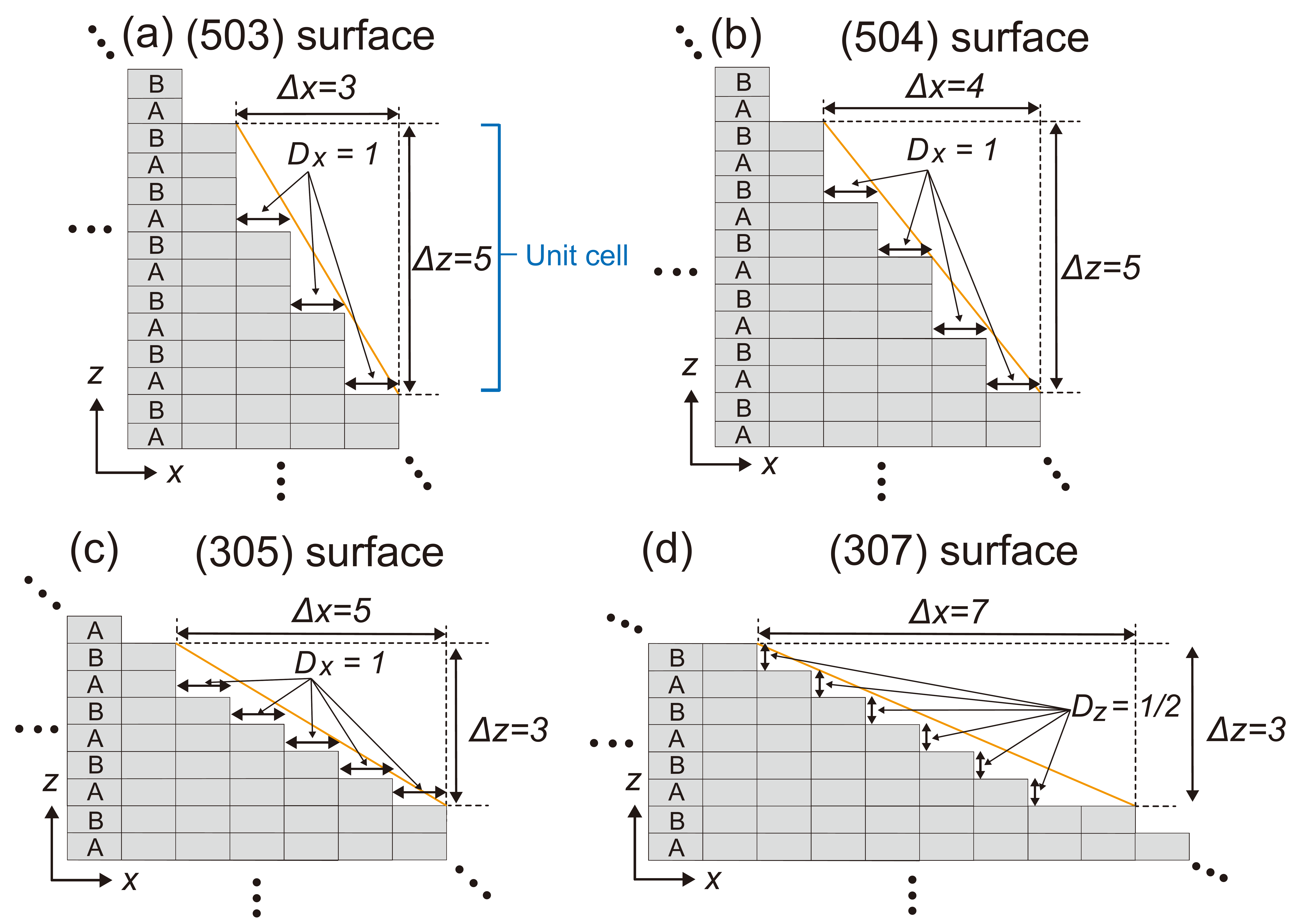}
	\caption{Examples of the steps for various surfaces. (a-c) Three examples of the configurations of the steps with $\gamma \leq 2 \alpha$. All the distances between the adjacent steps in the $x$ direction are $D_{x}=1$.  (d) An example of the configuration of the steps with $ 2 \alpha < \gamma$. All the distances between the adjacent steps in the $z$ direction are $D_{z}=1/2$. } 
	\label{fig:shorter_steps}
\end{figure}

First, we consider (i) $\gamma \leq 2 \alpha$ case. In this case, any distance between the adjacent steps in the $x$ direction is $D_{x}=1$  [Supplementary Figs.~\ref{fig:shorter_steps}(a-c)]. 
 Because  the inclination of the plane $\boldsymbol{n}\cdot \boldsymbol{r}=0$ is $-\alpha/\gamma$ ($\leq -1/2$) and $D_x=1$, the distances between the adjacent steps in the $z$ direction are $N_{z}/2$ or $(N_{z}+1)/2$, where $N_{z}$ is an integer satisfying $N_{z}/2 \leq \alpha/ \gamma < (N_{z}+1)/2$.  
Let  $N_{1}$ and $N_{2}$ denote the number of steps with the interval $D_{z}=N_{z}/2$ in the $z$ direction and that with the interval $D_{z}=(N_{z}+1)/2$, respectively. 
The surface in the unit cell has $N_1$ steps with $D_{z}=N_z / 2$ and $N_2$ steps with $D_{z}=(N_{z}+1)/2$.  
When $N_z$ is an even number, the energy of the steps with $D_{z}=N_z / 2$ can be expressed as
$E^{(\gamma=1)}_{{\rm Dirac, }\alpha=N_z /2}$, and the energy of the steps with $D_{z}=(N_z +1)/2$ is obtained by  $E^{(\gamma=2)}_{{\rm Dirac, }\alpha=N_z+1}$. In this case, the surface energy per unit area can be expressed as follows:
\begin{equation} 
	E^{(\alpha 0 \gamma )}_{{\rm Dirac}}=
	\frac{1}{S_{\rm cell}}\biggl[ N_{1}E^{(\gamma=1)}_{{\rm Dirac}, \frac{N_z}{2}}+N_{2}E^{(\gamma=2)}_{{\rm Dirac}, N_z +1}\biggr] \ \ (N_{z} \equiv 0\ {\rm mod}\ 2,\ \gamma \leq 2 \alpha) .
\end{equation}
On the other hand, when $N_{z}$ is an odd number, the energy of the steps with $D_{z}=N_z / 2$ can be expressed as
$E^{(\gamma=2)}_{{\rm Dirac, }\alpha=N_z}$, and the energy of the steps with $D_{z}=(N_z +1)/2$ is given by  $E^{(\gamma=1)}_{{\rm Dirac, }\alpha=(N_z+1)/2}$. In this case, we can write the surface energy as
\begin{equation}
	E^{(\alpha 0 \gamma)}_{{\rm Dirac}}=\frac{1}{S_{\rm cell}}\biggl[ N_1 E^{(\gamma=2)}_{{\rm Dirac, }N_z}+N_{2} E^{(\gamma=1)}_{{\rm Dirac, } \frac{N_z+1}{2}}  \biggr]  \ \ (N_{z} \equiv 1\ {\rm mod}\ 2, \ \gamma \leq 2 \alpha),
\end{equation}
where $S_{\rm cell}=\sqrt{\alpha^2 + \gamma^2}$ is the area of the unit cell.
Because the sum of the intervals $D_{x}$ of the steps in the unit cell is $\gamma$,  we get $N_{1}+N_{2}=\gamma$. In addition, from the similar condition in the $z$ direction, we can  obtain
$
	\frac{N_{z}}{2}N_{1}+\frac{N_{z}+1}{2}N_{2}=\alpha.
$
Thus, $N_1$ and $N_2$ are given by
\begin{gather}
	N_{1}=\gamma (N_z +1)- 2\alpha, \\
	 \ N_2 = 2\alpha -N_z \gamma,
\end{gather}
respectively. 

Next, we consider the case of (ii) $2\alpha < \gamma$ [Supplementary Fig.~\ref{fig:shorter_steps}(d)]. In this case, any distance between the adjacent steps in the $z$ direction is $D_{z}=1/2$ because of the inclination $-\alpha/\gamma$ ($>-1/2$).  The number of steps in the unit cell is $2\alpha$, and the energy of each step is  $E^{(\gamma =2)}_{{\rm Dirac}, 1}$. Therefore, the surface energy per unit area is given by 
\begin{equation} 
	E^{(\alpha 0 \gamma)}_{\rm Dirac}=\frac{2\alpha}{S_{\rm cell}} E^{(\gamma=2)}_{{\rm Dirac}, 1}\ \  (2 \alpha < \gamma).
\end{equation}
Thus, in both (i) $\gamma \leq 2 \alpha$ and (ii) $ 2 \alpha < \gamma$ cases, we can obtain the surface energy  on the ($\alpha 0 \gamma$) surface when $\gamma \neq 1,2$ by combining the surface energy with $\gamma = 1,2$.  

\section{Effects of surface reconstruction}
\begin{figure}[b]
	\includegraphics[width=0.8\columnwidth]{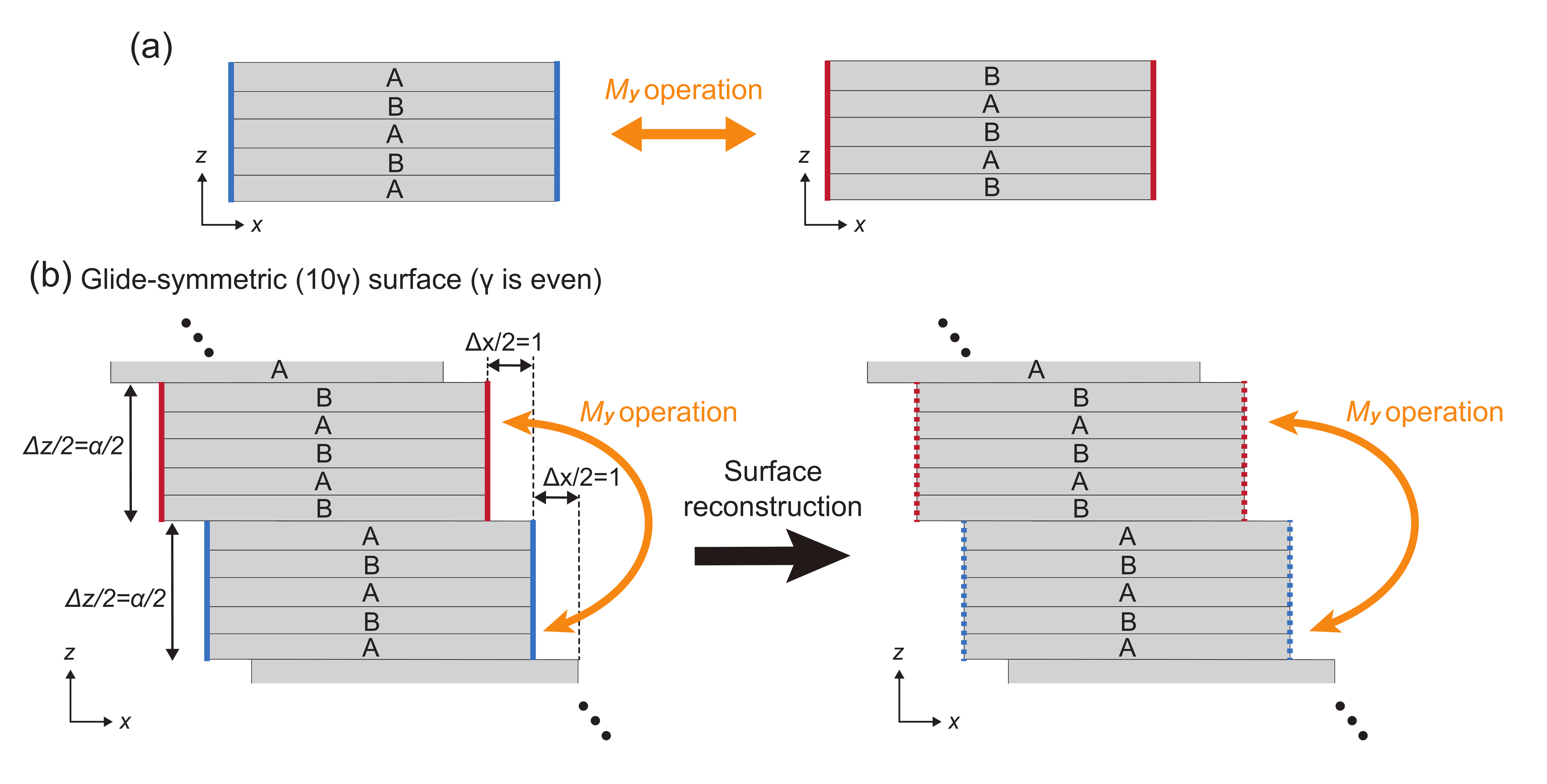}
	\caption{Surface reconstruction of the TCI constructed by  stacking two kinds of 2D TI layers,   $L_{A}$ and $L_{B}$. (a) Transformation between $L_{A}$ and $L_{B}$ under the mirror $M_{y}$ operation. (b) Glide-symmetric ($10\gamma$) surface with $\gamma$ being even and its reconstruction. } \label{fig:surface_reconst}
\end{figure}

In this section, we discuss the effects of surface reconstructions on our theory.
In general, surface reconstructions can occur, which will affect the surface electronic states, surface energies and crystal shapes. Since the surface reconstructions largely depend on details of the materials, we cannot develop a quantitative theory for general systems with surface reconstructions. 
Meanwhile, because our theory is based on the topology under glide symmetry, it remains valid if the glide symmetry is preserved after surface reconstruction. Meanwhile, as we discuss in the following, the glide symmetry is likely to be preserved through surface reconstruction. 
In order to discuss this point, we consider a layer construction, where 2D TI layers are periodically located along the out-of-plane direction. 
This consists of two kinds of layers $L_{A}$ and $L_{B}$, where $L_{A}$ and $L_{B}$ are interchanged by $M_{y}$ [Supplementary Fig.~\ref{fig:surface_reconst}(a)]. 
The layers $L_{A}$ are at $z=n$, and the layers $L_{B}$ are  at $z=n+\tfrac{1}{2}$ ($n$: integer).
Then we focus on  a glide-symmetric surface, a ($1 0 \gamma$) surface with $\gamma$ being an even number.  
When $\gamma$ is even, the ($10\gamma$) surface can be  divided into the two steps, and one consists of the layers  $L_{A}$ $L_{B}$ $\cdots$ $L_{A}$ and the other consists of the layers $L_{B}$ $L_{A}$ $\cdots$ $L_{B}$ [Supplementary Fig.~\ref{fig:surface_reconst}(b)]. 
The layers $L_{A}$ $L_{B}$ $\cdots$ $L_{A}$ and the layers $L_{B}$ $L_{A}$ $\cdots$ $L_{B}$ are interchanged under the $M_{y}$ operation. Therefore, the configuration of the layers is invariant under a combination between a translation $(-{1}/{2},0,{\alpha}/{2})$ and the $M_{y}$ mirror, which means that this surface has $G_{y}$ symmetry. When the surface reconstruction occurs, the reconstruction of $L_{A}$ $L_{B}$ $\cdots$ $L_{A}$ and $L_{B}$ $L_{A}$ $\cdots$ $L_{B}$ is likely to occur in such a way that they are mapped to each other by the $M_{y}$ operation [Supplementary Fig.~\ref{fig:surface_reconst}(c)]. Thus, the $(10\gamma)$ surface is likely to remain $G_{y}$ symmetric after the reconstruction, and the topological gapless states  remain. 
From this analysis, we can conclude that the topological gapless states are likely to survive surface reconstructions.

\clearpage

%